\documentclass[hyper]{prop2015}
\usepackage[all]{xy}
\usepackage{slashbox}
\usepackage{rotating}
\usepackage{multirow}
\usepackage[only,sslash]{stmaryrd}
\usepackage{array}
\usepackage{bbm}

\DeclareRobustCommand
  \joinrelaz{\mathrel{\mkern-3mu}}

\newcolumntype{L}[1]{>{\raggedright\let\newline\\\arraybackslash\hspace{0pt}}m{#1}}
\newcolumntype{C}[1]{>{\centering\let\newline\\\arraybackslash\hspace{0pt}}m{#1}}
\newcolumntype{R}[1]{>{\raggedleft\let\newline\\\arraybackslash\hspace{0pt}}m{#1}}

\newcommand{\EM}[1]{\mathfrak{l}K(\mathbbm{Z},#1)}

\category{Proceedings}
\keywords{M-theory, $p$-branes, supersymmetry, T-duality, gauge enhancement,
higher structures, rational homotopy theory, equivariant homotopy theory, cohomotopy}
\title{The Rational Higher Structure of M-theory}
\subtitle{\href{http://www.maths.dur.ac.uk/lms/109/index.html}{LMS/EPSRC Durham Symposium on Higher Structures in M-Theory}}
\author[D.~Fiorenza]{Domenico Fiorenza\inst{a}}
\author[H.~Sati]{Hisham Sati\inst{b}}
\author[U.~Schreiber]{Urs Schreiber\inst{c,}\footnote{Corresponding author e-mail: \href{mailto:urs.schreiber@googlemail.com}{\textsf{urs.schreiber@googlemail.com}}}}
\address[1]{Dipartimento di Matematica, La Sapienza Universita di Roma, Piazzale Aldo Moro 2, 00185 Rome, Italy}
\address[2]{Mathematics, Division of Science, New York University Abu Dhabi, United Arab Emirates}
\address[3]{Mathematics, Division of Science, New York University Abu Dhabi, United Arab Emirates, on leave from Czech Academy of Science}
\shortauthors{D.~Fiorenza, H.~Sati, U.~Schreiber}

\begin{abstract}
  We review how core structures of string/M-theory emerge
  as \emph{higher structures} in super homotopy theory;
  namely from systematic analysis of the \emph{brane bouquet}
  of universal invariant higher central extensions
  growing out of the superpoint.
  Since super homotopy theory is immensely rich,
  to start with we consider this
  in the rational/infinitesimal approximation
  which
  ignores torsion-subgroups in brane charges and
  focuses on tangent spaces of super space-time.
  Already at this level, super homotopy theory
  discovers all super $p$-brane species,
  their intersection laws, their M/IIA-, T- and S-duality relations,
  their black brane avatars at ADE-singularities, including their instanton contributions, and, last not least, Dirac charge quantization:
  for the D-branes it recovers twisted K-theory, rationally, but for the M-branes it gives \emph{cohomotopy cohomology theory}. We close
  with an outlook on the lift of these results beyond the
  rational/infinitesimal approximation to a candidate formalization
  of microscopic M-theory in super homotopy theory.
\end{abstract}
\shortabstract

\begin{document}

\maketitle

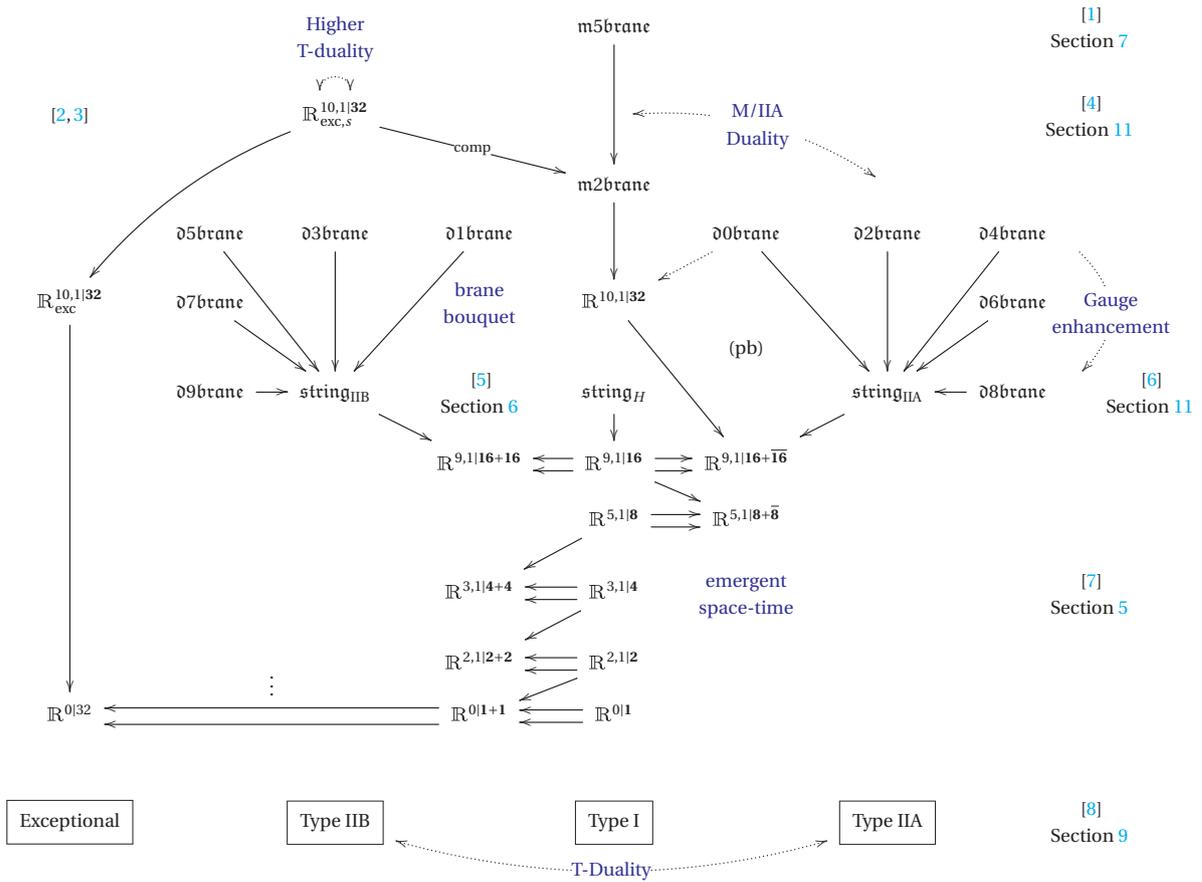
\begin{figure*}[htb]

\vspace{-.5cm}

\begin{center}
\hspace{-.4cm}
\scalebox{.78}{
{\small
  \xymatrix@=1em@R=5pt{
    &
    &&&& \mathfrak{m}5\mathfrak{brane}
     \ar[dd]
    &&&&
    \mathclap{
      \mbox{
        \begin{tabular}{c}
          \cite{Fiorenza:2015gla}
          \\
          Section \ref{ChargeQuantization}
        \end{tabular}
      }
    }
    \\
    &
    \mbox{ \cite{Fiorenza:2018ekd,Sati:2018tvj} }
    &&
    \mathbbm{R}^{10,1\vert \mathbf{32}}_{\mathrm{exc},s}
    \ar[rrd]|-{\mathrm{comp}}
    \ar@/_1.1pc/[dddll]
    \ar@{<..>}@(ul,ur)[]^{ \mbox{ \color{blue} \begin{tabular}{c} Higher \\ T-duality \end{tabular}  } }
    &&
    \;\;\;\;
    \ar@{<..>}@/^1.1pc/[drr]|{
      \mbox{
        \color{blue}
        \begin{tabular}{c}
          M/IIA
          \\
          Duality
        \end{tabular}
      }
    }
    &&
    &&
    \mathclap{
      \mbox{
        \begin{tabular}{c}
          \cite{Fiorenza:2016ypo}
          \\
          Section \ref{MIIADuality}
        \end{tabular}
      }
    }
    \\
    &
    &&
     && \mathfrak{m}2\mathfrak{brane}
    \ar[dd]
    &&
    &
    \\
    &
    &
    \mathfrak{d}5\mathfrak{brane}
    \ar[ddr]
    &
    \mathfrak{d}3\mathfrak{brane}
    \ar[dd]
    &
    \mathfrak{d}1\mathfrak{brane}
    \ar[ddl]
    &
    & \mathfrak{d}0\mathfrak{brane}
    \ar@{}[ddd]|{\mbox{(pb)}}
    \ar[ddr]
    \ar@{..>}[dl]
    &
    \mathfrak{d}2\mathfrak{brane}
    \ar[dd]
    &
    \mathfrak{d}4\mathfrak{brane}
    \ar[ddl]
    \ar@<+22pt>@{..>}@/^2.1pc/[dd]|{ \mbox{ \color{blue} \begin{tabular}{c}  Gauge \\  enhancement \\  \end{tabular} } }
    \\
    &
    \mathbbm{R}^{10,1\vert \mathbf{32}}_{\mathrm{exc}}
    \ar[dddddd]
    &
    \mathfrak{d}7\mathfrak{brane}
    \ar[dr]
    &
    &
    \mbox{
      \color{blue}
      \begin{tabular}{c}
        brane \\ bouquet
      \end{tabular}
    }
    & \mathbbm{R}^{10,1\vert \mathbf{32}}
      \ar[ddr]
    &&&
    \mathfrak{d}6\mathfrak{brane}
    \ar[dl]
    &
    \\
    &
    &
    \mathfrak{d}9\mathfrak{brane}
    \ar[r]
    &
    \mathfrak{string}_{\mathrm{IIB}}
    \ar[dr]
    &
    \mathclap{
      \mbox{
        \begin{tabular}{c}
          \cite{Fiorenza:2013nha}
          \\
          Section \ref{TheBraneBouquet}
        \end{tabular}
      }
    }
    & \mathfrak{string}_{H}
      \ar[d]
    &&
    \mathfrak{string}_{\mathrm{IIA}}
    \ar[dl]
    &
    \mathfrak{d}8\mathfrak{brane}
    \ar[l]
    &
    \mathrlap{
      \mbox{
        \begin{tabular}{c}
          \cite{Braunack-Mayer:2018uyy}
          \\
          Section \ref{MIIADuality}
        \end{tabular}
      }
    }
    \\
    &
    &
    &
    &
    \mathbbm{R}^{9,1 \vert \mathbf{16} + {\mathbf{16}}}
    \ar@{<-}@<-3pt>[r]
    \ar@{<-}@<+3pt>[r]
    & \mathbbm{R}^{9,1\vert \mathbf{16}} \ar[dr]
    \ar@<-3pt>[r]
    \ar@<+3pt>[r]
    &
    \mathbbm{R}^{9,1\vert \mathbf{16} + \overline{\mathbf{16}}}
    \\
    &
    &
    &
    &
    &
    \mathbbm{R}^{5,1\vert \mathbf{8}}
    \ar[dl]
    &
    \mathbbm{R}^{5,1 \vert \mathbf{8} + \overline{\mathbf{8}}}
    \ar@{<-}@<-3pt>[l]
    \ar@{<-}@<+3pt>[l]
    \\
    &
    &
    &
    &
    \mathbbm{R}^{3,1\vert \mathbf{4}+ \mathbf{4}}
    \ar@{<-}@<-3pt>[r]
    \ar@{<-}@<+3pt>[r]
    &
    \mathbbm{R}^{3,1\vert \mathbf{4}}
    \ar[dl]
    &
    \!\!\!\!\!\!\!
    \mbox{
      \color{blue}
      \begin{tabular}{c}
        emergent
        \\
        space-time
      \end{tabular}
    }
    \!\!\!\!\!\!\!
    &
    &&
    \mathclap{
      \mbox{
        \begin{tabular}{c}
          \cite{Huerta:2017utu}
          \\
          Section \ref{EmergentSuperspace-time}
        \end{tabular}
      }
    }
    \\
    &
    &
    &
    &
    \mathbbm{R}^{2,1 \vert \mathbf{2} + \mathbf{2} }
    \ar@{<-}@<-3pt>[r]
    \ar@{<-}@<+3pt>[r]
    &
    \mathbbm{R}^{2,1 \vert \mathbf{2}}
    \ar[dl]
    \\
    &
    \mathbbm{R}^{0\vert 32}
    \ar@{<-}@<-4pt>[rrr]
    \ar@{<-}@<+4pt>[rrr]^-{\vdots}
    &
    &
    &
    \mathbbm{R}^{0 \vert \mathbf{1}+ \mathbf{1}}
    \ar@{<-}@<-3pt>[r]
    \ar@{<-}@<+3pt>[r]
    &
    \mathbbm{R}^{0\vert \mathbf{1}}
    \\
    \\
    & \fbox{Exceptional} && \fbox{Type IIB} \ar@{<..>}@/_2pc/[rrrr]|{\mbox{\color{blue}T-Duality}} && \fbox{Type I} && \fbox{Type IIA}
    &&
    \mathclap{
      \mbox{
        \begin{tabular}{c}
          \cite{Fiorenza:2016oki}
          \\
          Section \ref{SuperTopologicalTDuality}
        \end{tabular}
      }
    }
  }
}
}
\end{center}

\vspace{-.9cm}

\caption{ {\bf The new brane bouquet.} Entries correspond to
\emph{higher central super $L_\infty$-extensions} which are classified
by invariant \emph{higher} cocycles (see Figure \ref{SomeHigherLieTheory}). On these higher extensions
(such as the $\mathfrak{m}2\mathfrak{brane}$ super Lie 3-algebra)
new $L_\infty$-cocycles appear (such as that for the super M5-brane)
which were missing from the old brane scan (Table \ref{TheOldBraneScan}).
We discuss review process in Section \ref{TheBraneBouquet}.
}
\label{Figure1}

\vspace{-.4cm}

\end{figure*}

\section{Introduction}

\noindent {\bf The open problem.}
The core open problem in string theory is still
\cite[Sec. 12]{Moore:2014aaa} the actual formulation
of the full non-perturbative theory -- ``M-theory''
\cite{Duff:1999baa} (see \cite[Sec. 2]{Huerta:2018xyh} for exposition).
While \emph{perturbative} string theory \cite{Green:1987sp,Polchinski:1998rq}
follows a clear principle
(perturbative scattering matrices in worldline formalism \cite{Schmidt:1994da,Schubert:1996jj} deformed to worldsheets, see \cite{Witten:2015ab} for exposition),
the heart of the problem of formulating non-perturbative M-theory
is that
even the underlying \emph{principles} have been unclear:
we are probably not looking either for a Langrangian density, nor
for a scattering matrix, nor for any other traditional structure in
quantum physics. How to proceed?

\begin{table}[htb]
\begin{center}
\scalebox{.8}{
\begin{tabular}{|c||c|c|c|c|c|c|cc|}
  \hline
\backslashbox{~~~$d+1$}{\raisebox{-9pt}{$\!\!\!\!\!\!\!\!p$}}
     &   & $1$ & $2$ & $3$ & $4$ & $5$ & $6$ & $\cdots$
	 \\[5pt]
	 \hline \hline
	 $\!\!\!\!\!\!\!\!\!\!\!\!\!\!\!\!\!\!\!\!\!10+1$ &   & &
	  $\mu_{{}_{M2}}$  & & & &&
	 \\[5pt]
	 \hline
	 $\!\!\!\!\!\!\!\!\!\!\!\!\!\!\!\!\!\!\!\!9+1$ &  		
	    &
        $\mu_{{}_{F1}}^{H/I}$
		&
		&
        $ \phantom{ {A \atop A} \atop {A \atop A}} $
		&
		&
        $\mu^{H/I}_{{}_{\mathrm{NS5}}}$ &&
	 \\[5pt]
	 \hline
	 $\!\!\!\!\!\!\!\!\!\!\!\!\!\!\!\!\!\!\!\!8 + 1$ & & & & & $\ast$ & &  &
	 \\[5pt]
	 \hline
	 $\!\!\!\!\!\!\!\!\!\!\!\!\!\!\!\!\!\!\!\!7 + 1$  &  & & & $\ast$ & & & &
	 \\[5pt]
	 \hline
	 $\!\!\!\!\!\!\!\!\!\!\!\!\!\!\!\!\!\!\!\!6 + 1$  &
     & & $\ast$ & & & & &
	 \\[5pt]
	 \hline
	 $\!\!\!\!\!\!\!\!\!\!\!\!\!\!\!\!\!\!\!\!5 + 1$  &  & $\ast$
		 & & $\ast$ & &&&
	 \\[5pt]
	 \hline
	 $\!\!\!\!\!\!\!\!\!\!\!\!\!\!\!\!\!\!\!\!4 + 1$ &  &  & $\ast$
&&& & &
	 \\[5pt]
	 \hline
	 $\!\!\!\!\!\!\!\!\!\!\!\!\!\!\!\!\!\!\!\!3 + 1$  & & $\ast$ & $\mu_{M2}^{D=4}$ && &&&
	 \\[5pt]
	 \hline
	 $\!\!\!\!\!\!\!\!\!\!\!\!\!\!\!\!\!\!\!\!2 + 1$  & $ \phantom{ {A \atop A} \atop {A \atop A}} $ & $\mu_{{}_{F1}}^{D=3}$ & &&&&&
     \\
     \hline
  \end{tabular}
  }
\end{center}

\vspace{-.6cm}

\caption{
\noindent {\bf The old brane scan.}
Entries represent the generators of the $\mathrm{Spin}(d,1)$-invariant
super Lie algebra cohomology groups
$H^{p+2}_{\mathrm{Lie}}\big( \mathbbm{R}^{d,1\vert \mathcal{N}=1} \big)^{\mathrm{Spin}(d,1)}$ of
super Minkowski space-times. These coincide
with the $\kappa$-symmetric WZW-term curvatures defining the
Green--Schwarz sigma models for super $p$-branes on these space-times
(``$\ast$'' indicates branes without a standard name).
\newline
However, the super D$p$-branes and the M-theory $p=5$-brane
 do not show up; and the web of dualities is not visible.
}
\label{TheOldBraneScan}
\end{table}

\smallskip
\noindent {\bf The old brane scan.}
  A suggestive insight came from investigation of the manifestly
  space-time supersymmetric formulation of the superstring in $D =10$ \cite{Green:1983wt} (see \cite[Sec. 5]{Green:1987sp}) and its generalization to other branes and space-time dimensions \cite{Bergshoeff:1985su,Bergshoeff:1987cm} (see \cite{Sorokin:1999jx}).
  These \emph{Green--Schwarz-type sigma models for super $p$-branes} turn out to
  have a deep supergeometric origin \cite{Henneaux:1984mh}
  which implies the remakable fact that they are mathematically classified
  by the $\mathrm{Spin}$-invariant super Lie algebra cohomology
  of super Minkowski space-times \cite{DeAzcarraga:1989vh}. This classification
  \cite{Achucarro:1987nc,DeAzcarraga:1989vh,Movshev:2010mf,Movshev:2011pr,Brandt:2010fa,Brandt:2010tz,Brandt:2013xpa}
  has come to be known as the \emph{old brane scan},
  reproduced in Table \ref{TheOldBraneScan} above.

  The old brane scan
  is striking in its combined
  success and shortcoming: On the one hand it shows that
  the mathematical principle of
  cohomology of super space-time governs the spectrum of
  super $p$-branes of various dimensions, notably of super membranes
  and thus of at least
  some shadow of M-theory; on the other hand
  various super $p$-branes fail to show up in the old brane scan:
  notably all the D$p$-branes in $D = 10$ (\cite{Cederwall:1996ri}, see \cite{Hashimoto:2012vsa})
  as well as the super 5-brane in
  $D = 11$ \cite{Bandos:9701149} \cite[Sec. 5.2]{Sorokin:1999jx}
   are missing from the old brane scan.
  Moreover, the old
  brane scan knows nothing about branes in their incarnation
  as space-time singularities (black branes \cite{Duff:9306052,Acharya:1998db}),
  nor about the web of dualities \cite[Ch. 6]{Duff:1999baa}.
This used to be an open problem \cite[p. 6-7]{Duff:1999baa}, \cite{Duff:2008pa}.

\smallskip
  \noindent {\bf The new brane bouquet.}
  In \cite{Fiorenza:2013nha} we pointed out (building on \cite{Chryssomalakos:1999xd} and \cite{Sati:2008eg}) that the problem is not
  with the basic principle of the old brane scan
  (cohomology of superspace), but with the mathematical generality
  in which this is understood:

  If we pass from classical Lie theory to the
  homotopy theory of \emph{homotopy} Lie algebras,
  also known as $L_\infty$-algebras or Lie $n$-algebras
  (see \cite[Sec . 6]{Sati:2008eg}) and further to $L_\infty$-algebroids \cite[Appendix]{Sati:2009ic},
  then a wealth of previously invisible \emph{higher structure}
  (see \cite{Jurco:2019aa}) emanate:

\smallskip
  \begin{enumerate}[i)]
    \item The previously missing super $p$-branes emerge
     to make a complete \emph{brane bouquet} (see Figure \ref{Figure1}),
     as does their
     \begin{enumerate}[a)]
    \item Dirac charge quantization (rationally) with:
      \begin{itemize}
        \item D-branes
          charged in twisted K-theory \cite[4]{Fiorenza:2016ypo},
        \item M2/M5-branes charged in Cohomotopy \cite{Fiorenza:2015gla};
      \end{itemize}
    \item brane intersection laws \cite[Sec. 3]{Fiorenza:2013nha},
    \item incarnation as black brane ADE-singularities \cite{Huerta:2018xyh}
    \item with their instanton contribution \cite[Sec. 6.2]{Huerta:2018xyh}.
    \end{enumerate}
    \item The duality relations among the branes emerge:
    \begin{enumerate}[a)]
    \item S-duality \cite[Sec. 4.3]{Fiorenza:2013nha},
    \item T-duality \cite{Fiorenza:2016oki},
    \item M/IIA-duality \cite{Fiorenza:2016ypo,Braunack-Mayer:2018uyy}.
    \end{enumerate}
    \item In fact, super space-time itself emerges \cite{Huerta:2017utu},
        as well as its exotica:
    \begin{enumerate}[a)]
      \item doubled super space-time \cite[Sec. 6]{Fiorenza:2016oki},
      \item 12d F-theory super space-time \cite[Sec. 8]{Fiorenza:2016oki}.
      \item exceptional super space-time \cite[Sec. 4.6]{Fiorenza:2018ekd},
    \end{enumerate}
    \item   Also novel phenomena are revealed,
        %which have not been part of the string theory folklore,
        such as higher topological T-duality
        of M5-branes \cite{Fiorenza:2018ekd,Sati:2018tvj}.
  \end{enumerate}

  \medskip

  \noindent Here we review this
  emergence of M-theoretic structure
  in rational super homotopy theory.

  \noindent We first give an exposition to some background in:

  \smallskip

  \hyperlink{SuperHomotopyTheory}{{\bf \ref{SuperHomotopyTheory}} {\it Super homotopy theory}}

  \hyperlink{RationalSuperHomotopyTheory}{{\bf \ref{RationalSuperHomotopyTheory}} {\it Rational super homotopy theory}}

  \hyperlink{HigherStructureFromHigherCocyles}{{\bf \ref{HigherStructureFromHigherCocyles}} {\it Higher structure form higher cocycles}}

\smallskip

  \noindent Then we review the brane bouquet:

\smallskip

  \hyperlink{EmergentSuperspace-time}{{\bf \ref{EmergentSuperspace-time}} {\it Emergent super space-time}}

  \hyperlink{TheBraneBouquet}{{\bf \ref{TheBraneBouquet}} {\it The brane bouquet}}

  \hyperlink{ChargeQuantization}{{\bf \ref{ChargeQuantization}} {\it Brane charge quantization}}

  \hyperlink{DoubleDimensionalReduction}{{\bf \ref{DoubleDimensionalReduction}} {\it Double dimensional reduction}}

  \hyperlink{SuperTopologicalTDuality}{{\bf \ref{SuperTopologicalTDuality}} {\it Super topological T-duality}}

  \hyperlink{BlackBraneScan}{{\bf \ref{BlackBraneScan}} {\it Black brane scan}}

  \hyperlink{MIIADuality}{{\bf \ref{MIIADuality}} {\it M/IIA-Duality and Gauge enhancement}}

\smallskip

  \noindent At the end we provide an outlook

\smallskip

  \hyperlink{SectionOutlook}{{\bf \ref{SectionOutlook}} {\it Outlook -- Beyond rational}}

  \noindent on construction and consistency checks of
  aspects of a formulation of microscopic M-theory
  suggested by the brane bouquet \cite{Braunack-Mayer:2019ip}.

\hypertarget{SuperHomotopyTheory}{
\section{Super homotopy theory}
  \label{SuperHomotopyTheory}
}

If history is anything to go by, understanding fundamental physics goes hand-in-hand with fundamental mathematics \cite{Galilei1623,Hilbert:1930aa,Wigner:1960:1-14}.
This is reminiscent of an old prophecy\footnote{
\cite{Witten:2003iaa}: ``{\it Back in the early '70s, the Italian physicist, Daniele Amati reportedly said that string
theory was part of 21st-century physics that fell by chance into the 20th century. I think it was a very wise remark.
How wise it was is so clear from the fact that 30 years later we're still trying to understand what string theory
really is.}''}
which suggests that unraveling the true nature of string/M-theory requires new concepts that would
become available only in the 21st century.
\smallskip

\begin{table}[htb]
\begin{center}
\begin{tabular}{llll}
  & {\bf Physics} & \phantom{AA} & {\bf Mathematics} \\
  \hline
  & gauge principle & & {homotopy theory}
  \\
  \& & Pauli exclusion  & & super geometry
  \\
  \hline
  \hline
  = & & & super homotopy theory
\end{tabular}
\end{center}

\vspace{-.9cm}

\caption{ Principles of fundamental physics and their
mathematical reflection.
}
\end{table}

\noindent{\bf Homotopy theory and the gauge principle.}
One development that the new millenium has brought is the
blossoming of \emph{homotopy theory} (see e.g.~\cite{Schreiber:2016laa})
into an immensely rich
(see e.g.~\cite{Ravenel:2003,Hill:0908.3724}),
powerful
(see \cite{Lurie:0608040,Lurie:book})
and foundational theory (see \cite{Shulman:1703.03007}). Homotopy theory indeed
serves as a new foundation of mathematics \cite{UFP2013:aa},
much like set theory, but with the difference
that
homotopy theory \emph{natively incorporates the gauge principle} \cite{Shulman:2014:109-126}:
by the gauge principle, no two things $x, y$ (e.g.~field histories) may
ever be assumed to be equal or not; instead the proposition of their equality is refined to the space of
\emph{gauge transformations} (homotopies)

\vspace{-5mm}
\begin{equation}
  \xymatrix{
    x
      \ar[rr]^-{\gamma}
      &&
    y
  }
\end{equation}
between them,
gauge-of-gauge transformations (homotopies of homotopies)
\begin{equation}
  \xymatrix{
    x
      \ar@/^1.5pc/[rr]^-{\gamma_1}_{\ }="s"
      \ar@/_1.5pc/[rr]_-{\gamma_2}^{\ }="t"
      &&
    y
    \ar@{=>}^{\kappa} "s"; "t"
  }
\end{equation}
between these, second order gauge transformations
\begin{equation}
  \xymatrix{
    x
      \ar@/^2pc/[rr]^-{\gamma_1}_{\ }="s"
      \ar@/_2pc/[rr]_-{\gamma_2}^{\ }="t"
      &&
    y
    \ar@/^1pc/@{=>}^{\kappa_1} "s"; "t"
    \ar@/_1pc/@{=>}_{\kappa_2} "s"; "t"
    \ar@{=>} (12,0); (16,0)
    \ar@{-} (12,0); (16,0)
  }
\end{equation}
between those, and so on.

If here $x,y$ are higher connections on higher bundles (see \cite{Sati:2008eg}),
then this precisely reproduces the familiar notion of higher gauge transformation
between higher gauge fields
\cite{Schreiber:0705.0452,Schreiber:2008aa,Fiorenza:2010mh}
(see \cite{Fiorenza:2013jz} for review); for example of the combined gauge field
and B-field in heterotic string theory \cite{Sati:2009ic}, or the
C-field in 11d supergravity \cite{Fiorenza:2012mr,Fiorenza:2012tb,Braunack-Mayer:2019ip}.

\begin{figure*}[htb]
  \begin{center}
   \includegraphics[width=.8\textwidth]{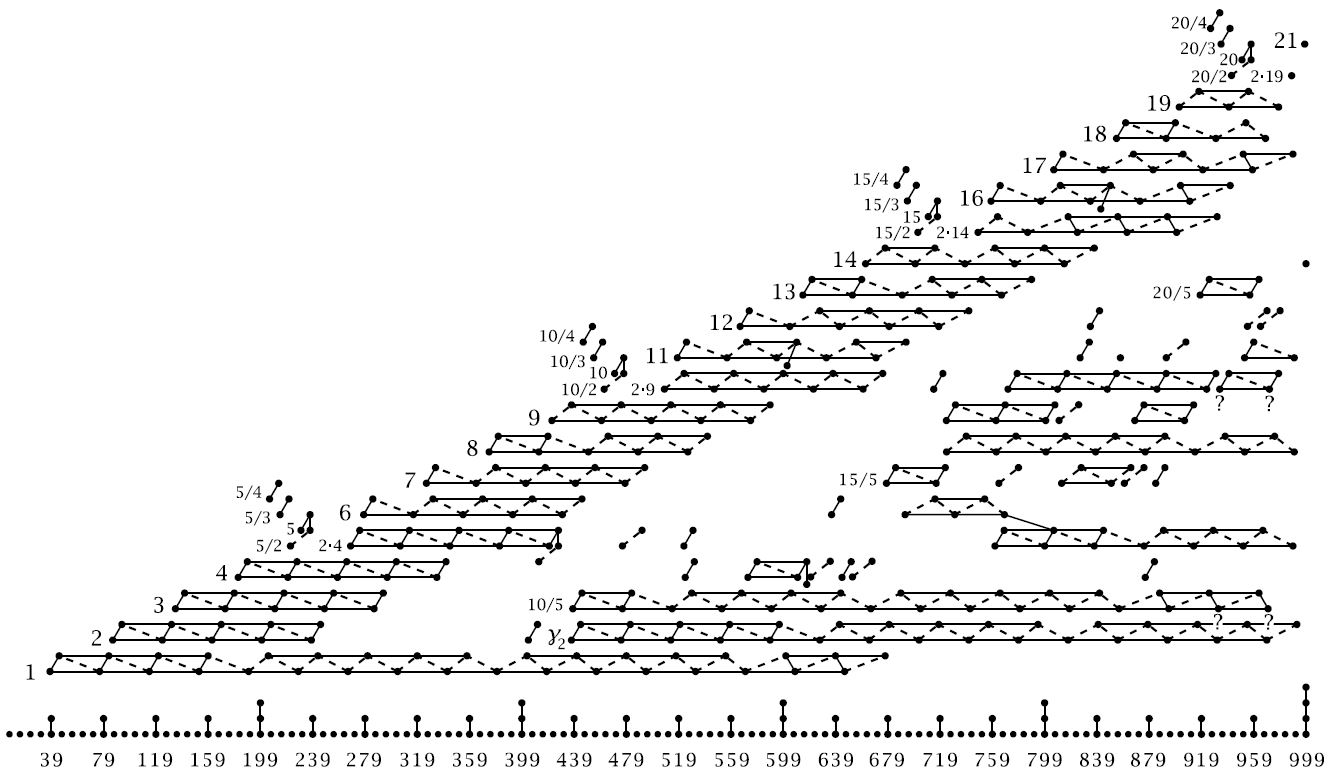}
 \end{center}

 \vspace{-.7cm}

 \caption{
   {\bf Order and chaos emerging from the sphere spectrum $\mathbbm{S}$}.
   Shown are the generators of
   the first 999 homotopy groups of $\mathbbm{S}$
   at prime 5 (graphics due to \cite{Hatcher:xx}, based on \cite{Ravenel:2003}).
   In order to manage this richness of homotopy theory,
   there are controlled approximations, such as
   \emph{rational} homotopy theory (Figure \ref{SullivanConstruction}).
 }
 \label{StableHomotopyGroupsOfSpheres}
\end{figure*}

More generally, the homotopy-theoretic gauge principle
captures also space-time re-identifications
in orbifolds and unifies these with
the higher gauge transformations of higher gauge fields
(see Table \ref{SuperHomotopyTheoryUnification})
to structures in differential \emph{equivariant} cohomology.
This turns out to capture M-theoretic ``hidden degrees of freedom''
inside orbifold singularities (Section \ref{BlackBraneScan} below.)
\smallskip

\begin{table}[htb]
%\vspace{-.5cm}
\begin{center}
\fbox{
\hspace{6pt}
$
  \begin{array}{c}
  \begin{array}{l}
 \hspace{-.6cm} \mathrm{SuperSingularities}
  :=
  \Bigg\{\!\!\!\!\!\!\!\!
    \raisebox{7pt}{
    $
    \underset{
      \mbox{
        \begin{tabular}{c}
          \tiny
          \color{blue}
          superspace
          \\
          $\phantom{A}$
        \end{tabular}
      }
    }{
    \underbrace{
      \mathbbm{R}^{d\vert N}
    }\
    }
  \!\!\!  \times \;\;
    \underset{
      \mbox{\setlength{\tabcolsep}{-20pt}
\renewcommand{\arraystretch}{.5}
     \tiny \color{blue}
  \begin{tabular}{c}
  {infinitesimal} \\ \tiny{disk}  \end{tabular} }
    }{
    \underbrace{
      \mathbbm{D}
    }
    }
    \;\;\;\times  \!\!\!
    \underset{
      \mbox{\setlength{\tabcolsep}{10pt}
\renewcommand{\arraystretch}{.5}
        \tiny
        \color{blue}
        \begin{tabular}{c}
          orbifold
          \\
          singularity
        \end{tabular}
      }
    }{
    \underbrace{
      \mathbbm{B}G
    }}
    $
    } \!\!\!\!\!\!\!\!\!\!
  \Bigg\}
  \end{array}
  \\
  \vspace{-3mm}
  \underset{
    \mathclap{
      \mbox{
        \tiny
        \color{blue}
        \begin{tabular}{c}
          super
          \\
          homotopy
          \\
          theory
        \end{tabular}
      }
    }
  }{
    \underbrace{
      \mathbf{H}
    }
  }
  \;:=\;
  \xymatrix@C=-7pt{
    \underset{
      \mathclap{
        \mbox{
          \tiny
          \color{blue}
          \begin{tabular}{c}
            generalized
            \\
             spaces
          \end{tabular}
        }
      }
    }{
      \underbrace{
        \mathrm{Sh}_\infty
      }
    }
  \ar@/_3pc/@{~>}[rr]_-{
    \mbox{
      \tiny
      \color{blue}
      \hspace{-10pt}
      \raisebox{10pt}{
      \begin{tabular}{c}
        probe-able
             by these
        \\
        local model spaces
      \end{tabular}
    }
    }
  }
  &&
  \big(
    \mathrm{SuperSingularities}
  \big)
  }
  \end{array}
$
}
\end{center}

\vspace{-.7cm}

\caption{
{\bf Super homotopy theory}
as: {\bf (I)}  the higher topos
over the site of super singularities.
}
\label{SuperTopos}
\vspace{-.5cm}
\end{table}

\noindent {\bf Supergeometry and the Pauli exclusion principle.} Another principle of
fundamental physics is the \emph{Pauli exclusion principle}, mathematically reflected
in the statement that the phase space of \emph{any} physical system with fermions (not necessarily a
supersymmetric theory!) is a supergeometric space (a ``superspace'' \cite{Gates:1983nr})) whose odd-graded
coordinates
\begin{equation}
  \label{Supercommutativity}
  \theta_i \theta_j = - \theta_j \theta_i
\end{equation}
reflect the fermionic field configurations (see \cite{Schreiber:2018lnaa}).

Such geometric qualities (differential geometry, supergeometry, etc.)
are naturally realized in homotopy theory in its refinement to
\emph{geometric homotopy theory} known as \emph{higher topos theory}
\cite{Rezk:2010aa}, \cite{Lurie:0608040}
(see also \cite{Schreiber:2018lab}).

\smallskip
\noindent {\bf Super homotopy theory.}
Hence we take super homotopy theory $\mathbf{H}$ to be the geometric homotopy
theory generated by infinitesimal thickenings of the Cartesian superspaces
$\mathbbm{R}^{d\vert N}$ (see \cite{Schreiber:2018lab,Schreiber:2018lac}),
or rather by orbifold singularities of this form.
By the general principles of higher topos theory
(see \cite{Schreiber:2018lab}) this means that
objects in super homotopy theory are those generalized
spaces which may be \emph{probed}, up to higher
gauge transformation, by mapping super singularities into them; see
Figure \ref{SuperTopos}. In other words, these are spaces as seen by classical
higher-gauged super sigma models.

One could think of this higher topos $\mathbf{H}$ of
super homotopy theory  as the natural theoretical context which
unifies orbi-space-time supergeometry (hence gravity)
with moduli stacks classifying differential equivariant generalized
cohomology theories (hence higher gauge fields), as
indicated in Table \ref{SuperHomotopyTheoryUnification}.

\begin{table}[htb]

%\vspace{-.7cm}

\begin{center}
 \fbox{
  \xymatrix@R=1em{
    \overset{
      \raisebox{7pt}{
      \mbox{
        \color{blue}
        space-times / gravity
      }
      }
    }{
    \left\{\hspace{-4mm}
      \mbox{\setlength{\tabcolsep}{10pt}
\renewcommand{\arraystretch}{.5}
        \begin{tabular}{l}
          super orbifolds
          with $G$-structure
          \\
          {\small ( e.g. super Riemannian, super conformal-, $\cdots$}
          \\
          {\small $\phantom{(}$ Spin-, String-, Fivebrane-, $\cdots$ structure )}
        \end{tabular}
      }
  \hspace{-4mm}  \right\}
    }
    \ar@{^{(}->}[d]
    \\
    \mathbf{H}
    \mathrlap{
      \;\;\;\;\;\;\;
      \mbox{
        \color{blue}
        Super homotopy theory
      }
    }
    \\
    \ar@{_{(}->}[u]
    \underset{
      \raisebox{-9pt}{
      \mbox{
        \color{blue}
        Higher gauge fields
      }
      }
    }{
    \left\{\hspace{-4mm}
      \mbox{\setlength{\tabcolsep}{10pt}
\renewcommand{\arraystretch}{.5}
        \begin{tabular}{c}
          universal moduli stacks for
          \\
          differential equivariant
          \\
          generalized cohomology theories
        \end{tabular}
      }
   \hspace{-4mm} \right\}
   \vspace{-3mm}   }
}
 }
\end{center}

\vspace{-.9cm}

\caption{ {\bf Super homotopy theory} as:
{\bf (II)} a unifying context.
}
\label{SuperHomotopyTheoryUnification}
\vspace{-.5cm}
\end{table}

\smallskip
\noindent {\bf The music of the spheres.}
Homotopy theory is immensely rich. Already the simplest
homotopy types by number of cells -- the spheres --
exhibit an endless richness of subtle patterns and apparent chaos
in their homotopy groups, even if they are
``stabilized'' and organized in the sphere spectrum $\mathbbm{S}$
(``the music of the spheres'' \cite{Ravenel:2003}; see Figure \ref{StableHomotopyGroupsOfSpheres}
for an impression).
At the same time, the
sphere spectrum is the most canonical object in homotopy mathematics:
equipped with its canonical ring spectrum structure (an appropriate way of doing
multiplication in (co)homology),
it is the homotopy-theoretic refinement of the ring of integers in classical
mathematics -- the ``real integers''.
Much of modern homotopy theory, notably chromatic homotopy theory, is
a grandiose attack on analyzing the sphere spectrum by approximating it
stagewise, such as by complex cobordism or topological modular forms,
all of which have striking but still somewhat mysterious relations to
string theory.

More concretely, in super homotopy theory we see the sphere spectrum emerge as
 a core ingredient of microscopic M-theory \cite{Braunack-Mayer:2019ip} as indicated below in
  Section \ref{SectionOutlook}.

Dealing with this richness leads, in a first step,
to \emph{rational} homotopy theory.

\hypertarget{RationalSuperHomotopyTheory}{
\section{Rational super homotopy theory \\ and higher super Lie theory}
 \label{RationalSuperHomotopyTheory}
}

\noindent {\bf Rational homotpy theory and higher Lie integration.}
Various tools
exist for extracting and analyzing certain
aspects of homotopy theory
in a controlled approximation. A basic such tool is
\emph{rational homotopy theory} (see \cite{hess2007rational,Braunack-Mayer:2018aa})
where torsion-subgroups of cohomology and of homotopy groups are
ignored (hence those Abelian groups that vanish under
rationalization, i.e., under tensor product
of Abelian groups with the additive group of rational numbers).
The key result of rational homotopy theory is that, in this
approximation, homotopy types with nilpotent fundamental
groups are entirely characterized by
differential-graded commutative algebras \cite{sullivan:infinites1977,Bousfield:1976:0-0}
of the kind familiar in modern mathematical physics.

The super algebraic version of these differential-graded algebras (or dg-algebras for short)
 have come to be known as ``FDAs'' in
the supergravity literature (\cite{D'Auria:1982nx,Castellani:1991et},
following \cite{vanNieuwenhuizen:1982zf}). Under Koszul duality, we
may equivalently think of these as being the Chevalley--Eilenberg algebras
of
super $L_\infty$-algebras (or, more generally, super $L_\infty$-algebroids) \cite{Fiorenza:2013nha}, see \cite{Schreiber:2018lac}:
\begin{equation}
  \hspace{-.4cm}
  \raisebox{30pt}{\xymatrix@C=2pt{
  \underset{
    \mbox{ \setlength{\tabcolsep}{2pt}
\renewcommand{\arraystretch}{.5}
      \tiny
      \color{blue}
      \begin{tabular}{c}
        terminology
        \\
        common in
        \\
        supergravity
        \\
        \cite{vanNieuwenhuizen:1982zf,D'Auria:1982nx,Castellani:1991et}
      \end{tabular}
    }
  }{
  \underbrace{
    \mbox{FDAs}
  }}
  \ar@{}[r]|-{:=}
  &
  \underset{
    \mathclap{
    \mbox{\setlength{\tabcolsep}{10pt}
\renewcommand{\arraystretch}{.5}
      \tiny
      \color{blue}
      \begin{tabular}{c}
        homotopy theory of
        \\
        differential
        \\
        graded-commutative
        \\
        superalgebras
      \end{tabular}
    }
    }
  }{
  \underbrace{
    \mathrm{dgcSuperAlg}
  }}
  \ar@{<-}[d]^-{
    \mathrm{CE}
    \;
    \mathrlap{
      \mbox{
        \tiny
        \begin{tabular}{c}
          \cite{Sati:2008eg,Sati:2009ic}
          \\
          \cite{Fiorenza:2013nha}
        \end{tabular}
      }
    }
  }_-{
    \simeq
  }
  \\
  &
  \Big(
    \underset{
      \mathclap{
      \mbox{\setlength{\tabcolsep}{10pt}
\renewcommand{\arraystretch}{.5}
        \tiny
        \color{blue}
        \begin{tabular}{c}
         homotopy theory of
          \\
         nilpotent
          \\
         super $L_\infty$-algebroids
        \end{tabular}
      }
      }
    }{
    \underbrace{
      \mathrm{Super}L_\infty\mathrm{Algbd}^{\mathrm{nil}}
    }
    }
  \Big)^{\mathrm{op}}
  }
  }
\end{equation}
In accord with the super $L_\infty$-algebraic interpretation of
``FDAs'', the Sullivan construction of rational homotopy theory
\cite{sullivan:infinites1977} may naturally be enhanced to a higher super analog
of the process of Lie integration of Lie algebras
to Lie groups \cite{Henriques:2006aa,Fiorenza:2010mh} \cite[Sec. 3.]{Braunack-Mayer:2018aa}
(see \cite{Schreiber:2018lac}). This is indicated in
Figure \ref{SullivanConstruction}, where the equivalence $\mathrm{Super}L_\infty\mathrm{Algbd}^{\mathrm{nil}}
      \underset{
        \simeq
      }{
      \xrightarrow{\mathrm{CE}}
      }
      \mathrm{FDAs}^{\mathrm{op}}$ appears on the left as part of a larger picture.

\medskip

We now say some of this in more detail.

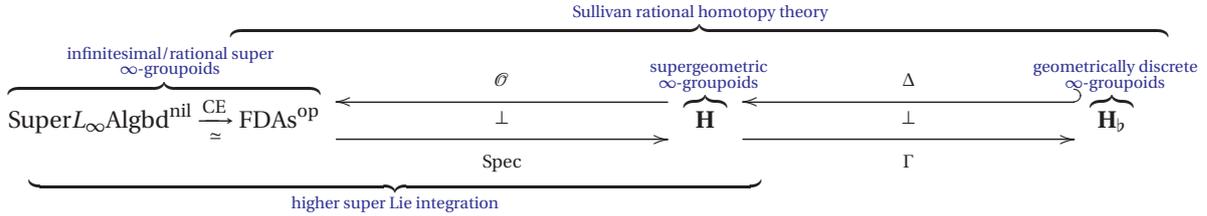
\begin{figure*}[htb]

$$
  \hspace{-2.4cm}
  \xymatrix@C=56pt@R=10pt{
    \ar@<-27pt>@{}[rrr]|>>>>>>>>>>>>>>>>>>{ \underset{ \mbox{\color{blue} \tiny higher super Lie integration} }{  \underbrace{\phantom{------------------------------------}} }  }
    &
    \overset{
      \mbox{
         \setlength{\tabcolsep}{10pt}
         \renewcommand{\arraystretch}{.5}
        \tiny
        \color{blue}
        \begin{tabular}{c}
          infinitesimal/rational super
          \\
          $\infty$-groupoids
        \end{tabular}
      }
    }{
    \overbrace{
      \mathrm{Super}L_\infty\mathrm{Algbd}^{\mathrm{nil}}
      \underset{
        \simeq
      }{
      \xrightarrow{\mathrm{CE}}
      }
      \mathrm{FDAs}^{\mathrm{op}}
    }}
    \ar@<+36pt>@{}[rrrr]|-{ \!\!\!\!\!\!
    \overset{  \mbox{ \color{blue} \tiny Sullivan rational homotopy theory } }{\overbrace{ \phantom{----------------------------------------------}} }  }
    \ar@{<-}@<+7pt>[rr]^-{\mathcal{O}}
    \ar@<-7pt>[rr]_-{\mathrm{Spec}}^-{\bot}
    &&
    \overset{
      \mathclap{
      \mbox{\setlength{\tabcolsep}{10pt}
\renewcommand{\arraystretch}{.5}
        \tiny
        \color{blue}
        \begin{tabular}{c}
          supergeometric
          \\
          $\infty$-groupoids
        \end{tabular}
      }
      }
    }{
    \overbrace{
      \mathbf{H}
    }
    }
    \ar@{<-^{)}}@<+7pt>[rr]^-{\Delta}
    \ar@<-7pt>[rr]_-{\Gamma}^-{\bot}
    &&
    \overset{
      \mathclap{
      \mbox{\setlength{\tabcolsep}{10pt}
\renewcommand{\arraystretch}{.5}
        \tiny
        \color{blue}
        \begin{tabular}{c}
          geometrically discrete
          \\
          $\infty$-groupoids
        \end{tabular}
      }
      }
    }{
    \overbrace{
      \mathbf{H}_{\flat}
    }
    }
  }
$$
\vspace{-.7cm}

\caption{  {\bf Rational approximation and higher Lie integration}
in super homotopy theory
\cite[Sec. 3.1]{Braunack-Mayer:2018aa} (following \cite{Severa:2001aa,Henriques:2006aa,Fiorenza:2010mh}, see \cite{Schreiber:2018lac}).
Shown is the
$\left\{
  { \mbox{\tiny infinitesimal} }
  \atop
  { \mbox{\tiny rational} }
\right\}
\mbox{approximation of super homotopy types via}
\left\{
  { \mbox{
    \tiny higher Lie integration} }
  \atop
  { \mbox{\tiny Sullivan construction} }
\right\}
$.
  The ``FDA''s or ``CIS''s in the supergravity literature
  \cite{vanNieuwenhuizen:1982zf,D'Auria:1982nx} are super Sullivan
  models for super rational homotopy types, or equivalently
  the super Chevalley--Eilenberg algebras of the corresponding
  super Quillen model super $L_\infty$-algebras \cite{Fiorenza:2013nha}.
}
\label{SullivanConstruction}
\end{figure*}

\smallskip
\noindent {\bf Super dg-algebra.}
A \emph{differential graded-commutative superalgebra}
is a $\mathbbm{Z} \times \mathbbm{Z}_2$-graded algebra,
hence with bigrading
\begin{equation}
  \Big(\hspace{-.5cm}
    \underset{
      \mbox{ \setlength{\tabcolsep}{2pt}
\renewcommand{\arraystretch}{.5}
        \tiny
        \color{blue}
        \begin{tabular}{c}
          cohomological
          \\
          degree
        \end{tabular}
      }
    }{
      \underbrace{
        n
      }
    }
    ,
    \underset{
      \mbox{ \setlength{\tabcolsep}{2pt}
\renewcommand{\arraystretch}{.5}
        \tiny
        \color{blue}
        \begin{tabular}{c}
          super
          \\
          degree
        \end{tabular}
      }
    }{
    \underbrace{
      \sigma
    }
    }\!\!\!
  \Big)
  \;\in\;
  \mathbbm{Z} \times \mathbbm{Z}_2
  \,,
\end{equation}
and equipped with a derivation  $d$ of degree
$(1,\mathrm{even})$, which is a differential, i.e., satisfies $d^2=0$.
Here ``graded-commutative''
may be interpreted with respect to either of two sign rules
used in the literature, as shown in Figure \ref{SignRules}.

\begin{table}[htb]

\begin{center}
\hspace{-.25cm}
\scalebox{.81}{
\begin{tabular}{|c|c|c|}
  \hline
  &
  Deligne's convention
  &
  Bernstein's convention
  \\
  \hline
  \hline
  $\alpha_i \cdot \alpha_j =$
  &
  $(-1)^{ n_i \cdot n_j + \sigma_i \cdot \sigma_j } \alpha_j \cdot \alpha_i$
  &
  $(-1)^{ (n_i + \sigma_i)\cdot (n_j + \sigma_j) } \alpha_j \cdot \alpha_i$
  \\
  \hline
  \begin{tabular}{l}
    common in
    \\
    discussion of
  \end{tabular}
  &
  supergravity
  &
  AKSZ sigma models
  \\
  \hline
  \begin{tabular}{l}
    representative
    \\
    references
  \end{tabular}
  &
  \begin{tabular}{c}
    \cite[Sec. 2]{Bonora:1987xn},
    \\
    \cite[II (2.109)]{Castellani:1991et}
    \\
    \cite[Appendix]{Deligne:1999ur}
  \end{tabular}
  &
  \begin{tabular}{l}
    \cite{Alexandrov:1995kv},
    \\
    \cite{Carchedi:1212.3745}
  \end{tabular}
  \\
  \hline
\end{tabular}
}
\end{center}

\vspace{-.7cm}

\caption{ {\bf Sign rules in super homotopy theory.}
The two sign rules are different, but equivalent
as symmetric monoidal structures on the category of
chain complexes of super vector spaces.}
\label{SignRules}

\end{table}

\smallskip
\noindent {\bf Example -- Super cartesian space}
A \emph{super Cartesian space} $\mathbbm{R}^{D,N}$
has algebra of functions
\begin{equation}
  \begin{array}{ccccc}
    C^\infty
    \big(
      \mathbbm{R}^{D\vert N}
    \big)
    &=&
    C^\infty
    \left(
      \mathbbm{R}^D
    \right)
    &\otimes_{\mathbbm{R}}&
    \left(
      \wedge^\bullet_{\mathbbm{R}} \mathbbm{R}^N
    \right)
    \\
    \mbox{
      \tiny
      \color{blue}
      coordinates:
    }
    &
    &
    \underset{
      (0,\mathrm{even})
    }{\underbrace{x^a}}
    &&
    \underset{
      (0,\mathrm{odd})
    }{\underbrace{\theta^\alpha}}
  \end{array}
\end{equation}
Derived from this, the algebra of
\emph{differential forms on Super cartesian space}
is the differential graded-commutative superalgebra
free over $C^\infty\left(\mathbbm{R}^{D\vert N}\right)$
on
\smallskip
\begin{enumerate}[i)]

\item $D$ generators $\mathbf{d} x^a$ in bi-degree $(1,\mathrm{even})$

\item $N$ generators $\mathbf{d} \theta^\alpha$ in bi-degree $(1,\mathrm{odd})$,
\end{enumerate}
\smallskip
i.e.,
\begin{equation}
  \Omega^\bullet
  \left(
    \mathbbm{R}^{D\vert N}
  \right)
  \;:=\;
  C^\infty
  \left[
    \left(
      \mathbbm{R}^{D\vert N}
    \right)
    \left\langle
        \mathbf{d}x^a
    \right\rangle_{a = 1}^N
   \; , \;
    \left\langle
      \mathbf{d}\theta^\alpha
    \right\rangle_{\alpha =1}^{N}
  \right].
\end{equation}

\noindent {\bf Example -- Super Lie algebras}
For $\mathfrak{g}$ a super Lie algebra of finite dimension,
its \emph{Chevalley--Eilenberg algebra}
is the differential graded-commutative superalgebra
\begin{equation}
  \mathrm{CE}(\mathfrak{g})
  \;\coloneqq\;
  \mathbbm{R}
  \big[
    \underset{
      (1,\bullet)
    }{
      \underbrace{
        \mathfrak{g}^\ast
      }
    }
  \big]/ {\rm d}_{\mathfrak{g}}
\end{equation}
equipped with the differential ${\rm d}_{\mathfrak{g}}$
which is the linear dual of the super Lie bracket
\begin{equation}
  {\rm d}_{\mathfrak{g}}
  \;\coloneqq\;
  [-,-]^\ast \;\colon\; \mathfrak{g}^\ast \to \mathfrak{g}^\ast \wedge \mathfrak{g}^\ast
  \,.
\end{equation}

\smallskip

\noindent {\bf Key Example -- Super Minkowski space-times.}
For $d \in \mathbbm{N}$
and $\mathbf{N}$ a real representation of $\mathrm{Spin}(d-1,1)$,
the super translation supersymmetry super Lie algebra
\begin{equation}
  \mathbbm{R}^{D-1,1\vert \mathbf{N}}
\end{equation}
has Chevalley--Eilenberg algebra given by
\begin{equation}
  \label{SuperMinkowskiCE}
  \mathrm{CE}
  \left(
    \mathbbm{R}^{D-1,1\vert \mathbf{N}}
  \right)
  \;=\;
  \mathbbm{R}
  \Big[
  \big\langle\;
    \underset{
      \mathclap{
        (1,\mathrm{even})
      }
    }{
      \underbrace{
        e^a
      }
    }\;
  \big\rangle_{a = 0}^{D-1}
  ,
  \;
  \big\langle\;
    \underset{
      \mathclap{(1,\mathrm{odd})}
    }{
      \underbrace{
        \psi^\alpha
      }
    } \;
  \big\rangle_{\alpha = 1}^N
  \Big]
  /{\rm d}_{\mathrm{CE}}
\end{equation}
with differential given by
\begin{equation}
  \begin{aligned}
    {\rm d}_{\mathrm{CE}} \, \psi_\alpha &\;=\;  0
    \\
    {\rm d}_{\mathrm{CE}}\, e^a &
      \;=\!\!\!\!
    \underset{
      \mbox{ \setlength{\tabcolsep}{2pt}
\renewcommand{\arraystretch}{.5}
        \tiny
        \color{blue}
        \begin{tabular}{c}
          spinor-to-vector
          \\
          pairing
        \end{tabular}
      }
    }{
    \underbrace{
      \overline{\psi} \wedge \Gamma^a \psi
    }} \hspace{-3mm}.
  \end{aligned}
\end{equation}
If we think of super Minkowski space-time
$\mathbbm{R}^{D-1,1\vert \mathbf{N}}$
as the supermanifold with coordinates
\begin{equation}
  \big\{\;
    \underset{(0,\mathrm{even})}{\underbrace{ \phantom{A}x^a\phantom{A} }}\;
  \big\}_{a = 0}^{D-1}
  \phantom{AAA}
  \big\{\;
    \underset{(0,\mathrm{odd})}{\underbrace{ \phantom{A}\theta^\alpha \phantom{A} }} \;
  \big\}_{\alpha = 1}^N
\end{equation}
then these generators correspond to
the super left invariant super vielbein
\begin{equation}
  \begin{aligned}
    \psi^\alpha & \;=\ {\rm d}_{\rm dR} \theta^\alpha
    \\
    e^a &
      \;=\;
    \underset{
      \mbox{ \setlength{\tabcolsep}{2pt}
\renewcommand{\arraystretch}{.5}
        \tiny
        \color{blue}
        \begin{tabular}{c}
          ordinary
          \\
          Minkowski vielbein
        \end{tabular}
      }
    }{
    \underbrace{
      \; {\rm d}_{\rm dR} x^a\;
    }}
      +
    \underset{
     \mbox{ \setlength{\tabcolsep}{2pt}
\renewcommand{\arraystretch}{.5}
        \tiny
        \color{blue}
        \begin{tabular}{c}
          correction term
          \\
          for left super invariance
        \end{tabular}
      }
    }{
      \underbrace{\;
        \overline{\theta} \Gamma^a {\rm d}_{\rm dR} \theta
      \;}
    } \hspace{-.7cm}.
  \end{aligned}
\end{equation}

Notice that ${\rm d}_{\rm dR} x^a$ alone
fails to be a left invariant differential form,
in that it is not annihilated by the supersymmetry
vector fields
$D_\alpha \;\coloneqq\; \partial_{\theta^\alpha} - \overline{\theta}_{\alpha'} \Gamma^a{}^{\alpha'}{}_\alpha \partial_{x^a}$\,.
Hence the appearance of the all-important correction term above.

\,

\noindent {\bf Homomorphisms of super Lie algebras} (of finite dimension) are in natural bijection with
the ``dual'' homomorphisms of dgc-superalgebras
between their Chevalley--Eilenberg algebras:
\begin{equation}
  \raisebox{10pt}{ \xymatrix@R=2pt{
    \mathfrak{g}_1
    \ar[rr]^-\phi
    &&
    \mathfrak{g}_2
    \\
    \mathrm{CE}(\mathfrak{g}_1)
    \ar@{<-}[rr]^-{ \phi^\ast }
    &&
    \mathrm{CE}(\mathfrak{g}_2)
  }
  }
\end{equation}
Technically this says that
forming $\mathrm{CE}$-algebras
constitutes a full embedding
\begin{equation}
  \mathrm{CE}
  \;\colon\;
  \mathrm{SuperLieAlg}
  \;\lhook\joinrelaz
  \longrightarrow  \;
  \mathrm{dgcSuperAlg}^{\mathrm{op}}
  =
  \mathrm{FDA}^{\mathrm{op}}
\end{equation}
of the category of super Lie algebras (of finite dimension)
into the opposite category of differential graded-commutative
superalgebras (in supergravity: ``FDA''s).
This makes the following definition immediate:

\smallskip
\noindent A {\bf super $L_\infty$-algebra} of finite type is
\smallskip
\begin{enumerate}[i)]
\item a $\mathbbm{Z}$-graded super vector space $\mathfrak{g}$,
  degreewise of finite dimension;
\item
    for all $n \geq 1$ a multilinear map
    \begin{equation}
      [-,\cdots, -] \;\colon\; \wedge^n \mathfrak{g}^\ast \longrightarrow \wedge^1 \mathfrak{g}^\ast
    \end{equation}
    of degree $(-1,\mathrm{even})$,
    such that the degree 1 graded derivation
    \begin{equation}
    \begin{aligned}
      &{\rm d}_{\mathfrak{g}}
      \;\coloneqq\;
      [-]^\ast + [-,-]^\ast +\\
      &\kern1.5cm+ [-,-,-]^\ast + \cdots \;\;\colon\;\; \wedge^1 \mathfrak{g}^\ast \longrightarrow \wedge^\bullet \mathfrak{g}^\ast
      \end{aligned}
    \end{equation}
    is a differential (i.e.~it squares to zero: ${\rm d}_{\mathfrak{g}} {\rm d}_{\mathfrak{g}} = 0$).
\end{enumerate}
\smallskip
These data define a dgc-superalgebra
\begin{equation}
  \mathrm{CE}(\mathfrak{g}) \coloneqq ( \wedge^\bullet \mathfrak{g}^\ast, {\rm d}_{\mathfrak{g}} )
  \,.
\end{equation}
Hence super $L_\infty$ algebras with (possibly ``curved'') $L_\infty$-morphisms between them form the larger full subcategory

\vspace{-8mm}
\begin{equation}
  \xymatrix{
    \mathrm{Super}L_\infty\mathrm{Alg}^{\mathrm{fin}}
    \ar[rr]^{
      \mathrm{CE}
    }
    &&
    \mathrm{dgcSuperAlg}^{\mathrm{op}}
  }
\end{equation}
obtained from that of plain super Lie algebras simply by
dropping the assumption that the
underlying super vector space is %cohomologically
concentrated
in degree zero \cite[Def. 13]{Sati:2008eg}.

\smallskip
\noindent {\bf Line Lie $n$-algebras.} A simple but
important example of $L_\infty$-algebras are the higher
analogs %$\mathfrak{l}B^n \mathbbm{R}$
$\mathbbm{R}[n]$
of the Abelian Lie algebra $\mathbbm{R}$.
These are defined as having Chevalley--Eilenberg algebra
generated from a single element in degre $(n+1, \mathrm{even})$
and vanishing differential. The corresponding $L_\infty$-algebras consist of the
vector space $\mathbbm{R}$ concentrated in degree $n$ endowed with the trivial
differential and brackets
%, and are denoted $\mathbbm{R}[n]$
; see Table \ref{RationalNotation}.

Now that we have introduced some terminology and seen a few basic examples,
we can understand algebraic models in rational homotopy theory
as providing for any connected topological space $X$ with nilpotent
fundamental group an infinitesimal
approximation of its loop $\infty$-group $\Omega X$
by an $L_\infty$-algebra (see \cite[Sec. 2]{Buijs:1209.4756}).
We will write $\mathfrak{l} X$ for
the \emph{minimal} such $L_\infty$-algebra,
hence for the one whose Chevalley--Eilenberg algebra
$\mathrm{CE}\big( \mathfrak{l}X\big)$ is the
Sullivan model dgc-algebra (``FDA'') of $X$. This is summarized in Table \ref{RationalNotation}.

\begin{table}[htb]
%\vspace{-.5cm}
\hspace{0cm}
\begin{tabular}{|c|c|c|c|}
  \cline{3-4}
  \multicolumn{2}{c|}{}
  &
  \multicolumn{2}{c|}{
    {\bf Examples}
  }
%  \\
%  \cline{3-4}
%  \multicolumn{2}{c|}{}
%  &
%  {\bf torus}
%  &
%  {\bf
%  \begin{tabular}{c}
%    classifying
%    \\
%    space
%  \end{tabular}
%  }
  \\
  \hline
  \hline
  {
  \bf
  \begin{tabular}{c}
    Topological
  \\     space
  \end{tabular}
  }
  &
  $X$
  &
  $\mathbbm{T}^D$
  &
  $K(\mathbbm{Z},n+1)$
  \\
  \hline
  {
  \bf
  \begin{tabular}{c}
    Loop
\\    $\infty$-group
  \end{tabular}
  }
  &
  $\Omega X$
  &
  $\mathbbm{Z}^D$
  &
  $K(\mathbbm{Z},n)$
  \\
  \hline
  {
  \bf
  \begin{tabular}{c}
    Minimal
    \\
    $L_\infty$-algebra
  \end{tabular}
  }
  &
  $\mathfrak{l} X$
  &
  $\mathbbm{R}^D$
  &
  $\mathbbm{R}[n]$
  \\
  \hline
  {\bf
  \begin{tabular}{c}
    Sullivan
    \\
    model
    \\
    (``FDA'')
  \end{tabular}
  }
  &
  $
  \mathrm{CE}
  \big(
    \mathfrak{l}X
  \big)
  $
  &
  $
  \mathbbm{R}
  \big[
    \underset{
      \mathclap{
        (1,\mathrm{even})
      }
    }{
      \underbrace{
        e^a
      }
    }
  \big]_{a=1}^D
  $
  &
  $
  \mathbbm{R}
  \big[
    \underset{
      \mathclap{
        (n+1,\mathrm{even})
      }
    }{
      \underbrace{
        c
      }
    }
  \big]
  $
  \\
  \hline
\end{tabular}

\vspace{-.7cm}

\caption{
{\bf Algebraic models in rational homotopy theory}
 with two explicit examples: in the third column we have a $D$-dimensional
 torus and in the fourth column the Eilenberg--MacLane space $K(\mathbbm{Z},n+1)$.
 }
\label{RationalNotation}

\end{table}

\hypertarget{HigherStructureFromHigherCocyles}{
\section{Higher structure from higher cocycles}
 \label{HigherStructureFromHigherCocyles}
 }

\noindent {\bf Homotopy theory as a microscope.}
A key aspect of homotopy theory
%(``higher structures'')
is that it serves as a
\emph{mathematical microscope} that reveals intrinsic structure invisible to
`non-homotopy theory'.
This extra information seen by homotopy theory is what is being
alluded to by the plethora of adjectives that
the literature uses for homotopy-theoretic improvements of
non-homotopy theoretic concepts,
such as
\emph{``enhanced'' triangulated category},
\emph{``derived'' functor},
\emph{``derived'' geometry},
\emph{``higher'' structure},
\emph{``higher'' geometry}.

The following is a simple but important example of this
phenomenon:

\medskip
  \noindent {\bf Cocycles and extensions.}
  Given a Lie algebra $\mathfrak{g}$,
  a classical fact of
  non-homotopy theory is that Lie algebra 2-cocycles
  $[\omega_2] \in H^2(\mathfrak{g})$
  classify Lie algebras $\widehat{\mathfrak{g}}$
  which are \emph{central extensions} of $\mathfrak{g}$
  \begin{equation}
    \begin{array}{ccc}
      H^2_{\mathrm{LieAlg}}(\mathfrak{g})
      &
      \simeq
      &
      \left\{\hspace{-3mm}
      {\raisebox{24pt}{
      {\xymatrix{
        \widehat{\mathfrak{g}}
        \ar[d]|<<<{
          \mbox{
            \tiny
            {\begin{tabular}{c}
            \color{blue}  central
              extension
            \end{tabular}}
          }
        }
        \\
        \mathfrak{g}
      }}
      }}
\hspace{-3mm}       \right\}
      \\
   \mu_2 \!\!\!
      &
    \!\!\!\!\!\!\!\!  \mapsto
      &
    \!\!\!\!\scalebox{0.96}{$\left(
      \mathfrak{g} \oplus \mathbbm{R}
      ,
      {\begin{array}{l}
        [(x_1,c_1), (x_2, c_2)]
        = \big( [x_1, x_2], \mu_2(x_1,x_2) \big)
      \end{array}}
      \!\right)$}.
    \end{array}
  \end{equation}
  While of course it is straightforward to check that there is this
  isomorphism, non-homotopy theory is speechless regarding its
  meaning.

%  (
%  \\
%  Under Lie integration, such a central extension corresponds to
%  a multiplicative line bundle over the Lie group $G$.
%  For example for $\mathfrak{g} = L\mathfrak{su}(n)$ the
%  the Lie algebra of the loop group of $\mathrm{SU}(n)$
%  (for $n \geq 2$), there is a 2-cocycle ....
%  The multiplicative line bundle on $\mathcal{L}\mathrm{SU}(n)$
%  which this classifies is the prequantum line bundle
%  for the WZW-model on $\mathrm{SU}(n)$.
%  \\
%  )

  This is reflected in the fact that non-homotopy theory has no
  answer to the evident followup question; if 2-cocycles
  classify central extensions, then:
  \begin{center}
    ``What do \emph{higher} cocycles classify?''
  \end{center}
  For example, the Lie algebra $\mathfrak{su}(n)$ itself
  (for $n \geq 2$)
  carries no non-trivial 2-cocycle, but it carries,
  a non-trivial 3-cocycle (in fact precisely one, up to rescaling)
  given on elements $x,y,z \in \mathfrak{su}(n)$ by
  \begin{equation}
    \mu_3(x,y,z)
    \;=\;
    \big\langle
      x, [y,z]
    \rangle
    \,,
  \end{equation}
  where $[-,-]$ is the Lie bracket, and $\langle -,-\rangle$ the Killing form. This cocycle controls both the $\mathrm{SU}(n)$ \emph{WZW-model}
  as well as $\mathrm{SU}(n)$p-Chern-Simons theory
  (see \cite{Fiorenza:2012ec,Fiorenza:2013jz,Fiorenza:2013kqa} for the higher-structure perspective
  on this phenomenon)
  and hence is crucial both in field theory (gauge instantons) as well as in string theory (rational  CFT compactifications).

  %The natural higher geometric object associated with the
  %WZW model is of course the WZW \emph{gerbe} on the Lie group $G$.
  %
  %Hence we expect that higher cocycles classify higher central
  %extensions corresponding to higher gerbes.

\medskip
  \noindent {\bf Classifying objects for cocycles.}
  The key to understanding higher cocycles on ordinary Lie algebras is
  to observe that they become \emph{representable} when
  regarded within $L_\infty$-algebra theory
  (see Figure \ref{SomeHigherLieTheory}):

  Unwinding of the definitions shows that
  a $(p+2)$-cocycle $\mu_{p+2}$ on a super Lie algebra $\mathfrak{g}$
  is equivalently an $L_\infty$-homomorphism to the line Lie $(p+2)$-algebra %$\mathfrak{l}B^{p+1}\mathbbm{R}$
  \begin{equation}
    \label{ClassifyingLieAlgebra}
    \EM{p+1} = \mathbbm{R}[p]
  \end{equation}
  from Table \ref{RationalNotation}.
  This means that the higher Lie algebra \eqref{ClassifyingLieAlgebra}
  plays the role of a
  \emph{classifying space} in higher Lie theory,
  as reflected in our notation.

  One checks that under this identification, the
  central extension classified by a 2-cocycle
  is equivalently just the \emph{homotopy fiber}
  of its classifying map. But the concept of homotopy
  fiber is defined generally, hence applies also to
  higher cocycles.

  We say that a \emph{higher central extension}
  of a super $L_\infty$-algebra
  is the homotopy fiber of a higher cocycle on it.
  We showed in
  \cite[Prop. 3.5]{Fiorenza:2013nha} \cite[Theorem 3.1.1.13]{Fiorenza:1304.6292}
  that, in coordinates, this reproduces just the
  construction of ``FDA''s in \cite{D'Auria:1982nx}.

  \noindent{\bf The string Lie 2-algebra.}
  In the example of the above 3-cocycle on $\mathfrak{su}(n)$,
  the higher central extension that it classifies is called
  the \emph{string Lie 2-algebra} (see \cite{Sati:2008eg}\cite[Appendix]{Fiorenza:2012tb})
  \begin{equation}
   \raisebox{30pt}{  \xymatrix@R=3em{
      \mathfrak{string}(\mathfrak{su}(2))
      \ar[d]|-{\color{blue}\mathrm{hofib(\mu_3)}}
      \\
      \mathfrak{su}(2)
      \ar[rr]_{\mu_3 := \langle -,[-,-]\rangle}
      &&
      \EM{3}
      %\mathbbm{R}[2]\;.
      %\mathfrak{l}\mathbf{B}^3 \mathbbm{Z}
    }
    }
  \end{equation}
  This is a Lie 2-algebra which is a $\mathbbm{R}[1]$-central extension of $\mathfrak{su}(2)$.
  The string Lie 2-algebra governs the Green--Schwarz mechanism
  of the heterotic string \cite{Sati:2009ic}, whence the name.
  See Table \ref{SomeHigherLieTheory}.

\begin{table}[htb]
  \begin{center}
  \scalebox{.9}{
  \begin{tabular}{|c|c|}
    \hline
    {\bf Traditional Lie theory} & {\bf Higher Lie theory}
    \\
    \hline
    \hline
    \begin{tabular}{c}
      cocycle
      \\
      \\
      $\mu_{p+2} \in \mathrm{CE}(\mathfrak{g})$
    \end{tabular}
    &
    \begin{tabular}{c}
    morphism
    \\
    $
      \xymatrix{
        \mathfrak{g}
        \ar[r]^-{\mu_{p+2}}
        &
        \EM{p+2}
        %\mathbbm{R}[p+1]
        %\mathfrak{l} \mathbf{B}^{p+2} \mathbbm{R}
      }
    $
    \end{tabular}
    \\
    \hline
    \begin{tabular}{c}
    coboundary
    \\
    \\
    \\
    $d \kappa = \mu'_{p+2} - \mu_{p+2} $
    \end{tabular}
    &
    \begin{tabular}{c}
    homotopy
    \\
    $
      \xymatrix{
        \mathfrak{g}
        \ar@/^1.2pc/[r]^{\mu_{p+2}}_{\ }="s"
        \ar@/_1.2pc/[r]_{\mu'_{p+2}}^{\ }="t"
        &
        \EM{p+2}
        %{\hskip .2 pt}\mathbbm{R}[p+1]
        %\mathfrak{l}\mathrlap{\mathbf{B}^{p+2} \mathbbm{R}}
        %
        \ar@{=>}|{\phantom{\int}\kappa\phantom{\int}} "s"+(0,2); "t"+(0,-2)
      }
    $
    \end{tabular}
    \\
    \hline
    \begin{tabular}{c}
      central extension
      \\
      \\
      $\widehat{\mathfrak{g}} := \mathfrak{g} \oplus \mathbbm{R} $,
      \\
      $
      \begin{aligned}
        & [(x_1,c_1), (x_2, c_2)]
        \\
        & = \big( [x_1, x_2], \mu_2(x_1,x_2) \big)
      \end{aligned}
      $
    \end{tabular}
    &
    \begin{tabular}{c}
      homotopy fiber
      \\
      \xymatrix{
        \widehat{g}
        \ar[d]|-{\color{blue} \mathrm{hofib}(\mu_2)}
        \\
        \mathfrak{g}
        \ar[r]^-{\mu_{2}}
        &
        \EM{2}
        %\mathbbm{R}[1]
        %\mathfrak{l}\mathbf{B}^2\mathbbm{R}
      }
    \end{tabular}
    \\
    \hline
    \begin{tabular}{c}
      higher
      \\
      central extension
      \\
      \\
      ?
      \\
    \end{tabular}
    &
    \begin{tabular}{c}
      homotopy fiber
      \\
      \xymatrix{
        \widehat{g}
        \ar[d]|-{\color{blue} \mathrm{hofib}(\mu_{p+2})}
        \\
        \mathfrak{g}
        \ar[r]^-{\mu_{p+2}}
        &
        \EM{p+2}
        %\mathbbm{R}[p+1]
        %\mathbf{B}^{p+1}\mathbbm{R}
      }
    \end{tabular}
    \\
    \hline
  \end{tabular}
  }
  \end{center}

  \vspace{-.7cm}

  \caption{
    {\bf Higher central extensions}
    of Lie algebras and of $L_\infty$-algebras
    are simply the homotopy fibers of
    higher cocycles, which in turn are simply
    $L_\infty$-homomorphisms to the
    \emph{line Lie $(p+2)$-algebras}
    $\EM{p+2}$
    %$\mathbf{B}^{p+1}\mathbbm{R}$.
    \cite[Prop. 3.5]{Fiorenza:2013nha} \cite[Theorem 3.1.1.13]{Fiorenza:1304.6292}.
  }
  \label{SomeHigherLieTheory}
\end{table}

  Hence in higher Lie theory, starting with a Lie algebra
  $\mathfrak{g}$, every higher cocycle gives a
  higher central extension $L_\infty$-algebra
  $\widehat{\mathfrak{g}}$; and then every higher $L_\infty$-cocycle
  on that gives a further higher central extension $L_\infty$-algebra
  $\widehat{\widehat{\mathfrak{g}}}$ and so ever on.

\vspace{-8mm}
\begin{equation}
  \raisebox{80pt}{
  \xymatrix@R=3em{
  \ar@{..}[d]
  &&
  \\
  \widehat{\widehat{\mathfrak{g}}}
    \ar[d]|-{\color{blue} \mathrm{hfib}(\mu_{p_2}+2)}
     \ar[rr]^-{\mu_{p_3+2}}
     &&
     \EM{p_3+2}
     %\mathbbm{R}[p_3+1]
     %\mathfrak{l}\mathbf{B}^{p_3+2}\mathbbm{R}
    \\
    \widehat{\mathfrak{g}}
      \ar[d]|-{\color{blue} \mathrm{hofib}(\mu_{p_1+2})}
      \ar[rr]^-{\mu_{p_2 + 2}}
      &&
      \EM{p_2+2}
      %\mathbbm{R}[p_2+1]
      %\mathfrak{l}\mathbf{B}^{p_2 + 2}\mathbbm{R}
	\\
	\mathfrak{g}
    \ar[rr]^-{\mu_{p_1 + 2}}
    &&
    \EM{p_1+2}
    %\mathbbm{R}[p_1+1]
    %\mathfrak{l}\mathbf{B}^{p_1+2} \mathbbm{R},
  }
  }
\end{equation}
Furthermore,  since there may be more than one cocycle in each step, there is
a whole \emph{bouquet} of higher central extensions \cite{Fiorenza:2013nha}:

  \begin{equation}
  \hspace{-1.2cm}
  \raisebox{30pt}{ \xymatrix@R=1em@C=1.9em{
     & \mathfrak{g}_{2,1} \ar[dr] & \cdots & \mathfrak{g}_{2,k} \ar[dl]	
	 && \mathfrak{g}_{3,1} \ar[dl]
     \\
     && \mathfrak{g}_{1,1} \ar[dr] && \mathfrak{g}_{1,2} \ar[dl] & \mathfrak{g}_{3,2} \ar[l]
     \\
     &&& \mathfrak{g} & & \mathfrak{g}_{3,3} \ar[ul]
	 \\
	 &&& \mathfrak{g}_3 \ar[u]
     \\
	 &&& \ar@{..}[u]
  }
  }
\end{equation}

\hypertarget{EmergentSuperspace-time}{
\section{Emergent super space-time}
  \label{EmergentSuperspace-time}
}

With the general principles of super homotopy theory
and of the process of higher central extensions in hand,
we now begin the discussion of the brane bouquet
(Figure \ref{Figure1}) by indicating how various
super space-times emerge out of the
superpoint \cite{Huerta:2017utu}, forming the ``trunk''
of the brane bouquet. We will spell this out in detail only in the first two
cases -- the emergence of the super line $\mathbbm{R}^{1\vert 1}$
and of the $D = 3$, $\mathcal{N}$ super Minkowski space-time
$\mathbbm{R}^{2,1\vert \mathbf{2}}$ --
which are immediate to see, but nicely illustrate
the general mechanism.
For more on this see the separate contribution \cite{contrib:huerta} to this collection.

\smallskip

\noindent {\bf The $\mathcal{N} = 1$ superpoint.}
By the general discussion in Section \ref{RationalSuperHomotopyTheory},
the algebra of functions on the $\mathcal{N} = 1$
superpoint $\mathbbm{R}^{0\vert 1}$
is generated by
one single odd-graded coordinate
\begin{equation}
  C^\infty\left(
    \mathbbm{R}^{0\vert 1}
  \right)
  \;=\;
  \mathbbm{R}
  \big[
    \underset{
      \mathclap{
        (0,\mathrm{odd})
      }
    }{
      \underbrace{
        \theta
      }
    }
  \big]
  \,.
\end{equation}
Regarded as the $D = 0$, $\mathcal{N} =1$ super Minkowski
space-time, the superpoint acts on itself by translational
supersymmetry. The corresponding
$D = 0$, $\mathcal{N} = 1$
supersymmetry super Lie algebra
(to be denoted by the same symbol $\mathbbm{R}^{0 \vert 1}$)
has a single generator $Q$ in odd degree, with vanishing
super Lie bracket
\begin{equation}
  [Q,Q] \;=\; 0\;.
\end{equation}
Therefore, its Chevalley--Eilenberg algebra has vanishing differential
and is generated by a single generator $d \theta$ (the linear dual to $Q$)
which is odd-graded but also carries unit cohomological
degree:
\begin{equation}
  \mathrm{CE}\left(
    \mathbbm{R}^{0\vert 1}
  \right)
  \;=\;
  \mathbbm{R}
  \big[
    \underset{
      (1,\mathrm{odd})
    }{
      \underbrace{
        \psi = {\rm d}\theta
      }
    }
  \big]/({\rm d}_{\mathrm{CE}} = 0)
  \,.
\end{equation}

\smallskip
\noindent{\mathversion{bold}\bf Emergence of $D = 1$ super space-time.}
By the sign rule of super homotopy theory
(Figure \ref{SignRules}, in either of its two versions)
this implies that the square of the generator
$\psi \in \mathrm{CE}\left( \mathbbm{R}^{0\vert 1}\right)$
does \emph{not} vanish.
Moreover, this square is also closed
and not exact in $\mathrm{CE}(\mathbbm{R}^{0\vert 1})$
-- trivially so, because ${\rm d}_{\mathrm{CE}}$ vanishes identically.
\begin{equation}
  \begin{aligned}
  \psi \wedge \psi
  & \neq
  0
  \\
  \psi \wedge \psi
  &
  \neq {\rm d}_{\mathrm{CE}}(\cdots)
  \\
  {\rm d}_{\mathrm{CE}}
  \big(
    \psi \wedge \psi
  \big)
  & =
  0
  \end{aligned}
  \;\;\;\in\;\;\;
  \mathrm{CE}
  \big(
    \mathbbm{R}^{0\vert 1}
  \big)
  \,.
\end{equation}
Hence $\psi \wedge \psi$ is a
non-trivial super Lie algebra 2-cocycle on the superpoint
super Lie algebra.
By the general formula (Figure \ref{SomeHigherLieTheory})
this means that this 2-cocycle $\psi \wedge \psi$
on $\mathbbm{R}^{0 \vert 1}$ classifies a
non-trivial super Lie algebra
extension by a single new bosonic element $P$,
with super Lie brackets given by
\begin{equation}
  \begin{aligned}
    [Q,Q] & = P\;,
    \\
    [Q,P] & = 0\;,
    \\
    [P,\; P] & = 0\;.
  \end{aligned}
\end{equation}
This is the translational part $\mathbbm{R}^{1|1}$
of the $D =1$, $\mathcal{N} = 1$ supersymmetry super Lie algebra,
hence the 1+0-dimensional super Minkowski space-time,
acting on itself by super translations.
\begin{equation}
  \raisebox{10pt}{ \xymatrix{
    \mathbbm{R}^{1\vert 1}
    \ar[d]|-{\color{blue}
      \mathrm{hofib}\left( \psi \wedge \psi\right)
    }
    \\
    \mathbbm{R}^{0\vert 1}
    \ar[rr]^-{ \psi \wedge \psi }
    &&
    \EM{2}
    \;.
  }
  }
\end{equation}
Dually, its Chevalley--Eilenberg algebra is
\begin{equation}
  \mathrm{CE}\big(
    \mathbbm{R}^{1\vert 1}
  \big)
  \;=\;
  \mathbbm{R}
  \big[
    \underset{
      (1,\mathrm{even})
    }{
      \underbrace{
        e = {\rm d}x
      }
    }\; , \;
    \underset{
      (1,\mathrm{odd})
    }{
      \underbrace{
        \psi = {\rm d}\theta
      }
    }
  \big]\big/
  \left(
    \begin{aligned}
      {\rm d}_{\mathrm{CE}} e & = \psi \wedge \psi
      \\
      {\rm d}_{\mathrm{CE}} \psi & = 0
    \end{aligned}
  \right)
  \,.
\end{equation}

\smallskip
\noindent {\mathversion{bold}\bf The $\mathcal{N} = 2$ superpoint.}
Similarly, the $\mathcal{N} = 2$ superpoint $\mathbbm{R}^{0 \vert 2}$
has algebra of functions
\begin{equation}
  C^\infty\left(
    \mathbbm{R}^{0\vert 2}
  \right)
  \;=\;
  \mathbbm{R}
  \big[
    \underset{
      \mathclap{
        (0,\mathrm{odd})
      }
    }{
      \underbrace{
        \theta^1
      }
    }\; ,\;\;\;
    \underset{
      \mathclap{
        (0,\mathrm{odd})
      }
    }{
      \underbrace{
        \theta^2
      }
    }
  \big]
  \,,
\end{equation}
where the sign rule (Figure \ref{SignRules}) says that
the two generators anti-commute:
\begin{equation}
  \theta^i \theta^j
  \;=\;
  -
  \theta^j \theta^i
  \,.
\end{equation}
The Chevalley--Eilenberg algebra of the corresponding
translational supersymmetry super Lie algebra is
\begin{equation}
  \mathrm{CE}\left(
    \mathbbm{R}^{0\vert 2}
  \right)
  \;=\;
  \mathbbm{R}
  \big[
    \underset{
      (1,\mathrm{odd})
    }{
      \underbrace{
        \psi^1 = {\rm d}\theta^1
      }
    } \; , \;
        \underset{
      (1,\mathrm{odd})
    }{
      \underbrace{
        \psi^2 = {\rm d}\theta^2
      }
    }
  \big]\big/({\rm d}_{\mathrm{CE}} = 0)
  \,.
\end{equation}
Due to the bi-grading,
the sign rule (Figure \ref{SignRules})
now says that these generators \emph{commute} with each other
in the Chevalley--Eilenberg algebra:
\begin{equation}
  \psi^i \wedge \psi^j
  \;=\;
  +
  \psi^j \wedge \psi^i
  \,.
\end{equation}

\smallskip
\noindent {\mathversion{bold}\bf Emergences of $D = 3$ super Minkowski space-time.}
By the same argument as before, this means that all
three wedge squares
\begin{equation}
  \raisebox{14pt}{
  \xymatrix@R=5pt@C=5pt{
    \psi^1 \wedge \psi^1,
    &
    \psi^1  \wedge \psi^2
    \ar@{=}[dl]
    \\
    \psi^2 \wedge \psi^1,
    &
    \psi^2 \wedge \psi^2
  }
  }
  \;\;\in\;
  \mathrm{CE}
  \left(
    \mathbbm{R}^{0\vert 2}
  \right)
\end{equation}
are nontrivial super Lie algebra 2-cocycles on $\mathbbm{R}^{0\vert 2}$.
Since there are three distinct such, the space of 2-cocycles
on $\mathbbm{R}^{0\vert 2}$ is 3-dimensional, and hence the
universal central extension of $\mathbbm{R}^{0\vert 1}$ has
three extra even generators $P_+$, $P_-$, $P_2$.
It remains to identify what the
general formula (Figure \ref{SomeHigherLieTheory})
for the super Lie bracket in this universal extensions gives.
We claim that if we identify
\begin{equation}
  P_0 \coloneqq \tfrac{1}{2}\left(P_+ - P_-\right)
  \quad
  \text{and}
  \quad
  P_1 \coloneqq \tfrac{1}{2}\left(P_+ + P_-\right)
\end{equation}
then this super Lie bracket is that of
the translational $D = 3$ $\mathcal{N} = 1$ supersymmetry
super Lie algebra, with non-trivial super Lie bracket the
usual
    \begin{equation}
      \{Q_\alpha, Q'_\beta\}
      =
      C_{\alpha \alpha'} \Gamma_a{}^{\alpha'}{}_\beta \,P^a
      \,,
    \end{equation}
   where $C_{\alpha \beta}$ is the charge conjugation matrix.
For this, observe (see \cite{Baez:2009xt} for review) the exceptional isomorphism
\begin{equation}
  \mathrm{Spin}(2,1)
  \;\simeq\;
  \mathrm{SL}(2,\mathbbm{R})
\end{equation}
under which the irreducible real $\mathrm{Spin}(2,1)$-representation
is simply
\begin{equation}
  \mathbf{2}
  \;\simeq\;
  \mathbbm{R}^2
\end{equation}
via the canonical action of $\mathrm{SL}(2,\mathbbm{R})$.
Moreover, using the identification of real inner product spaces
\begin{equation}
  \left(
    \mathbbm{R}^{2,1},\eta
  \right)
  \;\simeq\;
  \left(
    \mathrm{Mat}^{\mathrm{herm}}_{2\times 2}(\mathbbm{R}),
    -\mathrm{tr}
  \right)
\end{equation}
of 3d bosonic Minkowski space-time, with the symmetric
real $2\times 2$ matrices (the ``real Pauli matrices''),
the bilinear spinor-to-vector pairing is
given just by matrix multiplication of row vectors with
column vectors.

But since the 3-dimensional space of cocycles found above
is manifestly identified with that of symmetric $2 \times 2$
real matrices, the claim follows.
\begin{equation}
  \raisebox{20pt}{ \xymatrix@R=1.4em{
    \mathbbm{R}^{2,1\vert \mathbf{2}}
    \ar[d]_-{\color{blue}
      \mathrm{hofib}
    }
    \\
    \mathbbm{R}^{0\vert 2}
    \ar[rrr]|-{\tiny
        \;\left(
          \begin{array}{cc}
            \psi^1 \wedge \psi^1, & \psi^1 \wedge \psi^2
            \\
            & \psi^2 \wedge \psi^2
          \end{array}
        \right)
    \; }
    &&&
    \mathfrak{l}K\left(\mathbbm{Z}^{\oplus 3},2\right)\;.
  }
  }
\end{equation}

\smallskip
\noindent
{\bf Emergence of exceptional space-time.}
In the previous discussion we doubled supersymmetry
by passing from the $\mathcal{N} = 1$ superpoint to the
$\mathcal{N} = 2$ superpoint. In the same manner one
may discuss superpoints of higher supersymmetry and their
central extensions. Skipping ahead, consider the
superpoint $\mathbbm{R}^{0\vert 32}$.
By the directly analogous argument as before, it follows
that its maximal central extension is
by the space of symmetric real $32 \times 32$-matrices,
which is of dimension 528. The result is readily seen
\cite[Sec. 4.3]{Fiorenza:2018ekd}
to be an ``exceptional generalized geometry''-version
of $D = 11$ space-time in the sense of \cite{Hull:2007zu},
hence denoted $\mathbbm{R}^{10,1\vert \mathbf{32}}_{\mathrm{exc}}$
in Figure \ref{Figure1}. For further discussion of this
branch of the brane bouquet, we refer to \cite{Fiorenza:2018ekd,Sati:2018tvj}.

\smallskip
\noindent
{\bf The significance of universal invariant extensions.}
This last example is occasion to highlight the crucial
role of \emph{universal invariant} central extensions in the brane bouquet:
From the embedding \eqref{SuperMinkowskiCE} it is clear that
\emph{every} super Minkowski space-time is
\emph{some} central extension of one of lower dimension.
In particular every super Minkowski space-time is a
central extension of a superpoint
(as highlighted in \cite[Sec. 2.1]{Chryssomalakos:1999xd}).
What singles out the specific super space-times in the brane bouquet,
however, is that they are not random central extensions,
but rather universal invariant central extensions.

\begin{table}[htb]
%\vspace{-.3cm}
\begin{center}
\scalebox{.9}{
\begin{tabular}{|p{0.3cm}|p{4.8cm}|c|}
  \hline
  &
  {\bf
  \begin{tabular}{cc}
    Invariant
    higher
    \\
    central
    extension
  \end{tabular}
  }
  &
  \hspace{-6pt}
  {\bf
  \begin{tabular}{c}
    Cocycles /
    \\
    WZW-terms
  \end{tabular}
  }
  \hspace{-6pt}
  \\
  \hline
  \hline
  \multirow{3}{*}{
    \hspace{.1cm}
    \raisebox{
      -68pt
    }{
      \begin{rotate}{90}
        {\bf
          IIA superstring
        }
      \end{rotate}
    }
  }
  &
  \multirow{2}{*}{
    \xymatrix@C=17pt@R=3em{
      \mathfrak{string}_{\mathrm{IIA}}
      \ar[d]|-{\color{blue}
       \mathrm{hofib}
        \big(
          \mu^{\mathrm{IIA}}_{{}_{F1}}
        \big)
      }
      \\
      \mathbbm{R}^{9,1\vert \mathbf{16} + \overline{\mathbf{16}}}
      \ar[rr]^-{
        \mu^{\mathrm{IIA}}_{{}_{F1}}
      }_{
        =
        \left(\overline{\psi} \wedge \Gamma_a \psi\right) \wedge e^a
      }
      &&
      \EM{3}
    }
  }
  &
  \begin{tabular}{c}
    \hspace{-6pt}
    {\bf as WZW term}
    \hspace{-6pt}
    \\
    \cite{Green:1983wt}
  \end{tabular}
  \\
  \cline{3-3}
  & &
  \begin{tabular}{c}
    \hspace{-6pt}
    {\bf as cocycle}
    \hspace{-6pt}
    \\
    \cite{Henneaux:1984mh}
    \\
    \cite{DeAzcarraga:1989vh}
  \end{tabular}
  \\
  \cline{3-3}
  &
  \multicolumn{2}{c|}{
    \raisebox{-17pt}{
    $
      \mathrm{CE}
      \big(
        \mathfrak{string}_{\mathrm{IIA}}
      \big)
      =
      \mathrm{CE}
      \big(
        \mathbbm{R}^{9,1\vert \mathbf{16} + \overline{\mathbf{16}}}
      \big)
      \big[
        f_2
      \big]
      \big/
      \left(
        {\rm d}_{\mathrm{CE}}  f_2 = \mu^{\mathrm{IIA}}_{{}_{F1}}
      \right)
    $
    }
  }
  \\
  \hline
  \hline
  \multirow{3}{*}{
    \hspace{.1cm}
    \raisebox{
      -68pt
    }{
      \begin{rotate}{90}
        {\bf
          IIB superstring
        }
      \end{rotate}
    }
  }
  &
  \multirow{2}{*}{
    \xymatrix@C=17pt{
      \mathfrak{string}_{\mathrm{IIB}}
      \ar[d]|-{\color{blue}
        \mathrm{hofib}
        \left(
          \mu^{\mathrm{IIB}}_{{}_{F1}}
        \right)
      }
      \\
      \mathbbm{R}^{9,1\vert \mathbf{16} + {\mathbf{16}}}
      \ar[rr]^-{
        \mu^{\mathrm{IIB}}_{{}_{F1}}
      }_{
        =
        \left(\overline{\psi} \wedge \Gamma_a \psi\right) \wedge e^a
      }
     &&
     \EM{3}
    }
  }
  &
  \begin{tabular}{c}
    {\bf as WZW term}
    \\
    \cite{Green:1983wt}
  \end{tabular}
  \\
  \cline{3-3}
  & &
    \begin{tabular}{c}
      {\bf as cocycle}
      \\
      \cite{Henneaux:1984mh}
      \\
      \cite{DeAzcarraga:1989vh}
    \end{tabular}
  \\
  \cline{3-3}
  &
  \multicolumn{2}{c|}{
    \raisebox{-17pt}{
  $
    \mathrm{CE}
    \big(
      \mathfrak{string}_{\mathrm{IIB}}
    \big)
    \;=\;
    \mathrm{CE}
    \left(
      \mathbbm{R}^{9,1\vert \mathbf{16} + {\mathbf{16}}}
    \right)
    \big[
      f_2
    \big]
    \big/
    \left(
      {\rm d}_{\mathrm{CE}}  f_2 = \mu^{\mathrm{IIB}}_{{}_{F1}}
    \right)
  $
  }}
  \\
  \hline
  \hline
  \multirow{3}{*}{
    \hspace{.1cm}
    \raisebox{
      -68pt
    }{
      \begin{rotate}{90}
        {\bf
          super membrane
        }
      \end{rotate}
    }
  }
  &
  \multirow{2}{*}{
    \xymatrix@C=17pt{
      \mathfrak{m}2\mathfrak{brane}
      \ar[d]|-{\color{blue}
        \mathrm{hofib}
        \left(
          \mu_{{}_{M2}}
        \right)
      }
      \\
      \mathbbm{R}^{10,1\vert \mathbf{32}}
      \ar[rr]^-{
        \mu_{{}_{M2}}
      }_{
        =
        \tfrac{i}{2}\left(\overline{\psi} \wedge \Gamma_{a b} \psi \right) \wedge e^a \wedge e^b
      }
      &&
      \EM{4}
    }
  }
  &
  \begin{tabular}{c}
    {\bf as WZW term}
    \\
    \cite{Bergshoeff:1985su}
  \end{tabular}
  \\
  \cline{3-3}
  &&
  \begin{tabular}{c}
    {\bf as cocycle}
    \\
    \cite{D'Auria:1982nx}
    \\
    \cite{DeAzcarraga:1989vh}
  \end{tabular}
  \\
  \cline{3-3}
  &
  \multicolumn{2}{c|}{
    \raisebox{-17pt}{
    $
    \mathrm{CE}
    \big(
      \mathfrak{m}2\mathfrak{brane}
    \big)
    \;=\;
    \left(
      \mathrm{CE}
      \left(
        \mathbbm{R}^{10,1\vert \mathbf{32}}
      \right)
      \big[
        c_3
      \big]
    \right)
    \big/
    \left(
      {\rm d}_{\mathrm{CE}}  c_3
      =
      \mu_{{}_{M2}}
    \right)
    $
    }
  }
  \\
  \hline
\end{tabular}
}
\end{center}

\vspace{-.7cm}

\caption{
 {\bf The branes without tensor multiplets}
 (without gauge fields) on their world-volume
 arise in the brane bouquet
 as the invariant higher central extensions
 of super space-time itself. These are the
 branes captured also by the old brane scan
 (Figure \ref{TheOldBraneScan}).
 The remaining branes arise in the next higher
 stage of the brane bouquet (Figure \ref{HigherStageOfTheBraneBouquet}).
}
\label{FirstStageOfBraneBouquet}

\end{table}

\hypertarget{TheBraneBouquet}{
\section{The brane bouquet}
  \label{TheBraneBouquet}
}

With superspace-time having emerged from the superpoint
(Section \ref{EmergentSuperspace-time}) by a sequence
of ordinary invariant central extensions, terminating
in 11-dimensional super Minkowski space-time,
we may now look for further \emph{higher} invariant
central extensions (Table \ref{SomeHigherLieTheory}),
first of superspace-time itself, and then iterating.
We indicate here
(see Figure \ref{FirstStageOfBraneBouquet} and
Figure \ref{HigherStageOfTheBraneBouquet}) how this produces the brane bouquet shown in Figure \ref{Figure1}.

\begin{table}[htb]
\begin{center}
\scalebox{.9}{
\begin{tabular}{|p{.04cm}|p{5.1cm}|c|}
  \hline
   \multirow{3}{*}{
    \hspace{.1cm}
    \raisebox{
      -109pt
    }{\!\!\!\!
      \begin{rotate}{90}
        {\bf
          super D$p$-branes
        }
      \end{rotate}
    }
  }
  &
  {\bf
  \begin{tabular}{cc}
    Invariant
    higher
    \\
    central
    extension
  \end{tabular}
  }
  &
  \hspace{-6pt}
  {\bf
  \begin{tabular}{c}
    Cocycles /
    \\
    WZW-terms
  \end{tabular}
  }
  \hspace{-6pt}
  \\
  \hline
  \hline
  &
   \multirow{2}{*}{
     \!\!\!\!\!\!\!\!\! \xymatrix@R=4.5em@C=2.1em{
      \mathfrak{d}2p\mathfrak{brane}
      \ar[d]|-{\color{blue}
        \mathrm{hofib}
        \left(
          \mu_{{}_{D2p}}
        \right)
      }
      \\
      \mathfrak{string}_{\mathrm{IIA}}
      \ar[rrrrr]^-{
        \mu_{{}_{D2p}} =
      }_-{\tiny
        \sum_{k = 0}^{p+1}
        c_k^{2p}
        \big(
          \overline{\psi}
          \Gamma^{a_1 \cdots a_{2p-2k} (11)}
          \psi
        \big)
        e^{a_1}
        \wedge
          \cdots
        \wedge
        e^{a_{2p-2k}}
        \wedge f_2^{\wedge^k}
      }
      &&&&&
      \EM{2p+2}
    }
  }
  &
  \begin{tabular}{c}
    {\bf as WZW term}
    \\
    \cite{Cederwall:1996ri}
  \end{tabular}
  \\
  \cline{3-3}
  &
  &
  \begin{tabular}{c}
    {\bf as cocycle}
    \\
    \cite{Chryssomalakos:1999xd}
    \\
    \cite{Sakaguchi:1999fm}
  \end{tabular}
  \\
  \cline{3-3}
  &
  \multicolumn{2}{l|}{
    \raisebox{-30pt}{
      $
      \mathrm{CE}
      \big(
        \mathfrak{d}2p\mathfrak{brane}
      \big)
      =
      \mathrm{CE}
      \big(
        \mathfrak{string}_{IIA}
      \big)
      \big[
        c_{2p+1}
      \big]\big/
      \left(
        {\rm d}_{\mathrm{CE}}  c_{2p+1}
        =
        \mu_{D2p}
      \right)
      $
    }
  }
  \\
  \hline
  \hline
  \multirow{3}{*}{
    \hspace{.1cm}
    \raisebox{
      -80pt
    }{\!\!\!\!
      \begin{rotate}{90}
        {\bf
          super five-brane
        }
      \end{rotate}
    }
  }
  &
  \multirow{2}{*}{
    \xymatrix@R=6em@C=2.3em{
      \mathfrak{m}5\mathfrak{brane}
      \ar[d]|-{\color{blue}
        \mathrm{hofib}
        \left(
          \mu^{\mathrm{IIA}}_{{}_{F1}}
        \right)
      }
      \\
      \mathbbm{R}^{10,1\vert \mathbf{32}}
      \ar[rrrrr]^-{
        \mu_{{}_{M5}} =
      }_{\tiny
                \tfrac{1}{5!} \left( \overline{\psi} \Gamma_{a_1 \cdots a_5} \psi\right) \wedge  e^{a_1} \wedge
                \cdots \wedge e^{a_5}
       +
          c_3 \wedge \tfrac{i}{2} \left( \overline{\psi} \Gamma_{a_1 a_2} \psi \right) \wedge e^{a_1} \wedge e^{a_2}
            }
      &&&&&
      \EM{4}
    }
  }
  &
  \begin{tabular}{c}
    {\bf as WZW term}
    \\
    \cite{Bandos:9701149}
  \end{tabular}
  \\
  \cline{3-3}
  &&
  \begin{tabular}{c}
    {\bf as cocycle}
    \\
    \cite{D'Auria:1982nx}
    \\
    \cite{Chryssomalakos:1999xd}
  \end{tabular}
  \\
  \cline{3-3}
  &
  \multicolumn{2}{l|}{
    \raisebox{
      -50pt
    }
    {
      $
      \mathrm{CE}
      \big(
        \mathfrak{m}5\mathfrak{brane}
      \big)
      =
      \mathrm{CE}
      \big(
        \mathfrak{m}2\mathfrak{brane}
      \big)
      \big[
        c_3
      \big]
      \big/
      \left(
        {\rm d}_{\mathrm{CE}} c_3 =
        \mu_{{}_{M2}}
      \right)
      $
    }
  }
  \\
  \hline
\end{tabular}
}
\end{center}

\vspace{-.7cm}

\caption{
  {\bf The branes with tensor multiplets}
  (gauge fields or higher gauge fields)
  on their world-volume arise
  in the brane bouquet as the invariant higher
  central extensions of the
  higher extensions corresponding to the
  branes without tensor multiplets (Figure \ref{FirstStageOfBraneBouquet}).
  This relation also gives the intersection laws \cite[Sec. 3]{Fiorenza:2013nha}
  -- strings ending on D-branes and M2-branes ending on M5-branes,
  respectively.
}
\label{HigherStageOfTheBraneBouquet}

\end{table}

\smallskip
\noindent {\bf Old brane scan cocycles.}
At the first stage, the invariant higher central extensions
of super Minkowski space-time are classified by the invariant higher
Lie algebra cocycles of the corresponding ordinary translational
supersymmetry super Lie algebras. These have been classified
by a variety of methods
\cite{Achucarro:1987nc,DeAzcarraga:1989vh,Movshev:2010mf,Movshev:2011pr,Brandt:2010fa,Brandt:2010tz,Brandt:2013xpa}
and constitute the ``old brane scan'' (Table \ref{TheOldBraneScan}).

In particular, in the ``critical'' dimensions one finds:
\smallskip
\begin{enumerate}[i)]
\item The maximal invariant 3-cocycle on ten-dimensional super Minkowski space-time is

\vspace{-5mm}
\begin{equation}
  \mu_{F1} \;=\;
  \left(\overline{\psi} \wedge \Gamma_a \psi\right) \wedge e^a
  \;\;
  \in
  \mathrm{CE}
  \left(
    \mathbbm{R}^{9,1\vert \mathbf{16} + \overline{\mathbf{16}}}
  \right)
\end{equation}
and this is the curvature of the WZW term for the Green--Schwarz superstring \cite{Green:1983wt}.

\item
The maximal invariant 4-cocycle on super Minkowski space-time is
\begin{equation}
  \mu_{M2}
   \;=\;
   \tfrac{i}{2}\left(\overline{\psi} \wedge \Gamma_{a b} \psi \right) \wedge e^a \wedge e^b
   \;\;
   \in
   \mathrm{CE}
   \left(
     \mathbbm{R}^{10,1\vert \mathbf{32}}
   \right)
\end{equation}
and this the curvature of the higher WZE term for the supermembrane
\cite{Bergshoeff:1987cm}.
\end{enumerate}
\smallskip
This situation is summarized in Table \ref{FirstStageOfBraneBouquet}.

\smallskip
\noindent {\bf Higher super Minkowski space-time.}
That each of these cocycles in turn defines a new ``FDA''
with a higher-degree generator added has been emphasized
back in  \cite{D'Auria:1982nx} and developed into general theory
of higher dimensional supergravity \cite{Castellani:1991et}.
These ``FDA''-extensions were re-amplified in \cite{Chryssomalakos:1999xd}
and termed \emph{extended superspace-times} there.
\footnote{Beware the potential conflict of terminology with
``extended supersymmetry'', which refers to higher $\mathcal{N}$, instead.}
Finally the interpretation of these as the Chevalley--Eilenberg algebras
of corresponding higher super $L_\infty$-algebra extensions
(as per Table \ref{SomeHigherLieTheory}) is due to
\cite{Fiorenza:2013nha} (following \cite{Sati:2008eg} and using \cite[Theorem 3.1.1.13]{Fiorenza:1304.6292}).
In terms of higher homotopy-theoretic geometry
this means \cite{Nikolaus:1207ab}
that these are
 \emph{higher gerbes} on super space-time (here in their rational/infinitesimal approximation):

\medskip
\begin{equation}
  \raisebox{50pt}{ \xymatrix{
    \ar@{}[d]|-{
      \mbox{
        \raisebox{-75pt}{
          \begin{rotate}{90}
            \tiny
            \color{blue}
            \begin{tabular}{c}
              infinitesimal higher
              super gerbes
              \\
              over super space-time
            \end{tabular}
          \end{rotate}
        }
      }
    }
    &
    \widehat{\widehat{
      \mathbbm{R}^{d,1\vert \mathbf{n}}
    }}
    \ar[d]|-{\color{blue}
      \mathrm{hofib}
      \left(
        \mu_{p_2+2}
      \right)
    }
    &&
    \\
    &
    \widehat{
      \mathbbm{R}^{d,1\vert \mathbf{N}}
    }
    \ar[d]|-{\color{blue}
      \mathrm{hofib}
      \left(
        \mu_{p_1+2}
      \right)
    }
    \ar[rr]^-{ \mu_{p_2 + 2} }
    &&
    \EM{p_2 + 2}
    \\
    &
    \mathbbm{R}^{d,1\vert \mathbf{N}}
    \ar[rr]^-{\mu_{p_1+2}}
    &&
    \EM{p_1+2}
    \\
    &
   \ar@{}[rr]|-{ \vspace{-1cm}
      \underset{
        \mbox{
          \tiny
          \color{blue}
          \begin{tabular}{c}
            brane WZW terms /
            \\
            super cocycles
          \end{tabular}
        }
      }{
        \underbrace{
          \phantom{----------}
        }
      }
    }
    &&
  }
  }
\end{equation}

\noindent{\bf Branes with tensor multiplets.}
These higher super\linebreak Minkowski space-times $\mathfrak{string}_{\mathrm{IIA/B}}$
and $\mathfrak{m}2\mathfrak{brane}$
(from Figure \ref{FirstStageOfBraneBouquet})
now turn out to carry \emph{further} invariant cocycles,
corresponding to the D-branes and the M5-brane. This is how the brane
bouquet completes and then goes beyond the old brane scan.
This situation is shown in Table \ref{HigherStageOfTheBraneBouquet}.

\smallskip
\noindent {\bf Ends and loose ends of the brane bouquet.}
In closing the discussion of the brane bouquet itself,
we indicate some further points of interest and some open
questions:
\smallskip
\begin{enumerate}[i)]
  \item
  {\bf Further branes.}
  Not all branches of universal invariant higher central extensions
  are shown in Figure \ref{Figure1}.
  For instance:
  \begin{enumerate}[a)]
  \item
  {\bf Branes in lower dimensions.}
  The old brane scan (Table \ref{TheOldBraneScan})
  says that there are Green--Schwarz strings in
  also dimensions $D =3$, $D =4$ and $D = 6$. Their WZW curvatures
  are invariant cocycles which classify higher central extensions that
  would branch off, in Figure \ref{Figure1}, from the corresponding
  super space-time entries.
  These super cocycles in lower space-time dimension
  have traditionally been mostly ignored or dismissed
  (e.g. as not ``quantum consistent'' \cite[p. 15]{Duff:1987qa})
  since they do not directly match to the
  \emph{critical} NSR string in $D = 10$. But the little investigation
  that has been done here suggests that there is
  more to be said:
    \begin{itemize}
       \item {\bf string in $D =3$}: see \cite{Mezincescu:2010yp,Mezincescu:2011nh,Mezincescu:2013xza}.
       \item {\bf membrane in $D =4$}:
         see \cite{Achucarro:1988qb,Witten:1995ex,Bandos:2018gjp}.
    \end{itemize}
  \end{enumerate}

  \item {\bf Further super space-times.}
    There are also more super space-times in the brane
    bouquet than shown in Figure \ref{Figure1}:
    \begin{enumerate}[a)]
   \item {\mathversion{bold}\bf $D =1$, $\mathcal{N} =1 $ super space-time.}
     We have already seen in Section \ref{EmergentSuperspace-time},
   as the first trivial example given there, that
   also $D = 1$, $\mathcal{N}$ super Minkowski space appears in the
   brane bouquet as the universal central extension of the
   $\mathcal{N} =1$ superpoint $\mathbbm{R}^{0\vert 1}$.
     \item {\bf Exceptional space-times.}
     Similarly, at the other extreme of number of supersymmetries,
     we saw in Section \ref{EmergentSuperspace-time}
     the exceptional M-theory space-time emerges out of the
     $\mathcal{N}= 32$ superpoint. In between these two extremes,
     there will be branches of the brane bouquet emerging
     out of each of the intermediate superpoints $\mathbbm{R}^{0\vert 2^n}$.
     These remain to be investigated.
   \end{enumerate}

  \item
  {\mathversion{bold}\bf space-time progression terminates at $D = 11$, $\mathcal{N} = 1$.}
  However, the progression of space-times
  emerging out of $\mathbbm{R}^{0\vert 2}$ does stop after $D =11$, $\mathcal{N} = 1$
  superspace-time: If here we again double the fermions
  to pass to $D =11$, $\mathcal{N} = 2$ superspace-time,
  we find that this has a 1-dimensional invariant central extension,
  classified by the invariant 2-cocycle which is the invariant spinor
  pairing $\mathbf{32} \otimes \mathbf{32} \to \mathbbm{R}$ which
  appears inside $\mathrm{osp}(1\vert 64)$, corresponding to the
  dilatation operator \cite[Table 7]{vanHolten:1982mx}\cite[p 4-5]{Bars:9904063}.
\end{enumerate}

\hypertarget{ChargeQuantization}{
\section{Brane charge quantization}
 \label{ChargeQuantization}
}

After the bouquet has developed (Section \ref{TheBraneBouquet}),
we may, conversely,
ask whether its branches, corresponding to single isolated $p$-brane
species, may be descended back and unified into
a single cocycle in \emph{generalized cohomology}.
Here we review how this operation discovers,
that \emph{rationally}:
\smallskip
\begin{enumerate}[i)]
  \item the unified F1/D$p$-brane charge is in
twisted K-theory \cite[Sec. 4]{Fiorenza:2016ypo}
 \item the unified M2/M5-brane cocycle is in
 degree-4 \emph{Cohomotopy cohomology theory}
 \cite{Fiorenza:2015gla}.
\end{enumerate}
\smallskip

Before discussing the computation, the following comment is
in order.

\smallskip
\noindent {\bf Open question of brane charge quantization.}
The first statement above resonates with established folklore
\cite{Bouwknegt:2000qt,Witten:2000cn},
while the second matches with an observation
about the charge structure of the C-field in
11-dimensional supergravity that was made only more
recently, in \cite[Sec. 2.5]{Sati:2013rxa}.
The search for a generalized cohomology theory underlying
M-theory was advocated and initiated in \cite{Sati:2005hx,Sati:2005mi,Sati:2005wy,Sati:2010ss}
and formulated in terms of cohomotopy in \cite{Sati:2013rxa}.

Of course many different cohomology theories
share a given rationalization (Figure \ref{SullivanConstruction}),
so that the names
assigned to these rational cohomologies theories (``K-theory'', ``cohomotopy'')
a priori have a large degree of arbitrariness.

In fact, the seminal proposal that D-brane charge is quantized
in K-theory is based on differential form-level (hence rational)
computations \cite{Minasian:1997mm} combined with a plausible
but informal and unproven non-rational argument about tachyon condensation.
\cite[Sec. 3]{Witten:9810188}, and partial consistency checks \cite{Diaconescu:2000wy}.
But other plausibility aguments indicate, on the contrary, that
twisted K-theory can \emph{not} be quite the right choice, for instance since it apparently
produces spurious D-brane states \cite[around (137)]{deBoer:2001wca}, and
since it seems to be incompatible with S-duality \cite{Kriz:2004tu}\cite[8.3]{Evslin:2006cj}.

What has been missing here,
as throughout string/M-theory, is an actual formulation of the
ambient theory from which these questions could be decided
systematically, without relying on
educated guesswork and plausibility arguments.

The embedding of the question of D-brane charge
into the broader structure of the brane bouquet may provide just that.
Indeed, in Section \ref{MIIADuality} below we recall from \cite{Braunack-Mayer:2018uyy} that in a
full M-theoretic perspective rational D1/D$p$-brane is
equivalently in the fiberwise stabilization of the fiberwise looping of the
double dimensional reduction of the M2/M5-brane charge.
These are \emph{universal constructions} that apply equally
to any non-rational lift of M2/M5-brane charge. This
way the question about non-rational D-brane charge quantization reduces
to that of non-rational M-brane charge quantization.
We will turn to this below in Section \ref{ChargeQuantizationInEquivariantCohomotopy}.

Now we explain how the brane bouquet knows about
brane charge quantization.

\smallskip
\noindent {\bf Descent of cocycles.}
First consider the general situation:
Suppose a double stage in a bouquet of extensions,
hence (by Figure \ref{SomeHigherLieTheory}) a diagram
of super $L_\infty$-algebras of the form
\begin{equation}
   \raisebox{30pt}{
   \xymatrix@R=1.3em@C=10pt{
    \widehat{
      \mathfrak{g}
    }
    \ar[rr]^{ \mu_{{}_{p_2 + 2}} }
    \ar[d]_-{\color{blue}
      \mathrm{hofib}
      \left(
        \mu_{p_1+2}
      \right)
    }
    &&
    \EM{p_2+2}
    %\mathfrak{l}\mathbf{B}^{p_2+2}
    \\
    \mathfrak{g}
    \ar[dr]_{ \mu_{p_1 + 2} }
    \\
    &
    \EM{p_1 + 2}
  }
  }
\end{equation}
Here the left vertical morphism may equivalently be
regarded as a (multiplicative) $\EM{p_1}$-principal
$\infty$-bundle \cite{Nikolaus:1207ab} (a higher gerbe) over $\mathfrak{g}$,
so that the second cocycle $\mu_{{}_{p_2 + 2}}$ is a morphism on
a space with $\EM{p_1}$ $\infty$-action.
Therefore, a natural question is if this cocycle is equivariant
with respect to this action. For this to have content, we need
to specify also a $\EM{p_1}$-action on its codomain
$\EM{p_2+2}$. Again by the general results of \cite{Nikolaus:1207ab}, such a choice of
action is actually equivalent to there being a corresponding homotopy
fiber sequence as shown on the right here:
\begin{equation}
  \raisebox{50pt}{
  \xymatrix@R=3em@C=1.75em{
    \widehat{
      \mathfrak{g}
    }
    \ar[rr]^{ \mu_{{}_{p_2 + 2}} }
    \ar[d]|-{ \color{blue}
      \mathrm{hofib}(\mu_{p_1+2})
    }
    &&
    \EM{p_2+2}
    \ar[d]|-{\color{blue}
      \mathrm{hofib}(c)
    }
    \\
    \mathfrak{g}
    \ar[dr]_{ \mu_{p_1 + 2} }
    \ar@{-->}[rr]^-{
      \mu_{{}_{\mathrm{unified}}}
    }
    &&
    \mathfrak{l}
    \big(
      K(\mathbbm{Z}p_2+1)\sslash K(\mathbbm{Z},p_1)
    \big)
    \ar[dl]^{c}
    \\
    &
    \EM{p_1 + 2}
  }}
\end{equation}
and with this the equivariance of $\mu_{p_2 + 2}$
is equivalent to it descending to a
dashed horizontal morphism $\mu_{\mathrm{unified}}$,
as shown, which
makes the resulting square and the triangle commute up to homotopy
(we do not display these homotopies for simplicity, but they are there).

\smallskip
We now specify this general situation to the case of D-branes
and M-branes in the brane bouquet, Figure \ref{Figure1}.

\smallskip
\noindent {\bf The unified F1/D$p$-brane cocycle.} \cite[Sec. 4]{Fiorenza:2016ypo}
First of all, the collection of $D_{2p}$-brane super cocycles
\begin{equation}
  \mathfrak{string}_{\mathrm{IIA}}
  \xrightarrow{\;
    \mu_{2p}\;
  }
  \EM{2p+2}
\end{equation}
on the type IIA super string super Lie 2-algebra
is trivially summed up as a single cocycle with
coefficients in the Cartesian product of classifying spaces

\vspace{-1cm}
\begin{equation}
\raisebox{80pt}{  \xymatrix@R=2em@C=2em{
    \\
    \mathfrak{string}_{\mathrm{IIA}}
    \ar[rr]^{ \underset{p}{\prod}\mu_{{}_{D_{2p}}} }
    \ar[d]_-{\color{blue}
      \mathrm{hofib}\left( \mu^{\mathrm{IIA}}_{{}_{F1}}\right)
    }
    &&
    \mathfrak{l}\underset{p}{\prod}
    \,
    K(\mathbbm{Z},2p+2)
    \\
    \mathbbm{R}^{9,1\vert \mathbf{16} + \overline{\mathbf{16}}}
    \ar[dr]_{\mu_{F1}^{\mathrm{IIA}}}
    \\
    &
    \EM{3}
  }}
\end{equation}
Descent of this situation turns out to be given by
\begin{equation}
 \raisebox{80pt}{ \xymatrix@R=1.8em@C=2em{
    \\
    \mathfrak{string}_{\mathrm{IIA}}
    \ar[rr]^{ \underset{p}{\prod}\mu_{D_{2p}} }
    \ar[d]_-{\color{blue}
      \mathrm{hofib}
      \left(
        \mu^{\mathrm{IIA}}_{{}_{F1}}
      \right)
    }
    &&
    \mathfrak{l}\underset{p}{\prod}
    \,
    K(\mathbbm{Z},2p+2)
    \ar[d]
    \\
    \mathbbm{R}^{9,1\vert \mathbf{16} + \overline{\mathbf{16}}}
    \ar@{-->}[rr]^{\mu_{F1/Dp}}
    \ar[dr]_{\mu_{F1}^{\mathrm{IIA}}}
    &&
    \mathfrak{l}
    \big(
      \mathrm{KU}\sslash K(\mathbbm{Z},2)
    \big)
    \ar[dl]
    \\
    &
    \EM{3}
  }}
\end{equation}

\noindent {\bf The unified M-brane cocycle.} \cite{Fiorenza:2015gla}
The separate super cocycles for the M2-brane and the M5-brane
appear as

\vspace{-1.2cm}
\begin{equation}
\raisebox{80pt}{
  \xymatrix@R=1.8em{
    \\
    \mathfrak{m}2\mathfrak{brane}
    \ar[rr]^{ \mu_{M5}}
    \ar[d]_-{\color{blue}
      \mathrm{hofib}\left( \mu_{{}_{M2}}\right)
    }
    &&
    \mathfrak{l}S^7
    \\
    \mathbbm{R}^{10,1\vert \mathbf{32}}
    \ar[dr]_{\mu_{M2}}
    &&
    \\
    &
    \EM{4}
  } }
\end{equation}
where
\begin{equation}
   \mu_{{}_{M2}}
   \;=\;
   \tfrac{i}{2} \left( \overline{\psi} \Gamma_{a_1 a_2} \psi\right) \wedge  e^{a_1} \wedge e^{a_2}
\end{equation}
is the WZW-curvature of the Green--Schwarz-type sigma model
for the M2-brane, while
\begin{align}
  \mu_{{}_{\mathrm{M5}}}
 \! = \!
  \tfrac{1}{5!} \left( \overline{\psi} \Gamma_{a_1 \cdots a_5} \psi\right) \wedge  e^{a_1} \wedge \cdots \wedge e^{a_5}
  \nonumber \\
  \;+\;
  c_3 \wedge \tfrac{i}{2} \left( \overline{\psi} \Gamma_{a_1 a_2} \psi \right) \wedge e^{a_1} \wedge e^{a_2}
\end{align}
is the WZW-curvature of the Green--Schwarz-type sigma model
of the M5-brane. This has descent as follows

\vspace{-.5cm}
\begin{equation}
 \raisebox{50pt}{  \xymatrix{
    \mathfrak{m}2\mathfrak{brane}
    \ar[rr]^{ \mu_{M5}}
    \ar[d]_-{\color{blue}
      \mathrm{hofib}
      \left(
        \mu_{{}_{M2}}
      \right)
    }
    &&
    \mathfrak{l} S^7
    \ar[d]
    \\
    \mathbbm{R}^{10,1\vert \mathbf{32}}
    \ar@{-->}[rr]^{\mu_{M2/M5}}
    \ar[dr]_{\mu_{M2}}
    &&
    \mathfrak{l}S^4
    \ar[dl]^{c_2}
    \\
    &
    \EM{4}
  }
  }
\end{equation}
where on the right we have the
rational incarnation of the \emph{quaternionic Hopf fibration}.
A basic and classical fact of rational homotopy theory
gives that the  Sullivan model of the 4-sphere is
\begin{equation}
  \label{SullivanModelOf4Sphere}
 \mathcal{O}\left( S^4_{\mathbbm{R}}\right)
 \;\simeq\;
 \mathrm{CE}
 \big(
   \mathfrak{l}S^4
 \big)\;=\;
 \mathbbm{R}[G_4, G_7]\big/
 \left(
   {
   {\rm d} G_4  = 0
   \atop
   {\rm d} G_7  = - \tfrac{1}{2} G_4 \wedge G_4
   }
 \right)
\end{equation}
with a generator $G_4$ in degree 4 and a generator $G_7$
in degree 7.
In terms of this the unified M2/M5-brane cocycle
comes out to be
\begin{equation}
  \label{UnifiedM2M5braneCocycle}
   \raisebox{20pt}{\xymatrix@R=-4pt{
    \mathbbm{R}^{10,1\vert \mathbf{32}}
    \ar[rr]^-{\mu_{{}_{M2/M5}}}
    &&
    \mathfrak{l}S^4
    \\
    \tfrac{i}{2} \left(\overline{\psi}\Gamma_{a_1 a_2} \psi\right) \wedge e^{a_1} \wedge e^{a_2}
    &&
    G_4
    \ar@{|->}[ll]
    \\
    \tfrac{1}{5!}
    \left(\overline{\psi}\Gamma_{a_1 \cdots a_5}\psi\right)
    \wedge e^{a_1} \cdots e^{a_5}
    &&
    G_7
    \ar@{|->}[ll]
  }
  }
\end{equation}
and the fact that this is a homomorphism means that
the bifermionic expressions on the left do satisfy the
equation on the right in \eqref{SullivanModelOf4Sphere}.
That this is indeed the case is due to the Fierz identities
that were first presented in \cite{D'Auria:1982nx}.

These equations governing the Sullivan model of the 4-sphere
\begin{equation}
  \begin{aligned}
    {\rm d} G_4  &= 0
    \\
    {\rm d} G_7  & = - \tfrac{1}{2} G_4 \wedge G_4
  \end{aligned}
\end{equation}
are also precisely the equations of
motion of the $C_3/C_6$-field in $D =11$ supergravity.
This alone shows that, rationally, the unified M2/M5-brane charge
is in the non-Abelian generalized cohomology theory classified by the 4-sphere \cite[Sec. 2.5]{Sati:2013rxa}.
This cohomology theory is known as \emph{cohomotopy}
\cite{borsuk1936groupes,Spanier:1949:203-}.
We come back to this below in
Section \ref{ChargeQuantizationInEquivariantCohomotopy}.

But, moreover, when formulated on
superspace the torsion constraints of $D =11$ supergravity
say that the $C_3/C_6$-field is constrained to
have bifermionic components precisely as in \eqref{UnifiedM2M5braneCocycle}.
Hence the unified M2/M5-brane cocycle \eqref{UnifiedM2M5braneCocycle}
discovers the supergravity $C_3/C_6$-field in the case of
vanishing bosonic flux. We come back to this below in
Section \ref{OrbifoldSupergavity}.

\medskip

\hypertarget{DoubleDimensionalReduction}{
\section{Double dimensional reduction}
 \label{DoubleDimensionalReduction}
}

Underlying most of the dualities in string theory is
the phenomenon of \emph{double dimensional reduction}
(going back to \cite{Duff:1987bx})
so called because:
\smallskip
\begin{enumerate}[i)]
  \item the dimension of space-times is reduced
    by Kaluza-Klein compactification on a fiber $F$;

    \item in parallel, the dimension of branes is reduced
    if they wrap $F$.
\end{enumerate}

\smallskip

Here we explain the homotopy theory behind
double dimensional reduction
\cite[Sec. 3]{Fiorenza:2016oki}, \cite[Sec. 2.2]{Braunack-Mayer:2018uyy} (following \cite[Sec. 4.2]{Mathai:2003mu}, see also \cite{Figueroa-OFarrill:2002yel}).
 This turns out to be a beautiful application
of basic elements of homotopy theory (homotopy base change)
and serves as the central ingredient for T-duality and
M/IIA-duality discussed further below. Therefore we will
be more detailed here.

\medskip

For example, double dimensional reduction is supposed to underly
the duality between M-theory and type IIA string theory:
\smallskip
\begin{enumerate}[i)]
    \item space-time $X_{11}$ is an 11-dimensional circle-fiber bundle
    locally of the form $X_{11} = X_{10} \times S^1$
    over a 10-dimensional base space-time;

    \item an M2-brane with world-volume
    $\Sigma_3 = \Sigma_2 \times S^1$
    wraps the circle fiber if its trajectory
    $\phi_{M2} \;\colon\; \Sigma_3 \to X_{11}$
    is of the form
    \begin{equation}
      \phi_{F1} \times \mathrm{id}_{S^1}
      \;:\;
      \Sigma_2 \times S^1 \longrightarrow X_{10} \times S^1\;.
    \end{equation}
\end{enumerate}

\smallskip
As the Riemannian circumference of the circle fiber $S^1$ tends towards zero
this effectively looks like the 2-dimensional worldsheet $\Sigma_2$ of a string
 tracing out a trajectory in 10-dimensional space-time:
$\phi_{F1} : \Sigma_2 \to X_{10}$.

\smallskip
On the other hand,  there is also ``single dimensional reduction''
when the membrane does not wrap the fiber space:

\vspace{-7mm}
\begin{equation}
  \xymatrix@R=2em{
    \Sigma_3
    \ar[rr]^-{ \phi_{M2} }
    \ar[dr]_{\phi_{D2}}
    &&
    X_{11}
    \\
    &
    X_{10}
    \ar[ur]
  }
\end{equation}
In this case it looks like a membrane in 10-dimensional space-time (see \cite{Figueroa-OFarrill:2002yel}),
now called the D2-brane (see \cite{Johnson:2000ch} for an overview of D-branes).

Similarly the M5-brane in M-theory
$\phi_{M5} \;\colon\; \Sigma_{6} \longrightarrow X_{11}$
may wrap the circle fiber to yield a 4-brane in 10-dimensional, called the
\emph{D4-brane} or it may not wrap the circle fiber
to yield a 5-brane in 10-dimensional, called the \emph{NS5-brane}.

Beware the na{\"i}ve treatment of branes in this traditional argument.
And even naively, this is not the full story yet:
The $S^1$-fibration itself is supposed to re-incarnate in 10d
as the D0-brane and the D6-brane.

Hence double dimensional reduction from M-theory to type IIA string theory
is meant to, schematically, involve decompositions as follows
\begin{equation}
  \hspace{-.1cm}
  \xymatrix@C=-1.8pt@R=1.2em{
    X_{11}
    \ar[dd]^{\pi}
    & && &
    \mathrm{M2}
    \ar@{|->}[ddl]|{\mbox{\color{blue}\tiny wrapped}}
    \ar@{|->}[ddr]|{\mbox{\color{blue}\tiny \begin{tabular}{c} not \\ wrapped \end{tabular}}}
    && &&
    \mathrm{M5}
    \ar@{|->}[ddl]|{\mbox{\color{blue}\tiny wrapped}}
    \ar@{|->}[ddr]|{\mbox{\color{blue}\tiny \begin{tabular}{c}not \\ wrapped \end{tabular}}}
    &&
    \\
    \\
    X_{10}
    &
    \mathrm{D0}
    &&
    \mathrm{F1}
    &&
    \mathrm{D2}
    &&
    \mathrm{D4}
    &&
    \mathrm{NS5}
    %&&
    %{\color{gray} \mathrm{D6}}
    %&&
    %{\color{gray} \mathrm{D8}}
  }
\end{equation}
We saw above that all super $p$-branes are characterized by the
flux fields $H_{p+2}$ that they are charged under,
more precisely by the bispinorial component of $H_{p+2}$
which is constrained to be super tangent-space-wise the form
\begin{equation}
  H_{p+2}^{\mathrm{fermionic}}
    \;=\;
  \tfrac{i^{p(p-1)/2}}{p!} \, \left( \overline{\Psi} \wedge \Gamma_{a_1 \cdots a_p}\Psi \right) \wedge E^{a_1} \wedge \cdots \wedge E^{a_p}
\end{equation}
where $(E^a, \Psi^\alpha)$ is the super vielbein (graviton and gravitino).
Hence we will formalize double dimensional reduction in terms of these fields.

\,

Again there is a naive picture to help the intuition:
Let $G_4 \in \Omega^4_{\rm cl}(X_{11})$ be the differential 4-form flux field
strength of the supergravity C-field.
Under the Gysin sequence for the spherical fibration
\begin{equation}
  \xymatrix@R=2em{
    S^1
      \ar[r]
    &
    X_{11}
    \ar[d]^{\pi}
    \\
    & X_{10}
  }
\end{equation}
this decomposes in cohomology as \cite[Sec. 4.2]{Mathai:2003mu}
\begin{equation}
  G_4 = (d x^{10}) \wedge H_3 + \pi^\ast F_4
\end{equation}
thus giving rise in 10-dimensional to
\smallskip
\begin{enumerate}[i)]
\item a 3-form $H_3$, the Kalb-Ramond B-field field strength to which the string couples;
\item a 4-form $F_4$, the RR-field field strength in degree 4, to which the D2-brane couples.
\end{enumerate}

\smallskip
\noindent Similarly the dual 7-form field strength $G_7$ decomposes as
$G_7 = (d x^{10}) \wedge F_6 + \pi^\ast H_7$
thus giving rise in 10-dimensional to
\smallskip
\begin{enumerate}[i)]
    \item a 6-form $F_6$, the RR-field field strength in degree 6, to which the D4-brane couples;

    \item a 7-form $H_7$, the dual NS-NS field strength to which the NS5-brane couples.
\end{enumerate}

\smallskip
\begin{equation}
  \hspace{-.1cm}
  \raisebox{20pt}{
  \xymatrix@C=-.5pt@R=1.2em{
    X_{11}
    \ar[dd]^{\pi}
    && & &
    G_4
    \ar@{|->}[ddl]|{\mbox{\color{blue} \tiny wrapped}}
    \ar@{|->}[ddr]|{\mbox{\color{blue} \tiny \begin{tabular}{c} not \\ wrapped \end{tabular}}}
    && &&
    G_7
    \ar@{|->}[ddl]|{\mbox{\color{blue}\tiny wrapped}}
    \ar@{|->}[ddr]|{\mbox{\color{blue}\tiny \begin{tabular}{c}not \\ wrapped \end{tabular}}}
    &&
    \\
    \\
    X_{10}
    &
    F_2
    &&
    H_3
    &&
    F_4
    &&
    F_6
    &&
    H_7
    %&&
    %{\color{gray} \mathrm{D6}}
    %&&
    %{\color{gray} \mathrm{D8}}
  }
  }
\end{equation}

To first approximation
background fluxes represent classes in ordinary cohomology (their charges),
classified by the
Eilenberg--MacLane spaces $K(\mathbbm{Z}, \bullet)$,
\begin{equation}
  H^n
  \big(
    X, \mathbbm{Z}
  \big)
  \;\simeq\;
  \left\{
    \mbox{\!\!\!
      \begin{tabular}{c}
        continuous functions
        \\
        $X \longrightarrow K(\mathbbm{Z},n)$
      \end{tabular}
   \!\!\!\! }
  \right\}_{\Big/\mathrm{homotopy}}
\end{equation}
Hence the charge of $G_4$/$G_7$-flux, to first approximation, is represented by a classifying map
\begin{equation}
  ([G_4], [G_7])
  \;\colon\;
  X_{11}
    \longrightarrow
  K(\mathbbm{Z},4)
    \,\times\,
  K(\mathbbm{Z},7)
\end{equation}
and we saw that under double dimensional reduction
this is supposed to transmute into a map of the form
\begin{equation}
  X_{10}
    \xrightarrow{\big([F_2] , [H_3], [F_4], [F_6], [H_7]\big)}
  \begin{array}{c}
    K(\mathbbm{Z},2)
    \times
    K(\mathbbm{Z},4)
    \times
    K(\mathbbm{Z},6)
    \\
    \times
    K(\mathbbm{Z},3)
    \times
    K(\mathbbm{Z},7)
  \end{array}
  \,.
\end{equation}
Which mathematical operation could cause such a transmutation?

We will now find such an operation, which  knows about all the fine print of brane charges,
and then use it to give an improved definition of double dimensional reduction.

\subsection{Reduction via free looping (no 0-brane effect)}

We first record formally the state of affairs in the above story:
In the above double dimensional reduction
of the naive M-fluxes on a trivial 11-dimensional circle bundle
we used
\smallskip
\begin{enumerate}[i)]
\item the Cartesian product with the circle

\item functions out of the circle.
\end{enumerate}
\smallskip
Let us have a closer look at these two operations:
it is a classical fact about locally compact topological spaces
(which includes all topological spaces that one cares about in physics)
that
given topological spaces $\Sigma$, $X$ and $F$, then there is a natural bijection
\begin{equation}
  \hspace{-.05cm}
  \left\{\!\!
    \begin{array}{c}
      \mbox{\footnotesize continuous functions}
      \\
      \Sigma \times F \longrightarrow X
    \end{array}
  \!\! \right\}
  \xleftrightarrow{\!\!
    \mbox{
      \tiny
      \begin{tabular}{c}
        ``forming
        \\
        adjoints''
      \end{tabular}
    \!\!}
  }
  \left\{ \!\!
    \begin{array}{c}
      \mbox{\footnotesize continuous functions}
      \\
      \Sigma \longrightarrow \mathrm{Maps}(F,X)
    \end{array}
\!\!  \right\}
\end{equation}
where
\begin{enumerate}[i)]
    \item $F \times X$ is the product topological space of $F$ with $X$
    (the set of pairs of points equipped with the product topology);

    \item $\mathrm{Maps}(F,Y)$ is the mapping space from $F$ to $X$,
    (the set of continuous functions) $F \to X$ equipped with the compact-open topology).
\end{enumerate}
\smallskip
Except for the subtlety with the topology,
this bijection is just rewriting
a function of two variables as a function with values in a second function
\begin{equation}
  (\tilde f(a))(b) = f(a,b)
  \,.
\end{equation}
One says that the two functors
\begin{equation}
  \xymatrix{
    \mathrm{Top}_{\mathrm{cg}}
    \ar@{<-}@<+7pt>[rr]^-{F \times (-)}
    \ar@<-7pt>[rr]_-{\mathrm{Maps}(F,-)}^-{\bot}
    &&
    \mathrm{Top}_{\mathrm{cg}}
  }
\end{equation}
form a \emph{adjoint pair} or an \emph{adjunction}.
A remarkable amount of structure comes with every adjunction:
\smallskip
\begin{enumerate}[i)]
\item the adjunct of the identity $F \times X \overset{\rm id}{\to} F \times X$,
generally called the \emph{unit of the adjunction},
here is the wrapping operation
    $X \overset{}{\longrightarrow} \mathrm{Maps}(F, F \times X)$

\item the adjunct of the identity $\mathrm{Maps}(F,X) \overset{\rm id}{\to} \mathrm{Maps}(F,X)$
    generally called the \emph{counit of the adjunction},
    here is the evaluation map
    $X F \times {\rm Maps}(F, X) \overset{\rm ev}{\longrightarrow} X$
    that evaluates a function on an argument.
\end{enumerate}

\smallskip
\noindent We will now see that the following general fact
about adjoint functors
serves to implement the above physics story
of wrapped branes:

\smallskip
\noindent {\bf Fact} (e.g.~\cite[Prop. 1.38]{Schreiber:2018lab})
The adjunct of a map of the form
$G \;\colon\; F \times X \overset{}{\longrightarrow} A$
is the composite of its image under $\mathrm{Maps}(F,-)$ with the adjunction unit $\eta_X$:

\vspace{-9mm}
\begin{equation}
\scalebox{0.95}{
\xymatrix@C=1.3em{
  \mathrm{Maps}(F,A) \tilde G \colon X \! \ar[r]^-{\eta_X} & \!
  \mathrm{Maps}(F,F \times X)  \!\!\ar[rr]^-{\mathrm{Maps}(F,G)} && \!\! \mathrm{Maps}(F,A).
  }}
\end{equation}

Moreover, we will see that the following general fact in homotopy theory
accurately implements the idea of dimensional reduction of the brane dimensions:
Notice that for $F = S^1$ the circle, the mapping space
$\mathcal{L} X \;\coloneqq\; \mathrm{Maps}(S^1, X)$
is also called the \emph{free loop space of $X$}.

\smallskip
\noindent {\bf Proposition.}
For $G$ a topological group, the free loop space of its classifying space
is weakly homotopy equivalent to the
homotopy quotient of $G$ by its adjoint action:
\begin{equation}
  \mathrm{Maps}(S^1, B G) \;\simeq\; G\sslash_{\mathrm{ad}}G
\end{equation}

In the special case that $G$ is an Abelian topological group
this becomes a weak homotopy equivalence of following simple form
\begin{equation}
 \mathrm{Maps}(S^1 , B G)
 \; \simeq \;
 \underset{\mbox{\color{blue} \tiny wrapped} \atop \mbox{\color{blue} \tiny coefficient}}{\underbrace{G}} \; \times \; \underset{\mbox{\color{blue} \tiny plain} \atop \mbox{\color{blue} \tiny coefficient} }{\underbrace{ B G }}
 .
\end{equation}
In particular, if $G = K(\mathbbm{Z},n)$ then
\begin{equation}
  \mathrm{Maps}\big(S^1, K(\mathbbm{Z},n+1)\big)
  \;\; \simeq
  K(\mathbbm{Z},n) \;\times\; K(\mathbbm{Z},n+1)
  \,.
\end{equation}
These degrees capture the required reduction on brane dimensions!
In order to amplify the crucial higher group structure
on the classifying spaces,
we will now write
\begin{equation}
  B^{n}\mathbbm{Z} \coloneqq K(\mathbbm{Z},n)
  \,.
\end{equation}

\smallskip

\noindent {\bf Example.} Consider na{\"i}ve M-flux fields $G_4$ and $G_7$
on an 11d space-time that is a trivial circle bundle
$X_{11} = X_{10} \times S^1$.
Its charges is represented by a map of the form
$([G_4], [G_7]) \;\colon\; X_{10} \times S^1 \longrightarrow B^4 \mathbbm{Z} \times B^7 \mathbbm{Z}$.
By adjunction this is identified with a map of the form
\begin{equation}
  X_{10}
  \xrightarrow{\tiny
    \begin{array}{c}
      \mathclap{\left([H_3], [F_4], [F_6], [H_7]\right)}
      \\
      \coloneqq
      \\
      \widetilde{([G_4], [G_7])}
     \end{array}
  }\;\;
  \underset{
    \simeq
    \mathrm{Maps}
    \left(
      S^1, \; B^4 \mathbbm{Z} \times B^7 \mathbbm{Z} \;
    \right)
  }{
    \underbrace{
      B^3 \mathbbm{Z}
        \;\times\;
      B^4 \mathbbm{Z}
        \;\times\;
      B^6 \mathbbm{Z}
       \;\times\;
      B^7 \mathbbm{Z}
    }
  }
  \;,
\end{equation}
where on the right we have the transmuted coefficients
by the above proposition.
This is exactly the result we were after.

\medskip
Better yet, the adjunction yoga
accurately reflects the physics story:
Consider a $p$-brane propagating in 10d space-times along a trajectory
$\phi_p \;\colon\; \Sigma_p \longrightarrow X_{10}$
and coupled to these dimensionally reduced background fields

\vspace{-8mm}
\begin{equation}
\scalebox{0.95}{\xymatrix@=2.6em{
  \Sigma_p \!\!\!
    \ar[r]^-{\phi_p}&
\!\!\!  X_{10}
    \ar[rr]^-{([H_3], [F_4], [F_6], [H_7])} &&
  \underset{
    \simeq
    \mathrm{Maps}(S^1, B^4 \mathbbm{Z} \,\times\, B^7 \mathbbm{Z})
   }{
  \underbrace{
    B^3 \mathbbm{Z} \;\times\; B^4 \mathbbm{Z} \;\times\; B^6 \mathbbm{Z} \;\times\; B^7 \mathbbm{Z}
  }
  }
  }\!.}
\end{equation}
Then, by adjunction, this is identified with a map of the form

\vspace{-10mm}
\begin{equation}
\hspace{-2mm}
\xymatrix@=1.5em{
  \Sigma_p \times S^1 \!\!\ar[r]^-{\phi_p \times S^1} &
  X_{10} \times S^1 = X_{11}
  \ar[rr]^{([G_4], [G_7])} && B^4 \mathbbm{Z} \times B^7 \mathbbm{Z}
  }
\end{equation}
and this is exactly the coupling we saw in the story of double dimensional reduction.

So this works well as far as it goes, but so far it only applies to trivial circle fibrations and it does not see the D0-charge.
Next we discuss the improvement to the full formulation.

\medskip

\subsection{Reduction via cyclification (with 0-brane effect)}

In general the M-theory circle bundle
\begin{equation}
  \xymatrix@R=1.4em{
    S^1 \ar[r]
    &
    X_{11}
    \ar[d]
    \\
    & X_{10}
  }
\end{equation}
is only locally a product with of $X_{10}$ with $S^1$.
For example,
the complement of the locus of a KK-monopole space-time
is a circle principal bundle with first Chern class
equal to the charge carried by the KK-monopole
(which is the corresponding number of coincident D6-branes in type IIA).
Hence, in general, the above formulation of double dimensional reduction
via the pair of adjoint functors
$
  S^1 \times (-) \;\;\dashv\;\; \mathrm{Maps}(S^1, -)
$
applies only locally.

However, the problem to be solved is easily identified: essentially by definition, in a circle
principal bundle the fibers may all be identified with a fixed abstract circle $S^1$ only up to rigid rotation.
Hence while in general the above wrapping-map
$X_{10} \overset{}{\longrightarrow} {\rm Maps}(S^1, X_{11})$
given by sending each point of $X_{10}$ to
its fiber ``wrapping around itself''
does not exist, it does exist up to forgetting at which point in $S^1$ we start the wrapping,
hence the map that always exists lands in the quotient space
\begin{equation}
  \mathrm{Maps}
  \big(
    S^1, X_{11}
  \big)
  \sslash S^1
  \;=\;
  \frac{
    \left\{
      \begin{array}{c}
        \mbox{\small continuous functions}
        \\
        S^1 \longrightarrow X_{11}
      \end{array}
    \right\}
  }
  {
    \left\{
      \begin{array}{c}
        \mbox{\small rigid loop rotations}
        \\
        S^1 \xrightarrow{ t\mapsto (t + t_0) } S^1
      \end{array}
    \right\}
  }.
\end{equation}
There is then the following generalization of
the above proposition on
transmutation of coefficients under double dimensional reduction

\smallskip
\noindent {\bf Proposition.}
Let $G$ be an Abelian topological group.
Then there is a weak homotopy equivalence of the form
\begin{equation}
  \mathrm{Maps}(S^1 , B G)\sslash S^1
  \; \simeq \;
  \big(\!\!\!\!
    \underset{\mbox{\color{blue} \tiny wrapped} \atop \mbox{\color{blue} \tiny coefficient}}{\underbrace{G}}
      \times
    \underset{
      \mbox{\color{blue} \tiny plain}
      \atop
      \mbox{\color{blue} \tiny coefficient}
    }{
      \underbrace{B G}
    }
\!\!\!\!  \big) \;
  \underset{
    \mbox{\color{blue} \tiny twist}
  }
  {
    \underbrace{
      \times_{S^1}
    }
  }
  \underset{
    \mbox{\color{blue} \tiny D0-brane} \atop \mbox{\color{blue} \tiny coeff.}
  }{
    \underbrace{E S^1}
  } \!\!\!\!.
\end{equation}

Notice that a twisting appears. This is a general phenomenon.
We will see below that for the example of reduction of M-flux
the twist that appears is that in the twisted de Rham cohomology
$F_4 = H_3 \wedge F_2$
which connects RR-fields $F_{2p}$ with the H-flux $H_3$.

Indeed this dimensional reduction is again an equivalent way of regarding the higher dimensional situation:

\medskip
\noindent {\bf Proposition}
(Double dimensional reduction on topological flux fields)
There is a pair of adjoint $\infty$-functors
\begin{equation}
  \xymatrix{
    \mathrm{Spaces}
    \ar@{<-}@<+7pt>[rr]^-{\mathrm{hofib}}
    \ar@<-7pt>[rr]_-{ \mathrm{Maps}(S^1, -)\sslash S^1 }^-{\bot}
    &&
    \mathrm{Spaces}_{/B S^1}
  }
\end{equation}
or equivalently (by \cite{Nikolaus:1207ab}):
\begin{equation}
  \xymatrix{
    \mathrm{Spaces}
    \ar@{<-}@<+7pt>[rrr]^-{\mbox{ \tiny total space}}
    \ar@<-7pt>[rrr]_-{ \big[ \mathrm{Maps}(S^1,-) \to \mathrm{Maps}(S^1, -)\sslash S^1 \big] }^-{\bot}
    &&&
    S^1\text{Principal Bundles}
  }
\end{equation}
Hence for
\begin{equation}
  \xymatrix@=1.5em{
    S^1 \ar[r]
    &
    X_{d+1}
    \ar[d]^{\pi}
    \\
    & X_{d}
  }
\end{equation}
an $S^1$-principal bundle and $A$ some coefficients,
there is a natural equivalence
\begin{equation}
  \xymatrix@C=.7em{
    \mathrm{Hom}
    \big(
      X_{d+1}, A
    \big)
    \!\ar@{<-}@<+7pt>[rr]^-{\mbox{\color{blue} \tiny oxidation}}
    \ar@<-7pt>[rr]_-{\mbox{\color{blue} \tiny reduction}}^-{\simeq}
    &&
    \mathrm{Hom}_{/B S^1}
    \big(
      X_d, (\mathcal{L}A)\sslash S^1
    \big).
  }
\end{equation}
Accordingly we have the following generalization of
the previous example to the case with possibly non-trivial circle
fibration and non-trivial D0-flux:

\smallskip
\noindent {\bf Example.}
Consider naive M-flux fields $G_4$ and $G_7$  on an 11d space-time that is an $S^1$-principal bundle
\begin{equation}
  \xymatrix@R=1.5em{
    S^1 \ar[r]
    &
    X_{11}
    \ar[d]^{\pi}
    \\
    & X_{10}
  }
\end{equation}
Its charges is represented by a map of the form
\begin{equation}
  ([G_4], [G_7])
  \;\colon\;
  X_{11} \longrightarrow B^4 \mathbbm{Z} \;\times\; B^7 \mathbbm{Z} \,.
\end{equation}
By adjunction this is identified with a map of the form
\begin{equation}
 X_{10}
 \xrightarrow{\tiny
   \begin{array}{c}
     \mathclap{\left([F_2], [H_3], [F_4], [F_6], [H_7]\right)}
     \\
     \coloneqq
     \\
     \widetilde{([G_4], [G_7])}
   \end{array}
 }
 \underset{
  \mathrm{Maps}
  \left(
    S^1, \; B^4 \mathbbm{Z} \times B^7 \mathbbm{Z} \;
  \right)
 }{
    \underbrace{
     E S^1 \times_{S^1} \left( B^3 \mathbbm{Z} \times B^4 \mathbbm{Z} \times B^6 \mathbbm{Z} \times
      B^7 \mathbbm{Z} \right)
    }
  }
  \;,
\end{equation}
where on the right we transmuted the coefficients by the previous proposition.
Hence the D0-brane charge appears! It is the first Chern class of the M-theory circle bundle.

\smallskip
\noindent {\bf Conclusion.}
The double dimensional reduction of any flux field
$X_{d+1} \overset{G}{\longrightarrow} \mathcal{X}$
is
\begin{equation}
  \xymatrix@R=1.5em{
    X_d
    \ar[rr]^-{\widetilde G}
    \ar[dr]
    &&
    \mathrm{Maps}\big(S^1, \mathcal{X} \big) \sslash S^1 \;.
    \ar[dl]
    \\
    & B S^1
  }
\end{equation}
This can be written schematically as
\begin{equation}
[\text{Circle bundle}, \mathcal{X}]= [\text{Base}, \mathcal{L}_c (\mathcal{X})]\;,
\end{equation}
generalizing \cite{Mathai:2003mu}\cite{Figueroa-OFarrill:2002yel}\cite{Bergman:2004ne}.
The operation
\begin{equation}
 \mathcal{L}(-)_c= \mathcal{L}(-)/S^1
  \;\coloneqq\; \mathrm{Maps}(S^1, -)/S^1
\end{equation}
may be called \emph{cyclification}
because the cohomology of this quotient of the free loop space
is cyclic cohomology. Shadows of this construction appear prominently
also at other places in string theory
notably in discussion of the \emph{Witten genus}.
A closely related concept in mathematics involving this is the \emph{transchromatic character map}.

\,

In fact this formalization of double dimensional reduction
works also with geometry taken into account, notably it
works in full super homotopy theory.
The homotopy-cognescenti will realize that, abstractly, the
cyclification adjunction is nothing but the $\infty$-topos-theoretic
\emph{left base change} along $\mathbf{B}S^1 \to \ast$.

\subsection{Reduction on super $p$-brane cocycles}

\begin{figure*}[htb]

\vspace{-.7cm}

\begin{center}
  \xymatrix@C=2pt@R=1.2em{
    &
    \overset{
      p+1\text{-brane}
    }{
    \overbrace{
    {
      \mu^{d+1}_{(p+1)+2}
      =
    }
    \atop
    {
      \sum_{d=0}^{d}
      \left(
        \overline{\psi}
         \wedge
         \Gamma_{a_1 \cdots a_{p+1}}
         \psi
      \right)
      \wedge
      e^{a_1}
       \wedge
       \cdots
      e^{a_{p+1}}
      }
      }
    }
    \ar[dl]_{\color{blue}\mathrm{wrapped}}
    \ar[dr]^{\color{blue}\mathrm{non-wrapped}}
    \\
    \underset{
      \mbox{\tiny $p$-brane}
    }{
    \underbrace{
      {
        \mu^{d}_{p+2}
        =
      }
      \atop
      {
        {\sum}_{a_i = 0}^{d-1}
        \left(
        \overline{\psi}
        \wedge
          \Gamma_{a_1 \cdots a_{p}}
        \psi
        \right)
        \wedge
        e^{a_1}
         \wedge \cdots
        e^{a_p}
      }
      }
    }
    &&
    \underset{
      p+1\text{-brane}
    }{
    \underbrace{
      {
        \mu^{d}_{p+2}
        =
      }
      \atop
      {
        {\sum}_{a_i = 0}^{d-1}
        \left(
        \overline{\psi}
        \wedge
          \Gamma_{a_1 \cdots a_{p+1}}
        \psi
        \right)
        \wedge
        e^{a_1}
         \wedge \cdots
        e^{a_{p+1}}
      }
      }
    }
  }
\end{center}

\vspace{-1.2cm}

\caption{
  {\bf Double dimensional reduction}
  in the special case of single $\mathbbm{R}$-valued super cocycles.
  The wrapped part on the left is the reduction observed in the old brane scan \cite{Achucarro:1987nc} (see \cite[p. 15]{Duff:1987qa}).
  The non-wrapped part on the right is not a plain super cocycle, but a
  \emph{twisted} super cocycle, related to the appearance of D-branes
  in the brane bouquet (Figure \ref{Figure1}).
  Both the wrapped and the non-wrapped component are unified
  by dimensional reduction via cyclification in super homotopy theory \cite{Fiorenza:2016oki,Braunack-Mayer:2018uyy}.
}
\label{DDReduction}
\end{figure*}

By the discussion of rational homotopy theory above
we may think of $L_\infty$-algebras as rational topological spaces
and more generally as rational parameterized spectra.
For instance,  we found above that the coefficient space
for RR-fields in rational twisted K-theory is the
$L_\infty$ $\mathfrak{l}(\mathrm{ku}\sslash BU(1))$.
Hence in order to apply double dimensional reduction
to super $p$-brane
we now specialize the above formalization to
cyclification of super $L_\infty$-algebras \cite[Sec. 3]{Fiorenza:2016oki}.

\smallskip
\noindent{\bf Definition.}
For $\mathfrak{g}$ any super $L_\infty$-algebra of finite type, its \emph{cyclification}
\begin{equation}
  \mathfrak{L}\mathfrak{g}/\mathbbm{R} \in s L_\infty Alg_{\mathbbm{R}}
\end{equation}
is defined by having Chevalley--Eilenberg algebra of the form
\begin{equation}
  \begin{aligned}
  & \mathrm{CE}(\mathfrak{L}\mathfrak{g}/\mathbbm{R})
  \coloneqq
  \\
  &
  \scalebox{0.93}{\raisebox{0pt}{ 
  $\left(
    \wedge^\bullet
    \big( \!\!
      \underset{\mbox{\color{blue}\tiny original}}{\underbrace{\mathfrak{g}^\ast}}
      \!\!\! \oplus \!\!\!
      \underset{\mbox{\color{blue}\tiny shifted copy}}{\underbrace{s\mathfrak{g}^\ast}}
      \!\!\! \oplus \!\!\!
      \underset{\mbox{\color{blue}\tiny new generator}
      \atop
      \mbox{\color{blue}\tiny in degree 2}}{\underbrace{\langle \omega_2 \rangle}}
    \!\!\!\!\!\! \!\!\!\! \big)
    ,\;
    \mathrm{d}_{\mathfrak{d}\mathfrak{g}/\mathbbm{R}}
    \colon
    \!\!\!
  \left\{\setlength{\tabcolsep}{-10pt}
\renewcommand{\arraystretch}{.1}
    \begin{array}{rcl}
      \omega_2
      &\mapsto&
      0
      \\
      \alpha
      &\mapsto&
      \mathrm{d}_{\mathfrak{g}} \alpha + \omega_2 \wedge s \alpha
      \\
      s \alpha
      &\mapsto&
      -
      s d_{\mathfrak{g}} \alpha
    \end{array}
    \right.
  \right)$}}
  \end{aligned}
\end{equation}
where $\mathfrak{g}^\ast$
is a copy of $\mathfrak{g}^\ast$ with cohomological degrees shifted down by one,
 and where $\omega$ is a new generator in degree 2.
The differential is given for $\alpha \in \wedge^1 \mathfrak{g}^\ast$
by
\begin{equation}
  {\rm d}_{\mathfrak{d}\mathfrak{g}/\mathbbm{R}}
  \;\colon\;
  \left\{
    \begin{array}{rcl}
      \omega_2
      &\mapsto&
      0
      \\
      \alpha
      &\mapsto&
      {\rm d}_{\mathfrak{g}} \alpha \pm \omega_2 \wedge s \alpha
      \\
      s \alpha
      &\mapsto&
      - s {\rm d}_{\mathfrak{g}}
      \alpha
    \end{array}
  \right.
\end{equation}
where on the right we are extending as a graded derivation.

Define
$\mathfrak{L}\mathfrak{g} \in s L_\infty \mathrm{Alg}$
in the same way, but with $\omega_2 \coloneqq 0$.
For every $\mathfrak{g}$ there is a homotopy fiber sequence
\begin{equation}
 \raisebox{20pt}{ \xymatrix@R=1.5em{
    &
    \mathcal{L}\mathfrak{g}
    \ar[d]
    \\
    &
    \mathcal{L}\mathfrak{g}\sslash \mathbbm{R}
    \ar[dl]
    \\
    \mathbf{B}\mathbbm{R}
  }
  }
\end{equation}
which hence exhibits $\mathfrak{L} \mathfrak{g}/\mathbbm{R}$ as the homotopy quotient of $\mathfrak{L}\mathfrak{g}$ by an $\mathbbm{R}$-action.

The following says that the $L_\infty$-cyclification from above
indeed does model correspond to the topological cyclification from Prop. 4.4.

\smallskip
\noindent {\bf Proposition.} (\!\!\!\cite{Burghelea:1985:243-253,Vigue-Poirrier:1976:633-644})
If $\mathfrak{g} = \mathfrak{l}(X)$
is the $L_\infty$-algebra associated by
rational homotopy theory to a simply connected topological space $X$,
then $\mathfrak{L}( \mathfrak{l}(X) ) \simeq \mathfrak{l}( \mathcal{L}X )$
corresponds to the free loop space of $X$ and
$\mathfrak{L}( \;\mathfrak{l}( X )\; )/\mathbbm{R} \simeq \mathfrak{l}( \;\mathcal{L}X/S^1\; )$
corresponds to the homotopy quotient of the free loop space by the circle group action which rotates the loops.

Note that the cochain cohomology of the Chevalley--Eilenberg algebra
$\mathrm{CE}(\mathfrak{l}( \;\mathcal{L}X/S^1\; ))$
computes the cyclic cohomology of $X$ with coefficients in $\mathbbm{R}$
(whence ``cyclification'').
Moreover, the homotopy fiber sequence of the cyclification corresponds
to that of the free loop space:
\begin{equation}
  \left(
    \raisebox{49pt}{
    \xymatrix{
      \mathcal{L}X
      \ar[d]_{\color{blue} \mathrm{hofib}}
      \\
      \mathcal{L}X\sslash S^1
      \ar[d]
      \\
      K(\mathbbm{Z},2)
    }
    }
  \right)
  \;\;\overset{\mathfrak{l}}{\longmapsto}\;\;
  \left(
    \raisebox{49pt}{
    \xymatrix{
      \mathcal{L}\mathfrak{l}X
      \ar[d]_{\color{blue} \mathrm{hofib}}
      \\
      \mathcal{L}\mathfrak{l}X\sslash \mathbbm{R}
      \ar[d]
      \\
      \EM{2}
    }
    }
  \right)
\end{equation}

The following gives the super $L_\infty$-theoretic formalization
of ``double dimensional reduction''
by which both the space-time dimension is reduced
while at the same time the brane dimension
reduces (if wrapping the reduced dimension).

\smallskip
\noindent {\bf Proposition. } (\!\cite[Prop. 3.5]{Fiorenza:2013nha})
For
\begin{equation}
  \xymatrix@R=1.5em{
    \widehat{\mathfrak{g}}
    \ar[d]_-{\color{blue}\mathrm{hofib}}
    \\
    \mathfrak{g}
    \ar[r]^{\mu_2}
    &
    \EM{2}
  }
\end{equation}
a central extension of super Lie algebras, the operation
of sending a super $L_\infty$-homomorphsm of the form
$\widehat{\mathfrak{g}} \overset{\phi}{\longrightarrow} \mathfrak{h}$
to the composite
$\mathfrak{g} \longrightarrow \mathfrak{L}\widehat{\mathfrak{g}}/\mathbbm{R} \overset{\mathfrak{L}\phi\sslash\mathbbm{R}}{\longrightarrow} \mathfrak{L}\mathfrak{h}\sslash\mathbbm{R}$
produces a natural bijection
\begin{equation}
  \xymatrix{
    \underset{
      \mathclap{
        \mbox{\color{blue}\tiny original cocycles}
      }
    }{
    \underbrace{
    \mathrm{Hom}
    \big(
      \widehat{\mathfrak{g}}, \mathfrak{h}
    \big)
    }
    }
    \ar@{->}@<+7pt>[rr]^-{\color{blue}\mathrm{reduction}}
    \ar@{<-}@<-7pt>[rr]_-{\color{blue}\mathrm{oxidation}}^-{\simeq}
    &&
    \mathrm{Hom}_{/B \mathbbm{R}}
    \big(
      \mathfrak{g},
      \mathcal{L}\mathfrak{h} \sslash \mathbbm{R}
    \big)
  }
\end{equation}
between $L_\infty$-homomorphisms out of the exteded super
$L_\infty$-algebra $\widehat{\mathfrak{g}}$
and homomorphism out of the base $\mathfrak{g}$
into the cyclification of the original coefficients
with the latter constrained so that
the canonical 2-cocycle on the cyclification is taken to the 2-cocycle classifying the given extension.

\smallskip
\noindent {\bf Example.}
Let
\begin{equation}
  \left(
    \raisebox{49pt}{
    \xymatrix@C=11pt{
      \widehat{\mathfrak{g}}
      \ar[d]
      \\
      \mathfrak{g}
      \ar[dr]
      \\
      &
      \EM{2}
    }
    }
  \right)
  \;\;\coloneqq\;\;
  \left(
    \raisebox{49pt}{
    \xymatrix@C=11pt{
      \mathbbm{R}^{d,1\vert \mathbf{N}_{d+1}}
      \ar[d]
      \\
      \mathbbm{R}^{d-1,1\vert \mathbf{N}_d}
      \ar[dr]_{ \overline{\psi} \Gamma^d \psi }
      \\
      &
      \EM{2}
    }
    }
  \right)
\end{equation}
be the extension of a super Minkowski space-time from dimension $d$ to dimension $d+1$.
Let, moreover,
$\mathfrak{h} \coloneqq b^{(p+1)+1} \mathbbm{R}$
be the line Lie $(p+3)$-algebra
and consider any super $(p+1)$-brane cocycle from the old brane scan in dimension $d+1$
\begin{equation}
\hspace{-2mm}
  \xymatrix@C=13pt{
  \mathbbm{R}^{d,1\vert N_{d+1}}\ar[rrrrr]^
  {\tiny
     \mu_{(p+1)+2} \coloneqq
      {\sum}_{a_i = 0}^{d} \left( \overline{\psi} \wedge \Gamma_{a_1 \cdots a_{p+1}} \psi \right) \wedge e^{a_1} \wedge \cdots \wedge e^{a_{p+1}}
  }\!\!\!
&&&&&
\!\!  \EM{p+2}.
  }
\end{equation}
Then the cyclification $\mathfrak{L}(b^{p+1}\mathbbm{R})\sslash\mathbbm{R}$ of the coefficients is
\begin{equation}
  \mathrm{CE}
  \left( \, \mathfrak{L}(b^{p+2}\mathbbm{R})/\mathbbm{R} \, \right)
  \;=\;
  \left\{ \setlength{\tabcolsep}{-20pt}
\renewcommand{\arraystretch}{.5}
    \begin{array}{c}
       {\rm d} \omega_2 = 0
       \\
       {\rm d} \omega_{p + 2} = 0
       \\
       {\rm d} \omega_{(p+1)+2} = \omega_{p+1} \wedge \omega_2
    \end{array}
  \right\}
\end{equation}
and the dimensionally reduced cocycle
has the components shown in Figure \ref{DDReduction}.
But there is more: the un-wrapped component of the dimensionally reduced cocycle
 satisfies the twisted cocycle condition
\begin{equation}
{\rm d} \, \mu^d_{(p+1)+2} \;=\; \mu^d_{p+2} \wedge \mu^d_{0+2}.
\end{equation}
We will study there  relations next.

\medskip

\hypertarget{SuperTopologicalTDuality}{
\section{Super topological T-duality}
 \label{SuperTopologicalTDuality}
}

Among all dualities of string theory, T-duality is
still the archetypical one. What has come to be known
as \emph{topological T-duality} (see \cite{Bouwknegt:2003vb}) is the proposal that
the global topological and cohomological aspects of
T-duality between two space-times $X$ and $\widetilde X$
carrying B-field strengths $H_3$ and $\widetilde H_3$, respectively,
should be reflected` by a \emph{correspondence} \cite{Hori:1999me,Bouwknegt:2003vb}
\begin{equation}
 \raisebox{50pt}{ \xymatrix{
    &
    X \underset{B}{\times} \widetilde X
    \ar[dl]_-{p}
    \ar[dr]^-{\widetilde p}
    \\
    X
    \ar[dr]_-{\pi}
    &&
    \widetilde X
    \ar[dl]^-{\widetilde \pi}
    \\
    & B
  }
  }
\end{equation}
where
\smallskip
\begin{enumerate}[i)]
\item the projections $\pi$ and $\widetilde \pi$ exhibit
both space-times as being circle bundles over a common base $B$,
with Chern classes $c_1$ and $\widetilde c_1$, respectively,
in the cohomology of $B$;
\item
 the B-field strengths are related to these by
 \begin{equation}
   \pi_\ast H_3 = \widetilde c_1
   \,,
   \phantom{AAA}
   \widetilde \pi_\ast \widetilde H_3 = c_1
 \end{equation}
\end{enumerate}
\smallskip
The \emph{correspondence space} $X \times_B \widetilde X$ is the corresponding \emph{fiber product};
we pointed out \cite{Fiorenza:2016oki} that this is what elsewhere came to be
called the corresponding \emph{doubled space-time}.

A core success of this proposed formalization of
cohomological T-duality is that it implies
an isomorphism between the twisted K-theory of $X$
with the twisted K-theory of $\widetilde X$ \cite{Bouwknegt:2003vb}
\begin{equation}
  \xymatrix{
    \mathrm{KU}^{0 + H_3}(X)
    \ar[rr]^-{ \widetilde p_\ast \circ p^\ast }_-{\simeq}
    &&
    \mathrm{KU}^{1 + \widetilde H_3}(X)
  }
\end{equation}
which may
be interpreted as a globalized version of Hori's formula
for the Buscher rules of RR-fields under T-duality \cite{Hori:1999me}.
A grand generalization of this statement
is constructed in \cite{Lind:2016gsh}.

\begin{figure*}[htb]

\vspace{-.7cm}

\begin{center}
\includegraphics[width=.65\textwidth]{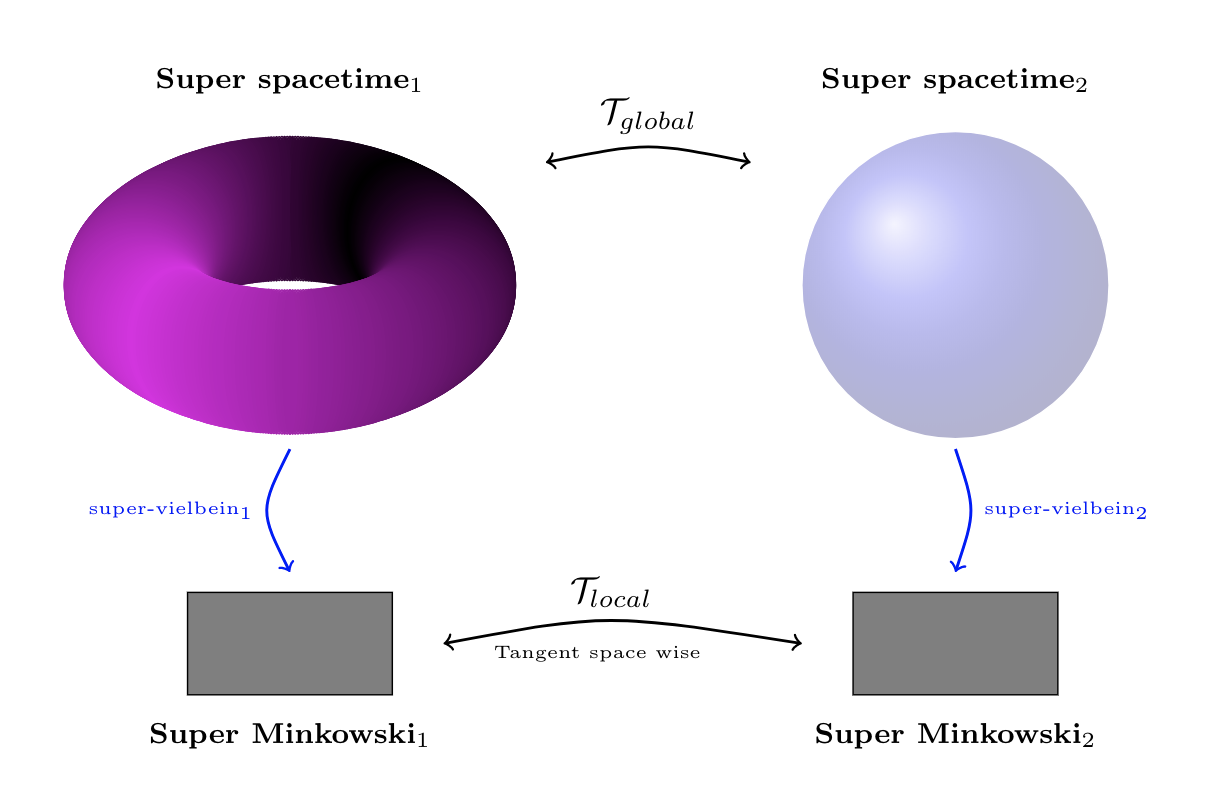}
\end{center}

\vspace{-.7cm}

\caption{
  {\bf Duality and the supergravity torsion constraints.}
  The torsion constraints of supergravity
  (Figure \ref{TorsionConstraintsOf11dSupergravity})
  fully constrain the bifermionic
  super cocycle component of all flux fields
  on superspace-time in each super tangent space.
  Therefore all global operations, such as dualities,
  when formulated in superspace need to be
  such as to preserve this local structure.
  This is a strong constraint, which allows to derive the
  cohomological rules of ``topological T-duality''
  from analysis of the D$p$-brane's supercocycles
  \cite{Fiorenza:2016oki,Fiorenza:2018ekd,Sati:2018tvj} (exposition in\cite{Fiorenza:2017jqx}).
}
\label{SuperLocalToGlobal}
\end{figure*}

While these results strongly suggested that the rules
of topological T-duality are a correct reflection
of T-duality in string theory, there has not been an actual derivation of
these rules from string theory.  This used to be an open problem.

In \cite{Fiorenza:2016oki,Fiorenza:2018ekd,Sati:2018tvj} (for exposition see \cite{Fiorenza:2017jqx}) we showed that
when passing from plain space-time to super space-time
and incorporating there the supergravity super torsion constraints,
which constrain the bifermioninc components of the RR-forms to be
given by twisted super cocycles as in the previous sections,
then the rules of topological T-duality are indeed implied by the structure
of these super cocyles.
More explicitly,  the super torsion constraints imply
the rules of topological T-duality \emph{super tangent-space-wise}
and thus globally, see Figure \ref{SuperLocalToGlobal}.
Here we briefly review this.

First of all, one finds that the type IIA/IIB $D=10$ super Minkowski
space-times are both fibered as central extensions over
$D =9$ super space-time \cite[Prop. 2.14]{Fiorenza:2016oki} in the sense
discussed above, and hence define a correspondence super space-time,
which we denote $\mathbbm{R}^{8+(1,1),1 \vert 32}$ \cite[Def. 6.1]{Fiorenza:2016oki}
  \begin{equation}
  \raisebox{70pt}{  \xymatrix@C=4pt{
      & \mathbbm{R}^{8+(1,1),1 \vert 32}
      \ar@{}[dd]|{ \mbox{\tiny (pb)} }
      \ar[dr]^{p_A}
      \ar[dl]_{p_B}
      \\
      \mathbbm{R}^{9,1\vert \mathbf{16} + \mathbf{16}}
      \ar[dr]|-{\color{blue}
        \pi_9^{\mathrm{IIB}}
        =
        \mathrm{hofib}\left( c_2^{\mathrm{IIB}} \right)
      }
      & &
      \mathbbm{R}^{9,1\vert \mathbf{16}  + \overline{\mathbf{16}}}
      \ar[dl]|-{\color{blue}
        \pi_9^{\mathrm{IIA}}
        =
        \mathrm{hofib}
        \left(
          c_2^{\mathrm{IIA}}
        \right)
      }
      \\
      & \mathbbm{R}^{8,1 \vert \mathbf{16} + \mathbf{16}}
      \ar[dl]_{c_2^{\mathrm{IIB}}}
      \ar[dr]^{c_2^{\mathrm{IIA}}}
      \\
      \EM{2}
      &&
      \EM{2}
    }
    }
  \end{equation}
To see what this implies for the super RR-charges,
hence the super cocycles for the D-branes, we
may hence apply
double dimensional reduction (from Section \ref{DoubleDimensionalReduction}) in two ways:
for the type IIA supercocycles along $\pi_9^{\mathrm{IIA}}$,
and for the type IIB supercocycles alont $\pi_9^{\mathrm{IIB}}$.
By the rules for double dimensional reduction, this process yields
cocycles in $\mathcal{L}\mathfrak{l}\big(\mathrm{KU}\sslash B U(1)\big)\sslash \mathbbm{R}$
and $\mathcal{L}\mathfrak{l}\big(\Sigma^{1}\mathrm{KU}\sslash B U(1)\big)\sslash \mathbbm{R}$,
respectively.

The key result now is
that there is an isomorphism $\phi_T$ which
identifies these two double dimensional reductions in a compatible fashion
\cite[Theorem 5.3]{Fiorenza:2016oki}
\begin{equation}
  \hspace{-.1cm}
 \scalebox{0.95}{\raisebox{40pt}{ \xymatrix@C=-14pt{
    &&
    \mathbbm{R}^{8,1\vert \mathbf{16} + \mathbf{16}}
    \ar[drr]^-{c_2^{\mathrm{IIA}}}
    \ar[dll]_-{c_2^{\mathrm{IIB}}}
    \ar[ddl]|{
      \mathcal{L}
      \left(
        \mu^{\mathrm{IIA}}_{F1/Dp}
      \right)\sslash \mathbbm{R}
    }
    \ar[ddr]|{
      \mathcal{L}
      \left(
        \mu^{\mathrm{IIB}}_{F1/Dp}
      \right)\sslash \mathbbm{R}
    }
    \\
    B \mathbbm{R}
    && &&
    B \mathbbm{R}
    \\
    &
    \mathcal{L}\mathfrak{l}
    \big(
      \Sigma \mathrm{KU} \sslash BU(1)
    \big)\sslash \mathbbm{R}
    \ar[ul]|{\omega_2}
    \ar[rr]^-{\simeq}_-{\phi_T}
    \ar@{}@<-18pt>[rr]|{
      \underset{
        \mathclap{
          \mbox{
            \tiny
            \color{blue}
            super topological T-duality
          }
        }
      }{
        \underbrace{
          \phantom{--------------------}
        }
      }
    }
    &&
    \mathcal{L}\mathfrak{l}
    \big(
      \mathrm{KU} \sslash BU(1)
    \big)\sslash \mathbbm{R}
    \ar[ur]|{\omega_2}
  }
  }}
\end{equation}

Keeping in mind that our super homotopy-theoretic
double dimensional reduction does not lose information,
since it is adjoint to ``oxidation'', we may oxidize this
situation back to a statement on the correspondence space.
There we find it is exactly the kind of pull-push isomorphism
representing Hori's formula for the Buscher rules of the R-fields
\cite[Prop. 6.4]{Fiorenza:2016oki}:
  \begin{equation}
    \hspace{-.4cm}
   \scalebox{1}{\raisebox{50pt}{ \xymatrix@C=-54pt{
    &
    H_{ \left(\pi_9^{\mathrm{IIA}}\right)^\ast\mu_{F1}^{\mathrm{IIA}}}
    \big(
      \underset{
        \mathclap{
          \mbox{
            \tiny
            \color{blue}
            \begin{tabular}{c}
              correspondence space
              \\
              doubled super space-time
            \end{tabular}
          }
        }
      }{
        \underbrace{
          \mathbbm{R}^{8+(1,1),1\vert 32}
        }
        ,
        \mathfrak{l}(\mathrm{KU})
      }
    \big)
    \ar@(ul,ur)^{\nu^\ast}
    \ar[ddr]^-{
      (\pi_9^{\mathrm{IIA}})_\ast
    }
    \\
    \\
    H_{\mu_{F1}^{\mathrm{IIA}}}
    \big(
      \underset{
        \mathclap{
        \mbox{
          \tiny
          \color{blue}
          \begin{tabular}{c}
            type IIA
            \\
            super space-time
          \end{tabular}
        }
        }
      }{
      \underbrace{
        \mathbbm{R}^{9,1\vert \mathbf{16} + \overline{\mathbf{16}}}
      }
      },
      \mathfrak{l}(\mathrm{KU})
    \big)
    \ar[rr]_-{
      \underset{
        \mathclap{`
          \mbox{
            \tiny
            \color{blue}
            \begin{tabular}{c}
              Hori's formula
              \\
              Buscher rules for RR-fields
            \end{tabular}
          }
        }
      }{
        \underbrace{
          \;\;\;\;\;\;\;\;\simeq\;\;\;\;\;\;\;\;
        }
      }
    }
    \ar[uur]^-{
      (\pi_9^{\mathrm{IIB}})^\ast
    }
    &&
    H_{\mu_{F1}^{\mathrm{IIB}}}
    \big(
      \underset{
        \mathclap{
        \mbox{
          \tiny
          \color{blue}
          \begin{tabular}{c}\
            type IIB
            \\
            super space-time
          \end{tabular}
        }
        }
      }{
        \underbrace{
          \mathbbm{R}^{9,1\vert \mathbf{16} + {\mathbf{16}}}
        }
      }
      ,
      \mathfrak{l}(\Sigma \mathrm{KU})
    \big)
    }
    }}
  \end{equation}

\hypertarget{BlackBraneScan}{
\section{Black brane scan}
 \label{BlackBraneScan}
}

\begin{figure*}[htb]

\vspace{-1.4cm}
\hspace{3cm}
\raisebox{-190pt}{
\includegraphics[width=.6\textwidth]{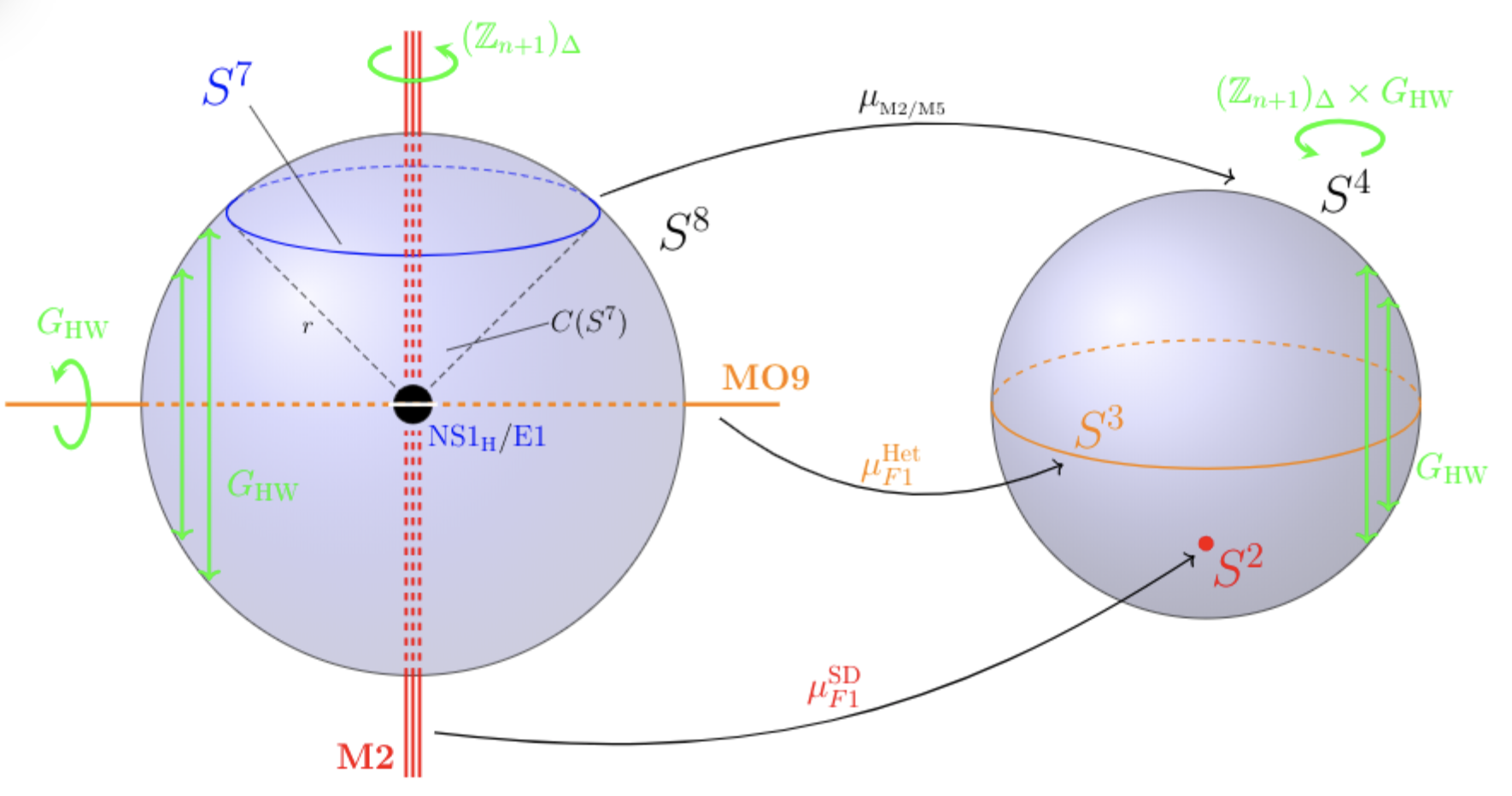}
}

\vspace{.3cm}
\begin{tabular}{|C{6.4cm}  C{2.8cm}  C{7cm} |}
%\begin{tabular}{|ccc|}
\hline
  $G$-equivariance &
  ${\xymatrix{\ar@{<->}[rrr]^{\mbox{\footnotesize Elmendorf's theorem}}&&&}}$  & $G$-fixed points
  \\
  \hline
  \hline
  \begin{tabular}{c}
    Fundamental M2/M5-branes
    \\
    on 11d superspace-time with
    \\
    real ADE-equivariant sigma model
  \end{tabular}
    & ${\xymatrix{\ar@{<->}[rrr]^{\mbox{\footnotesize
        \cite[Theorem 6.1]{Huerta:2018xyh}}}&&&}}$ &
  \begin{tabular}{c}
    Fundamental F1/M2/M5-branes
    \\
    on intersecting black M-branes
    \\
    at real ADE-singularities
    \\
  \end{tabular}
  \\
  \hline
\end{tabular}

\vspace{-.7cm}

\caption{
  {\bf The interpretation of Elmendorf's theorem}
  in equivariant homotopy theory (e.g. \cite[Thm. 1.3.6, 1.3.8]{Blumberg:2017aa}) as providing the
  missing connection in M-theory between orbifold group actions
  and hidden degrees of freedom at
  the fixed point singularities \cite{Huerta:2018xyh}.
  Specifically, the equivariant enhancement of the
unified M2/M5-brane cocycle $\mu_{{}_{M2/M5}}$ \eqref{UnifiedM2M5braneCocycle} makes appear
 BPS-branes at orbifold singularities,
 complete with their GS-instanton contributions \cite[Sec. 6.2]{Huerta:2018xyh}.
}
\label{ElmendorfConnection}
\label{EquivariantEnhancedCocycle}

\end{figure*}

Plausible but informal folklore has it that M-theory must exhibit some
``hidden degrees of freedom'' inside orbifold singularities
\cite[Sec. 4.6]{Witten:1995ex} (see \cite{Acharya:2004qe}).
The question of how to fill this idea with formal life
had been completely open.
We review here how the results of \cite{Huerta:2018xyh}
suggest that, rationally, the right kind of extra
degrees of freedom appear when enhancing
the unified M2/M5-cocycle \eqref{UnifiedM2M5braneCocycle}
from plain to \emph{equivariant homotopy theory}
(see \cite{Blumberg:2017aa}).

The key observation here is that we may
interpret a core theorem of equivariant homotopy theory,
\emph{Elmendorf's theorem} (see \cite[Thm. 1.3.6, 1.3.8]{Blumberg:2017aa}),
as providing exactly the
missing connecton between orbifold geometry and
``hidden degrees of freedom'' localized inside the fixed point singularities. See Figure \ref{ElmendorfConnection}.

\begin{figure*}[htb]

\begin{center}
\hspace{-.1cm}
\includegraphics[width=.9\textwidth]{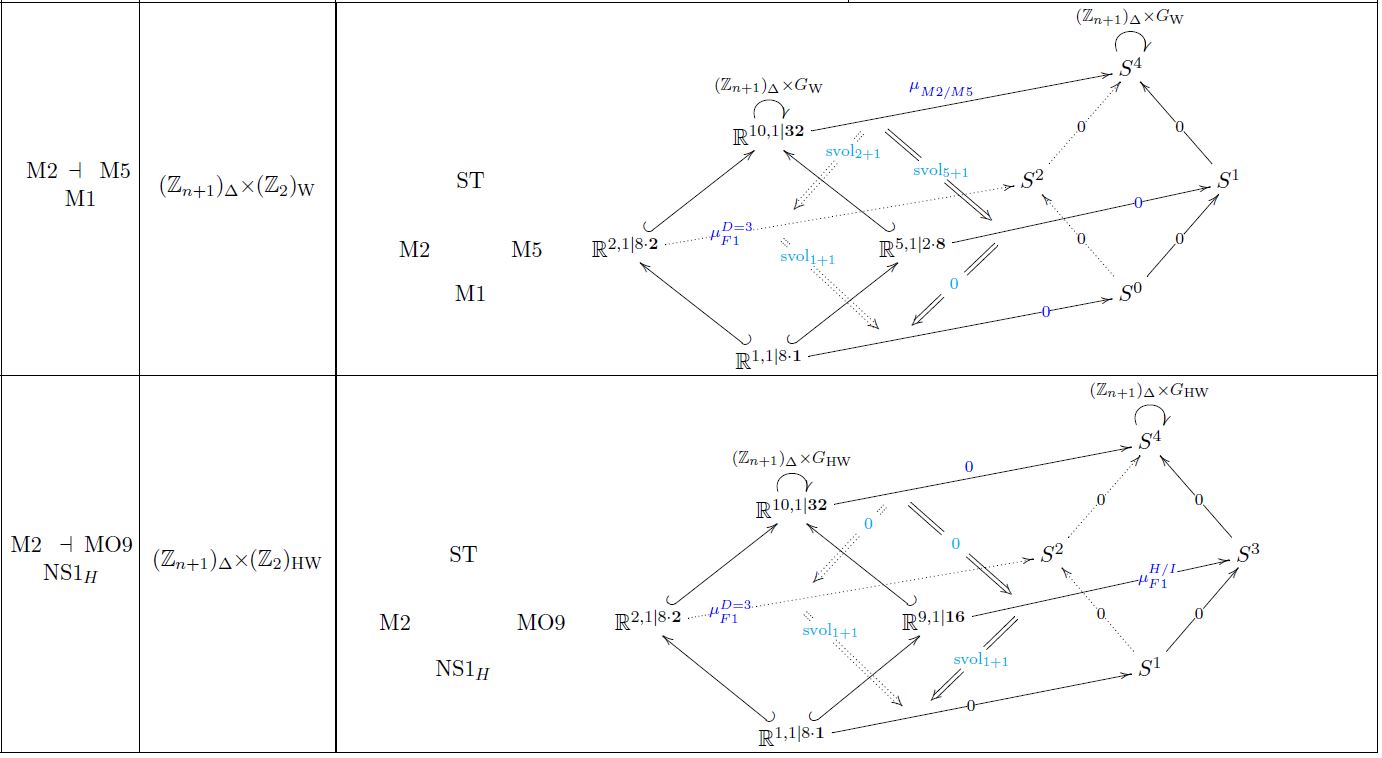}
\end{center}

\vspace{-.7cm}

\caption{
  {\bf Intersecting black branes}
  with fundamental $p$-branes propagating on them, obtained
  by enhancing the fundamental M2/M5-brane cocycle $\mu_{{}_{M2/M5}}$
  \eqref{UnifiedM2M5braneCocycle} to equivariant homotopy theory
  (via Elmendorf's theorem, Figure \ref{EquivariantEnhancedCocycle})
  \cite[Fig. 3]{Huerta:2018xyh}.
  This reveals \emph{super embeddings} of the branes
  into super space-time as in
  \cite{Sorokin:1999jx} (shown in the middle).
  The homotopies that appear filling the diagram
  turn out to be given by the super volume
  $\mathrm{svol}_{p+1}$,
  which, by 1/2 BPS super embedding, turns out to
  equal the full Green-Scharz action functional,
  of the embedded brane, hence its
  brane instanton contribution \cite[Sec. 6.2]{Huerta:2018xyh}.
}
\label{InstantonContributions}
\end{figure*}

\smallskip
\noindent {\bf Instanton contributions.}
So far we have entirely been considering the
super cocycles of the super $p$-branes, which are the
(curvatures of) their WZW-terms. Of course this is just one
term in the Lagrangian density for the
Green--Schwarz-type sigma models for these super sigma models,
the other being the kinetic NG-action, proportional to the
proper super world-volume.
Exactly that appears now in the equivariant enhancement
at the given black brane's embedding locus \cite[Sec. 6.2]{Huerta:2018xyh},
see Figure \ref{InstantonContributions}.

This exhibits in fact the \emph{superembedding} perspective on
super $p$-branes \cite{Sorokin:1999jx}, where not just the target space is a supermanifold, but also the world-volume of the super $p$-brane is, and where
the embedding fields are super embeddings picking half-BPS loci.

\hypertarget{MIIADuality}{
\section{M/IIA duality and gauge enhancement}
 \label{MIIADuality}
}

Applying double dimensional reduction (Section \ref{DoubleDimensionalReduction}) to the
M2/M5-brane charge in cohomotopy yields
the F1/D0/D2/D4-brane cocycle in IIA in
truncated twisted K-theory, rationally \cite[Sec 3.]{Fiorenza:2016ypo}.

We may then invoke fiberwise Goodwillie linearization to
parameterized stable homotopy theory. This induces
the missing $D6/D8$ brane charges, completes the cocycle
to a cocycle in un-truncated twisted K-theory (rationally)
and hence exhibits gauge enhancement \cite{Braunack-Mayer:2018uyy}.
This is reviewed in a separate contribution to this
collection \cite{contrib:braunackmeyer} and hence we will not further
discuss it here.

\medskip

\begin{figure}

\vspace{.7cm}

$$
  \hspace{-.4cm}
  \xymatrix@C=-14pt@R=8pt{
    \overset{
      \mbox{
        \tiny
        \color{blue}\
        \begin{tabular}{c}
          $C_3/C_6$-field
          \\
          in rational Cohomotopy
        \end{tabular}
      }
    }{
      \overbrace{
        \Big(
          \mathbbm{R}^{10,1\vert \mathbf{32}}
          \xrightarrow{\tiny \mu_{M2/M5} }
          S^4_{\mathbbm{R}}
        \Big)
      }
    }
    \ar@{|->}@/^2pc/[drr]|>>>>>{
      \mbox{
        \tiny
        \color{blue}
        \begin{tabular}{c}
          double dimensional reduction
          \\
          \& gauge enhancement
        \end{tabular}
      }
    }
    \\
    &&
    \underset{
      \mbox{
        \tiny
        \color{blue}
        \begin{tabular}{c}
          type IIA NS- \& RR-fields
          \\
          in rational twisted K-theory
        \end{tabular}
      }
    }{
      \!\!\!\!\underbrace{
        \Big(
          \mathbbm{R}^{9,1\vert \mathbf{16} + \overline{\mathbf{16}}}
          \xrightarrow{
            \mu_{ F1/D_{2p} }
          }
          \mathrm{ku}_{\sslash B U(1)}
        \Big)
      }
    }
    \\
    \raisebox{-11pt}{
      \hspace{-27pt}
      \begin{rotate}{90}
        \mbox{
          \tiny
          \color{blue}
          \hspace{7pt} brane bouquet
        }
      \end{rotate}
      \includegraphics[width=.18\textwidth]{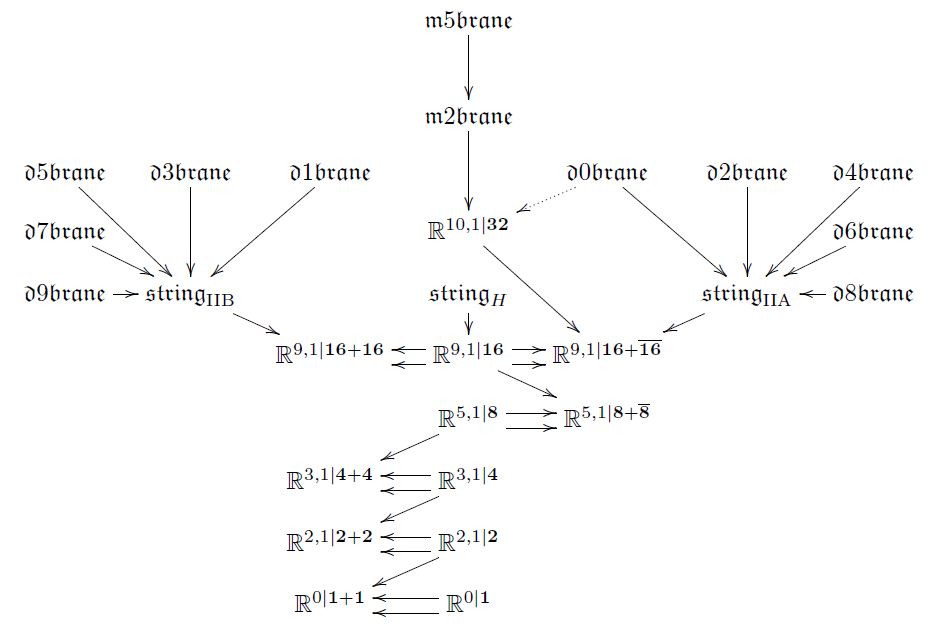}
    }
    \ar@{|->}@/^2pc/[uu]|-{
      \mbox{
        \tiny
        \color{blue}
        \begin{tabular}{c}
          descend
          \\
          M5-cocycle
        \end{tabular}        \\
      }
    }
    \ar@<+7pt>@{|->}@/_2.6pc/[urr]|-{
      \mbox{
        \tiny
        \color{blue}
        \begin{tabular}{c}
          descend
          \\
          $D_{2p}$-cocycles
        \end{tabular}
      }
    }
    \\
    \mathbbm{R}^{0 \vert 1}
    \ar@{|->}@/^{2pc}/[u]^-{
      \mbox{
        \tiny
        \color{blue}
        \begin{tabular}{c}
          progression of
          \\
          universal
          \\
          invariant
          \\
          higher
          \\
          central extensions
        \end{tabular}
      }
    }
    \\
    \\
    \varnothing
    \ar@{|->}@/^2pc/[uu]^-{
      \mbox{
        \tiny
        \color{blue}
        \begin{tabular}{c}
          progression of
          \\
          modalities
        \end{tabular}
      }
    }
  }
$$

\vspace{-.9cm}

\caption{
  {\bf Emergence of rational higher structure of M-theory}
  in super homotopy theory. Shown is part of the structure
  and their progression as reviewed above.
}
\label{Conclusions}
\end{figure}

\hypertarget{SectionOutlook}{
\section{Outlook -- Beyond rational}
 \label{SectionOutlook}
}
In conclusion, we find that natural progressions
in super homotopy theory discover at least the rational/infinitesi\-mal
core structures of M-theory; see Figure \ref{Conclusions}.

\smallskip
\noindent {\bf Reduction to mathematical classification.}
Like other classifications in pure mathematics,
for instance that of finite groups, these are god-given structures that pure
homotopy theorists could have and eventually would have discovered by themselves,
even if no hints from perturbative string scattering had been available.
This suggests that super homotopy theory holds the key principle
for unraveling M-theory, and that further refinement of these
classifications, beyond the infinitesimal/rational approximation,
should reveal it.

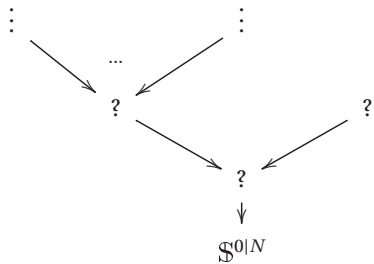
\begin{figure}[htb]

\vspace{-.7cm}

$$
  \xymatrix@R=1em{
    \vdots
    \ar[dr]
    &
    \ar@{}[d]|-{\cdots}
    &
    \vdots
    \ar[dl]
    \\
    &
    ?
    \ar[dr]
    &&
    ?
    \ar[dl]
    \\
    &
    &
    ?
    \ar[d]
    \\
    &
    &
    \mathbbm{S}^{0\vert N}
  }
$$

\vspace{-.7cm}

\caption{ {\bf The bouquet of universal invariant extensions
of the ``absolute superpoint'' $\mathbbm{S}^{0\vert N}$}
in spectral algebraic geometry would suggest itself as the god-given
completion of the brane bouquet in Figure \ref{Figure1}, beyond
the infinitesimal/rational approximation. This remains to be worked
out.
}

\end{figure}

\smallskip
\noindent {\bf M-theory from the \emph{absolute} superpoint?}
For example, in the homotopy-theoretic enhancement of
algebraic geometry to \emph{spectral algebraic geometry}
\cite{Lurie:xx}
it is natural to regard the superpoint, which as
a graded scheme is
\begin{equation}
  \mathbbm{R}^{0\vert N}
  \;=\;
  \mathrm{Spec}
  \Big(
    \mathrm{Sym}_{\mathbbm{R}}
    \big(
      \underset{
        \mathclap{
          \mbox{\color{blue}
            \tiny
            $N$ direct summands
          }
        }
      }
      {
      \underbrace{
        \mathbbm{R}[1]
        \oplus
         \cdots
        \oplus
        \mathbbm{R}[1]
      }
      }
    \big)
  \Big)
  \,,
\end{equation}
instead as a \emph{spectral scheme} (which are automatically graded! see also \cite[Sec. 2]{Rezk:0902.2499})
\begin{equation}
  \begin{aligned}
    R^{0\vert 1}
    & :=
    \mathrm{Spec}
    \Big(
      \mathrm{Sym}_{R}
      \big(
        \underset{
          \mathclap{
            \mbox{\color{blue}
              \tiny
              $N$ wedge summands
            }
          }
        }{
          \underbrace{
            \Sigma R \vee \cdots \vee \Sigma R
          }
        }
      \big)
    \Big)
    \\
    & \simeq
    \mathrm{Spec}
    \Big(
      R
        \wedge
      \mathrm{Sym}_{\mathbbm{S}}
      \big(
        \underset{
          \mathclap{
            \mbox{\color{blue}
              \tiny
              $N$ wedge summands
            }
          }
        }{
        \underbrace{
          \Sigma \mathbbm{S}
          \vee \cdots \vee
          \Sigma \mathbbm{S}
        }
        }
      \big)
    \Big)\;,
  \end{aligned}
\end{equation}
where $\mathbbm{S}$ denotes the sphere spectrum
(as in Figure \ref{StableHomotopyGroupsOfSpheres}),
$R$ is some ring spectrum serving as the ground ring,
and where $\wedge$ denotes smash product
and $\Sigma(-)$ denotes suspension of spectra.
In fact, the only canonical choice at this point seems to be
$R = \mathbbm{S}$ itself (the ``real integers''),
which suggests that the \emph{absolute superpoint} should
be the spectral scheme
\begin{equation}
  \mathbbm{S}^{0\vert 1}
  \;\coloneqq\;
  \mathrm{Spec}
  \Big(
    \mathrm{Sym}_{\mathbbm{S}}
    \big(
      \Sigma\mathbbm{S}
    \big)
  \Big)
  \,.
\end{equation}
It would be interesting to work out the bouquet
of universal invariant higher central extensions
in spectral algebraic geometry that grows out of this
absolute superpoint, in direct analogy to
the bouquet growing out of $\mathbbm{R}^{0\vert 1}$
in Figure \ref{Figure1}.
Since $\mathbbm{S}^{0\vert N}$
is a highly non-rational version of $\mathbbm{R}^{0\vert N}$,
it would be a plausible conjecture that this spectral bouquet
discovers M-theoretic structure beyond the rational approximation.
But working this out is mighty hard and will need to be done on another day.

\medskip

\begin{figure*}[htb]

\vspace{-.7cm}

\begin{center}
 \includegraphics[width=.8\textwidth]{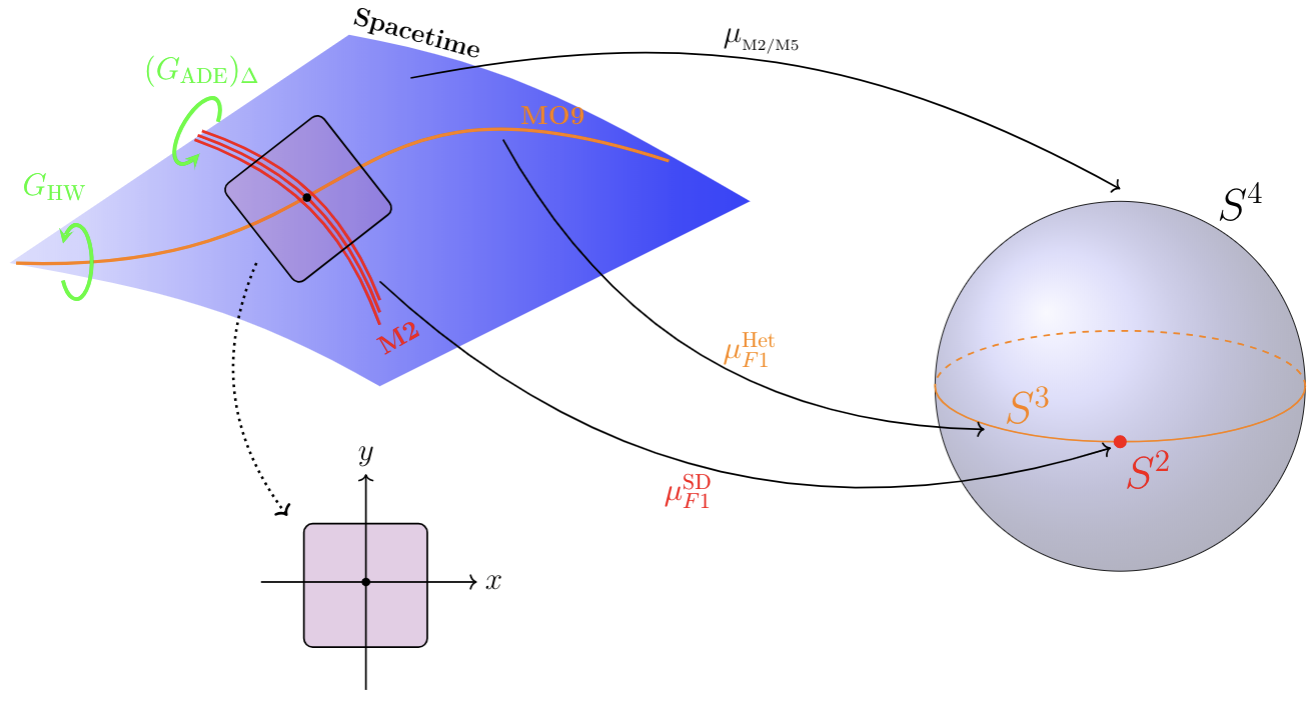}
\end{center}

\vspace{-.7cm}

\caption{
  {\bf The global structure of microscopic M-theory},
  as suggested via Cartan geometry
  (Figure \ref{SuperCartanGeometry})
  by the infinitesimal/rational brane bouquet
  (Figure \ref{Figure1}, and Figure \ref{Conclusions}),
  is a theory of orbifold supergravity with
  Dirac charge quantization of the
  $C_3/C_6$-field in some version of
  differential equivariant Cohomotopy, constrained
  on each super tangent space to reduce rationally
  to the canonical M2/M5-supercocycle $\mu_{{}_{M2/M5}}$
  from Section \ref{ChargeQuantization}.
  We construct and analyze this in \cite{Braunack-Mayer:2019ip}.
  Sections \ref{OrbifoldSupergavity} \& \ref{ChargeQuantizationInEquivariantCohomotopy}
  give some indications
}
\label{TowardsMicroscopicMTheory}
\end{figure*}

\noindent {\bf Towards microscopic M-theory.}
In the meantime, we may try to climb down from the heavens
of god-given structures with what we already managed
to grasp there, and see if with some educated guesswork we
may complete the infinitesimal/rational higher structure
of M-theory, as in Figure \ref{Conclusions}, to a
global and torsionful structure that passes some consistency
checks of a putative formulation of M-theory.
We will discuss this in \cite{Braunack-Mayer:2019ip}. Here we just close with
some brief indications.

Looking at Figure \ref{Conclusions} and in view of
the discussion in Section \ref{BlackBraneScan},
the task is to consistently  define a theory of supergravity
coupled to higher gauge fields subject to the following:
We require that on each equivariant super tangent space
$G \hspace{.25cm}\raisebox{-1pt}{\begin{rotate}{90}
$\curvearrowright$\end{rotate}} \; \mathbbm{R}^{10,1\vert \mathbf{32}} $
equipped with its canonical super vielbein, the theory is
given in rational approximation by an equivariant enhancement
of the
canonical M2/M5-brane super cocycle
\begin{equation}
  \xymatrix@R=-1.5em{
  \overset{G}{\curvearrowright} && \overset{G}{\curvearrowright}\\
   { \mathbbm{R}^{10,1\vert \mathbf{32}}}
   % \ar@(ul,ur)^{G}
    \ar[rr]^-{ \mu_{{}_{M2/M5}} }
    &&
    S^4_{\mathbbm{R}}
   % \ar@(ul,ur)^{G}
  }
\end{equation}
as indicated in Figure \ref{EquivariantEnhancedCocycle}.
But the required refinement of both sides is fairly clear
(see Figure \ref{TowardsMicroscopicMTheory}):
\smallskip
\begin{enumerate}[i)]
  \item on the left we are led to
  super torsion-free orbifold supergravity
  (Section \ref{OrbifoldSupergavity});
  \item on the right we are led to
   differential equivariant Cohomotopy
   (Section \ref{ChargeQuantizationInEquivariantCohomotopy}).
\end{enumerate}

\begin{figure}[htb]

\includegraphics[width=.49\textwidth]{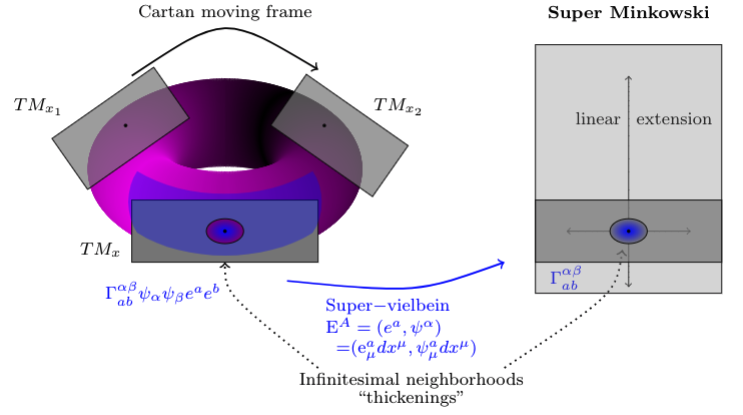}

\vspace{-.5cm}

\caption{
  {\bf Super Cartan geometry} is, mathematically,
  the most natural formulation
  of supergravity \cite{Lott:2001st,Egeileh:2012bca}. It is
  also implicit in the powerful ``geometric'' approach to supergravity
  in the physics literature \cite{D'Auria:1982nx,Castellani:1991et}.
  Last not least, its local-to-global principle
  is just the kind of principle needed to obtain from the
  rational/infinitesimal brane bouquet
  (Figure \ref{Figure1} and Figure \ref{Conclusions})
  a global and torsionful theory of supergravity/M-theory.
}
\label{SuperCartanGeometry}
\vspace{-.5cm}

\end{figure}

\subsection{Orbifold supergravity}
 \label{OrbifoldSupergavity}

\noindent {\bf Local-to-global principle of Cartan geometry.}
The way that the rational/infinitesimal analysis of the
brane bouquet (Figures \ref{Figure1} \& \ref{Conclusions})
connects to global curved space-time geometry should be that
the super Minkowski space-times
  $G \;\;\; \;\raisebox{-1pt}{\begin{rotate}{90} $\curvearrowright$\end{rotate}}
  \; \mathbbm{R}^{10,1\vert\mathbf{32}}$ be the super tangent spaces to
  super orbifold space-times $\mathcal{X}$.
  Moreover, the canonical super vielbein fields $(e^a, \psi^\alpha)$
  on $\mathbbm{R}^{10,1\vert \mathbf{32}}$
  should be the restriction, up to $\mathrm{Spin}(10,1)$-gauge transformation, of a super vielbein field $(E,\Psi)$
  on all of $\mathcal{X}$  on the infinitesimal neighborhood of every point (see Figure \ref{SuperCartanGeometry}).

Note that this is the perspective of \emph{super Cartan geometry}
  on supergravity \cite{Lott:2001st,Egeileh:2012bca}, which in
  the physics literature
  is the ``geometric perspective on supergravity'' due
  to \cite{D'Auria:1982nx,Castellani:1991et}.

\begin{figure*}[htb]

\begin{center}
\begin{tabular}{|c|l|l|}
  \hline
  \multirow{2}{*}{
    \bf
    \begin{tabular}{c}
      Constraints on
      \\
      11d supergravity
    \end{tabular}
  }
  &
  \multicolumn{2}{c|}{ \bf \hspace{-2.2cm} Torsion constraints }
  \\
  \cline{2-3}
  &
  \begin{tabular}{l}
    {\bf Supertorsion tensor}
    \\
    {\bf vanishes\dots}
  \end{tabular}
  &
  \begin{tabular}{l}
    {\bf Geometric $\mathrm{Spin(10,1)}$-structure}
    \\
    {\bf coincides with that of $\mathbbm{R}^{10,1\vert \mathbf{32}}$\dots }
  \end{tabular}
  \\
  \hline
  \hline
  \begin{tabular}{c}
    \begin{tabular}{l}
       \!\!\!\! equations of motion \!\!
    \end{tabular}
  \end{tabular}
   &
  \begin{tabular}{l}
     \dots in its bosonic components \dots
  \end{tabular}
 &
  \\
  \hline
  \begin{tabular}{l}
 equations of motion \!\!
    \\
      \&
    vanishing flux
  \end{tabular}
  &
  \begin{tabular}{l}
    \dots completely
  \end{tabular}
  &
  \begin{tabular}{l}
  \dots on all first-order infinitesimal neighborhoods
  \end{tabular}
  \\
  \hline
  \begin{tabular}{l}
     vanishing curvature \!\!
    \\
    \& vanishing flux
  \end{tabular}
  &&
  \begin{tabular}{l}
  \dots on all infinitesimal neighborhoods
  \end{tabular}
  \\
  \hline
\end{tabular}
\end{center}
\vspace{-.7cm}
\caption{
  {\bf Torsion constrains in 11d-supergravity.}
  Regarding supergravity as super Cartan geometry (Figure \ref{SuperCartanGeometry}),
  classical results of \cite{Guillemin:1965maa}
  (see \cite[Sec. 3]{Lott:2001st})
  imply that vanishing
  of the super torsion tensor $\tau$ is equivalent to space-time
  super geometry being equivalent to that of super Minkowski space-time
  $\mathbbm{R}^{d,1\vert \mathbf{N}}$
  on the first order infinitesimal neighborhood of every space-time
  point (Figure \ref{SuperCartanGeometry}).
  Remarkably, for $D = 11$ and $\mathcal{N} = 1$
  this condition is equivalent to the
  equations of motion of 11-dimensional supergravity
  ($\tau^a = 0$) subject to the constraint of vanishing
  bosonic 4-form flux ($\tau^\alpha = 0$)
  \cite{Candiello:1993di,Howe:1997he} (see \cite[Sec. 2.4]{Cederwall:2004cg}).
  }
  \label{TorsionConstraintsOf11dSupergravity}
\end{figure*}

 \noindent{\bf Supergravity equations of motion from torsion constraints.}
  In particular, by the classical result of \cite{Guillemin:1965maa}, the
  condition that the global supergravity geometry coincides with that
  of super Minkowski space-time on the infinitesimal neighborhood of
  each point is equivalently the condition that the
  \emph{super torsion tensor vanishes}.
  Moreover, by the striking
  result of \cite{Candiello:1993di,Howe:1997he} (see \cite[Sec. 2.4]{Cederwall:2004cg}) in
  $D =11$, $\mathcal{N} = 1$ the vanishing of
  the bosonic components of the supertorsion tensor
  ($\tau^a =0$) is already equivalent to the equations of motion
  of 11-dimensional supergravity, which implies that the full vanishing of
  the super torsion tensor (also $\tau^\alpha = 0$) is equivalent
  to 11-dimensional supergravity with vanishing bosonic 4-form flux.

  Hence when viewed through the lens of
  higher Cartan geometry, the brane bouquet naturally
  leads to on-shell 11-dimensioanl supergravity.

  We could take this one step further and demand that the
  super orbifold space-time geometry $\mathcal{X}$ is equivalent
  to that of $\mathbbm{R}^{10,1\vert \mathbf{32}}$ not only
  on each first-order infinitesimal neighborhood, but
  on the full formal neighborhood.
  Again by \cite{Guillemin:1965maa}\cite[Sec. 3]{Lott:2001st},
  this is now equivalent to the further constraint that
  in addition to the flux also the
  Einstein curvature tensor vanishes.

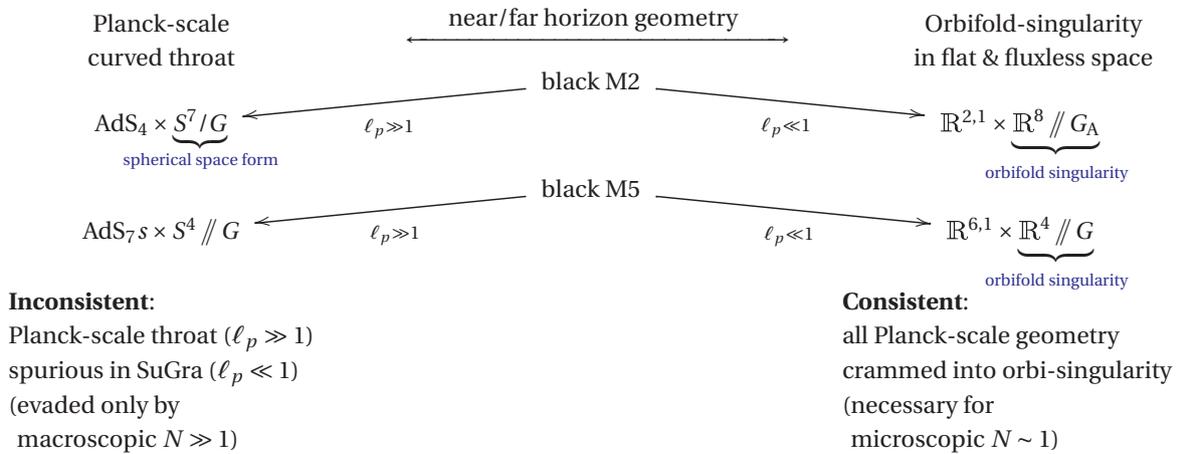
\begin{figure*}[htb]

\vspace{-.1cm}

$$
  \hspace{-.6cm}
  \raisebox{51pt}{
  \xymatrix@R=-14pt@C=9pt{
    \mathclap{
    \mbox{
      \begin{tabular}{c}
        Planck-scale
        \\
        curved throat
      \end{tabular}
    }}
    %\ar@{~>}@<-35pt>[dd]
    &
   \;\;\; \xleftrightarrow{\;\;\;\;
      \mbox{
        \begin{tabular}{c}
          near/far horizon geometry
        \end{tabular}
      \;\;\;\;}
    }
    &
\;\;\; \;\;\;  {
   \mbox{
      \begin{tabular}{c}
        Orbifold-singularity
        \\
        in flat \& fluxless space
      \end{tabular}
    }
    }
    \\
    &
    \mbox{black M2}
    \ar[dr]_-{ \;\ell_p \ll 1\; }
    \ar[dl]^-{ \;\ell_p \gg 1\; }
    \\
    \mathrm{AdS}_{4} \times
    \underset{
      \mathclap{
        \mbox{
          \tiny
          \color{blue}
          spherical space form
        }
      }
    }{
      \underbrace{
        S^{7}/ G
      }
    }
    &&
    \mathbbm{R}^{2,1}
    \times
    \underset{
      \mathclap{
        \mbox{
          \tiny
          \color{blue}
          \begin{tabular}{c}
            orbifold singularity
          \end{tabular}
        }
      }
    }{
      \underbrace{
        \mathbbm{R}^{8}\sslash G_{\mathrm{A}}
      }
    }
    \\
    {\phantom{A}}
    \\
    {\phantom{A}}
    \\
    &
    \mbox{black M5}
    \ar[dr]_-{ \;\ell_p \ll 1\; }
    \ar[dl]^-{ \;\ell_p \gg 1 \;}
    \\
    \mathrm{AdS}_{7}
      s\times
    S^{4}\sslash G
    &&
    \mathbbm{R}^{6,1}
    \times
    \underset{
      \mathclap{
        \mbox{
          \tiny
          \color{blue}
          \begin{tabular}{c}
            orbifold singularity
          \end{tabular}
        }
      }
    }{
      \underbrace{
        \mathbbm{R}^{4} \sslash G
      }
    }
    \\
    {\phantom{A}}
    \\
    {\phantom{A}}
    \\
    \\
    \mbox{
      \begin{tabular}{l}
        {\bf Inconsistent}:
        \\
        Planck-scale throat ({$\ell_p \gg 1$})
        \\
        spurious in SuGra
        ({$\ell_p \ll 1$})
        \\
        (evaded only by
        \\
        $\phantom($macroscopic $N \gg 1$)
      \end{tabular}
    }
    &&
    \!\!\!\!\!\!\!\!\!\!
    \mbox{
      \begin{tabular}{l}
        {\bf Consistent}:
        \\
        {all Planck-scale geometry}
        \\
        {crammed into orbi-singularity}
        \\
        (necessary for
        \\
        $\phantom{(}$microscopic  $N \sim 1$)
      \end{tabular}
   }
  }
  }
$$

\vspace{-.7cm}

\caption{
  {\bf Microscopic M-theory on flat and fluxless orbifold space-times.}
  In the microscopic ``small $N$-limit''
  black branes are consistent only as
  cone branes
  \cite[Sec. 2 \& 3]{Acharya:1998db}, \cite[Sec. 8.3]{deMedeiros:2010dn}.
  These are flat and fluxless orbifolds
  with ``hidden degrees of freedom'' inside the orbifold
  singularities -- which by \cite{Huerta:2018xyh} is taken care of by
  equivariant homotopy theory, via Elmendorf's theorem
  (Figure \ref{ElmendorfConnection}).
}
\label{ConeBranes}

\end{figure*}

  \noindent{\bf Flat orbifolds are universal quantum geometries.}
  The resulting
  \emph{flat and fluxless supergravity} would be essentially
  trivial in ordinary geometry, but here, in the higher geometry
  of super orbifolds, it is not only highly non-trivial, but also
  curiously relevant for M-theory:

  On the one hand, a ``flat orbifold''
  (often: ``Euclidean orbifold'', e.g. \cite[Sec. 13]{Ratcliffe:2006})
  really means that it is flat \emph{away from the orbifold singularities},
  while curvature is concentrated singularly inside the
  orbifold singularities. In particular there are ``flat orbifolds''
  whose underlying topological space is an $n$-sphere
  (the simplest of these being the pillowcase orbifold structure on the
  2-sphere).

  This is noteworthy because flatness away from the singularities
  means that we have a ``universal space-time without quantum corrections''
  (in the sense of \cite{Coley:2008th}) which is thus guaranteed to be a
  solution to 11-dimensional supergravity with
  \emph{all M-theoretic higher curvature corrections included}, which is
  otherwise a wide open problem (see e.g. \cite{Cederwall:2004cg}). At the
  same time, our ambient \emph{equivariant} super homotopy theory
  ensures that M-theoretic degrees of freedom hidden inside the
  orbifold/curvature singularities are being accounted for
  (see Figure \ref{ElmendorfConnection}).

\smallskip
  \noindent{\mathversion{bold}\bf Microscopic M-theory: the ``small $N$-limit''.}
   In the microscopic ``small $N$-limit''
  black branes are consistent only as
  cone branes (\cite[Sec. 2 \& 3]{Acharya:1998db}\cite[Sec. 8.3]{deMedeiros:2010dn})
  which are flat and fluxless orbifolds
  with ``hidden degrees of freedom'' inside the orbifold
  singularities -- which here are taken care of by
  equivariant homotopy theory, via Elmendorf's theorem
  (Figure \ref{ElmendorfConnection}). See Figure \ref{ConeBranes}.

\smallskip
  \noindent {\bf Toroidal orbifolds.}
  Finally, under mild conditions all flat orbifolds are global quotients
  of flat $n$-tori \cite[Theorem 13.3.10]{Ratcliffe:2006},
  and these \emph{toroidal orbifolds}
  constitute most of the examples of orbifolds considered in the string/M-theory literature, anyway.

  \smallskip

  In summary this means that equivariant super homotopy theory
  of flat superorbifold space-time does have a plausible chance to
  know about microscopic M-theory.

\begin{figure*}[htb]

\begin{center}
\begin{tabular}{ll}
  \raisebox{60pt}{
  \xymatrix@C=4pt@R=4pt{
  \ar@{}[r]^-{
    \!\!\!\!\!\!\!\!\!\!
    \overset{
      \mbox{\cite{Huerta:2018xyh}}
    }{
    \overbrace{
      \phantom{----------------}
    }
    }
  }
  &
  \\
  \overset{
    \mbox{
      \begin{tabular}{c}
        {ADE-singularity}
        \\
        {\cite{deMedeiros:2009pp,deMedeiros:2010dn}}
      \end{tabular}
    }
  }{
  \overbrace{
  \left(
    G_{\mathrm{ADE}} \times_{Z} G'_{\mathrm{ADE}}
  \right)
  }}
  \ar@{}[r]|>>{\times}
  \ar@{^{(}->}[d]
  &
  \overset{
    \mbox{
      \begin{tabular}{c}
        {O-plane}
        \\
        $\phantom{A}$
      \end{tabular}
    }
  }{
  \overbrace{
    \mathbbm{Z}_2
  }
  }
  \ar@{=}[d]
  \\
  \mathrm{Spin}(4)
  \ar@{}[r]|-{\times}
  &
  O(1)
  \ar@{^{(}->}[rr]
  &&
  \mathrm{Pin}(5)
  }
  }
  &
 \includegraphics[width=.3\textwidth]{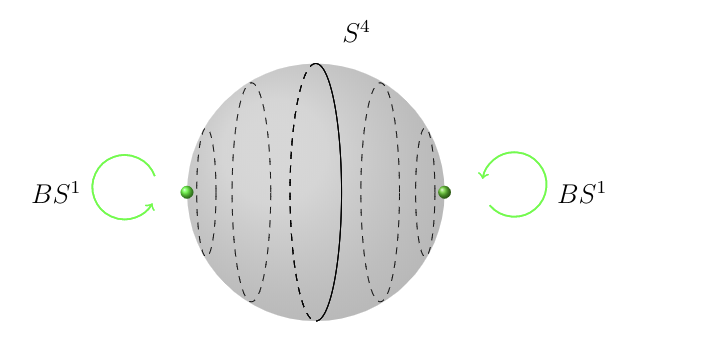}
\end{tabular}
\end{center}
\vspace{-.7cm}
\caption{
  {\bf The canonical $\mathrm{Pin}(5)$-equivariant
  enhancement of the 4-sphere} yields
  equivariant Cohomotopy cohomology theory which
  accounts both for the
  combined $\mathrm{ADE} \times \mathrm{ADE}$-singularities
  of M2- and M5-branes \cite{deMedeiros:2009pp,deMedeiros:2010dn}
  as well as the for the $\mathbbm{Z}_2$-equivariance
  of the MO9-, MO5-planes \cite[Sec. 3]{Hanany:2000fq}.
  Shown on the right is a schematic indication of
  just the $S^1 \subset \mathrm{SU}(2) \subset \mathrm{Pin}$
  equivariant structure, relevant for KK-monopoles.
}
\label{Equivariant4Sphere}

\end{figure*}

\subsection{M-brane charge quantization in cohomotopy}
 \label{ChargeQuantizationInEquivariantCohomotopy}

We had seen in Section \ref{ChargeQuantization}
that analysis in rational homotopy theory (Figure \ref{SullivanConstruction}) reveals the unified M2/M5 super cocycle,
hence the $C_3/C_6$-field
\eqref{UnifiedM2M5braneCocycle}, to have coefficients that
are rationally the 4-sphere, hence that the M-brane charge
quantization is rationally given by \emph{cohomotopy} in degree 4.

While there are many non-rational lifts of the
rational 4-sphere, one immediately stands out as being \emph{minimal}
with respect to number of cells: the actual 4-sphere, classifying
actual \emph{cohomotopy} cohomology theory
in degree 4 \cite{borsuk1936groupes,Spanier:1949:203-}. This is noteworthy,
since homotopy-theoretic formulation of M/IIA-duality \cite{Braunack-Mayer:2018uyy},
reviewed in Section \ref{MIIADuality}, applies also to
any non-rational lift of M-brane charge quantization and hence
implies a non-rational lift of D-brane charge quantization.

More precisely, in view of the discussion in Section \ref{BlackBraneScan}
we are to ask for a lift beyond the rational approximation in
\emph{equivariant} homotopy theory, which similarly leads to
consideration of the equivariant 4-sphere under its
canonical $O(5)$-action, see Figure \ref{Equivariant4Sphere}.

Hence the rational analysis of the brane bouquet suggests that
the correct charge quantization of the $C_3/C_6$-field in
in some version of \emph{differential $\mathrm{Pin}(5)$-equivariant 4-cohomotopy}.
 We will describe this cohomology theory in
\cite{Braunack-Mayer:2019ip} and discuss various consistency and plausibility checks.

Here we close with sketching one of these checks.

\smallskip
\noindent {\bf Open problem of M5-branes at ADE-singularities.}
By \cite{Acharya:1998db} the general form of a black M5-brane solution to 11d supergravity is a metric
that has the following two limits in terms of the Riemannian scale, expressed in units of the Planck length $\ell_P$
times the cube root of the number $N \in \mathbbm{N}$ of ``coincident'' M5-branes:
\smallskip
\begin{enumerate}[i)]

\item  in the {\bf near horizon/large $N$-limit} ($\ell_P N^{1/3} \gg 1$) it is the product of
an $\mathrm{AdS}_7$-space-time with
  the quotient $S^4 \!\sslash\! G$ of the 4-sphere $S^4$ by the group action;

\item  in the {\bf far horizon/small $N$-limit} ($\ell_P N^{1/3} \gg 1$) it is the
world-volume $\mathbbm{R}^{5,1}$ of the 5-brane
 times the \emph{metric cone} on $S^4 \!\sslash\! G$
\end{enumerate}

\vspace{-7mm}
\begin{equation}
  \hspace{-.3cm}
 \xymatrix@C=2.5pt{
    &
    \mbox{\color{blue}
      \tiny
      \begin{tabular}{c}
        full
        \\
        black M5-brane
        \\
        space-time
      \end{tabular}
    }
    \ar[dr]^{\ell_P N^{1/3} \ll 1 }
    \ar[dl]_{ \ell_P N^{1/3} \gg 1 }
    \\
    \mathrm{AdS}_7 \times (S^4 \!\sslash\! G)
    &&
    \mathbbm{R}^{5,1} \times \mathrm{C}(S^4 \!\sslash\! G)
  }
\end{equation}

The tacit assumption would be that the action of $G$ on the 4-sphere is free,
hence that the homotopy quotient coincides with the usual quotient,
$S^4 \!\sslash \!G = S^4/G$. The analogue of this statement does hold for the
M2-brane space-times, as  long as they are $> 1/4$ BPS \cite{deMedeiros:2009pp}.
But this does \emph{not} actually hold for M5-branes \cite[Sec. 8.3]{deMedeiros:2010dn}:
In that situation the action is in fact the one
induced from the left action of $\mathrm{SU}(2)$ on $\mathbbm{H}$ via the
following identification:
\begin{equation}
  S^4 = S(\mathbbm{R} \oplus \mathbbm{H}) \simeq S^{\mathbbm{H}}
  \,.
\end{equation}
With this 4-sphere we have of course that there are fixed points on the 4-sphere itself.
This indicates that, contrary to what may have been anticipated, the fixed point locus of
the near-horizon geometry is not empty,
meaning that we did not actually remove the full fixed M-brane form the space-time:
\begin{equation}
  \left(
    \mathrm{AdS}_7 \times S^4
  \right)^G
  \;=\;
  \mathrm{AdS}_7 \times S^0
  \,.
\end{equation}
If we choose a local chart in which AdS-space-time is topologically $\simeq \mathbbm{R}^{5,1} \times \mathbbm{R}_{> 0}$, then we
see that we did remove a 5-brane world-volume $\mathbbm{R}^{5,1}$ at the origin, but that spreading out from this removed locus are
two rays of fixed stratum in the directions $S^0 \subset S^4$ (thinking of $S^4$ as the unit sphere of ``directions'' away from the M5 locus).

This situation becomes clearer/more pronounced as we go to the far horizon limit, because there we get the identifications
shown on the right in the following:
\begin{equation}
  \hspace{-.4cm}
  \raisebox{45pt}{\xymatrix@C=0pt{
    &
    \ar[dr]^{ \ell_P \ll 1 }
    \ar[dl]_{ \ell_P \gg 1 }
    \\
    \mathrm{AdS}_7 \! \times  \! S^4\!\! \sslash \!\! G
    &&
    \mathbbm{R}^{5,1} \! \times \! \mathrm{C}\big( S^4\!\! \sslash \!\! G \big)=
  \!\!\!\!\!\!\!\! \ar@{}[r]
    & \!\!\!\!
    \mathbbm{R}^{5,1} \! \times \! (\mathbbm{R} \oplus \mathbbm{H})\sslash G
    \ar@{=}[d]
    \\
    &&&
    \mathbbm{R}^{6,1} \times \left(  \mathbbm{H}\sslash G \right)
  }
  }
\end{equation}
We see on the right that, in the far horizon limit, what started out seeming to be a black M5-brane ends up being
an MK6-monopole space-time!

But in fact, in other parts of the literature it is well-known that the M5-brane is a ``domain wall'' inside the MK6,
pertinent literature is referenced in ``Table L'' in \cite{Huerta:2018xyh}, see the rows with Examples 2.7 and Examples 2.8,
and see the illustrating graphics in Example 2.7:
\begin{center}
\raisebox{-30pt}{
\includegraphics[width=.4\textwidth]{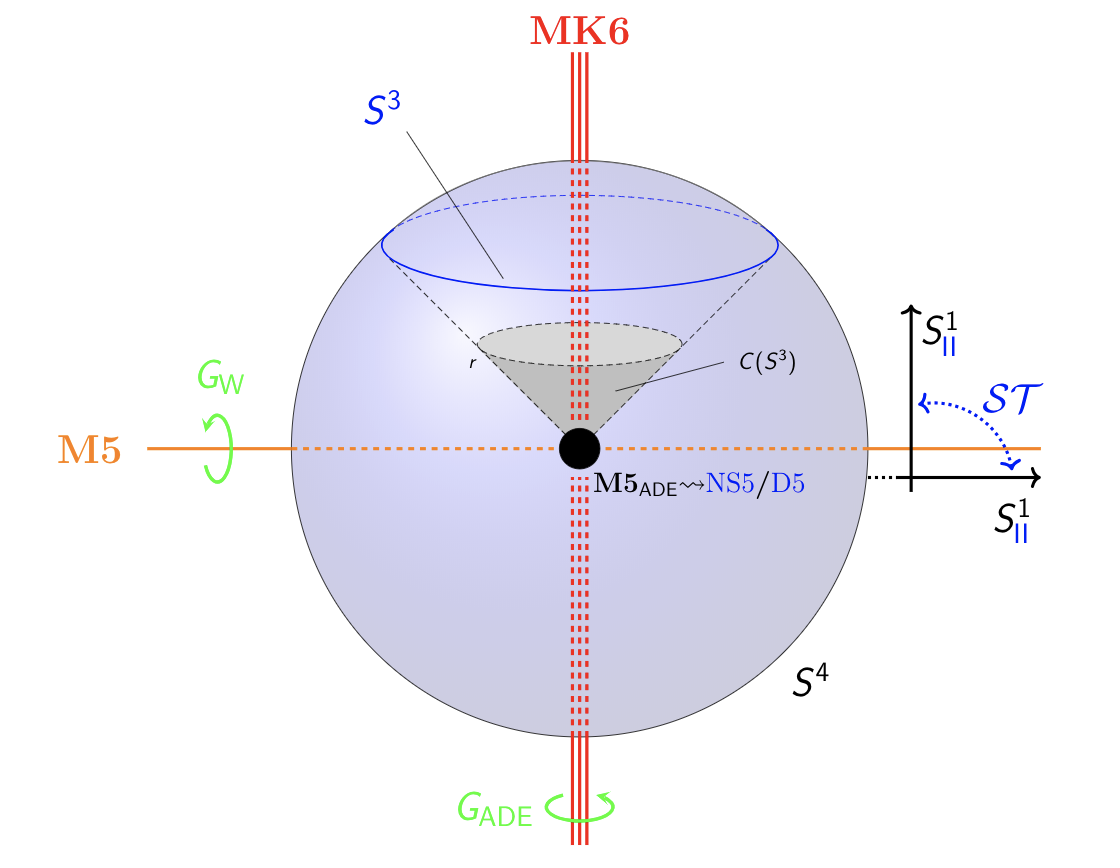}
}
\end{center}
The green label $G_W$ in this graphics shows how the actual M5-brane-locus inside the
MK6-brane locus was mathematically isolated in
\cite{Huerta:2018xyh}: namely the MK5-singularity is intersected there with another singularity,
such that the joint fixed locus is just $\mathbbm{R}^{5,1}$ instead of $\mathbbm{R}^{6,1}$.

This works, but (thinking now of super space-times) the extra intersection also reduces the fermionic space-time directions
by half and hence actually identifies the ``$\mathcal{N} = (1,0)$''-supersymmetric M5-brane (as per the last three rows in the table of Prop. 4.19 in \cite{Huerta:2018xyh})
but not the $\mathcal{N} =(2,0)$-supersymmetric M5-brane.

\smallskip

It remained an open problem how to pick, in a mathematically systematic way the, the codimension-1 sublocus of $\mathbbm{R}^{6,1}$,
which is really codimension-1 also as supermanifolds, hence which does not restrict the fermionic dimensions.

\smallskip
\noindent {\bf Solution in equivariant cohomotopy.}
We close by indicating that this open problem is resolved
if M-brane charge is quantized in equivariant cohomotopy.

First observe the classical fact that, by Pontryagin--Thom theory
plain 4-cohomotopy
(before equivariant enhancement) classifies cobordism classes of
co-dimension 4 submanifolds in space-time \cite{Sati:2013rxa} (see e.g. \cite[Ch. IX]{Kosinski:Differential} for background).
After equivariant enhancement, this statement refines to produce
information about ``hidden degrees of freedom inside singularities'':
it now says that $G$-equivariant cohomotopy classifies
submanifolds \emph{inside the $G$-singularities of space-time}
of co-dimension the dimension of the $G$-fixed points $(S^4)^G$. Now for
$G_{\mathrm{ADE}} \subset \mathrm{SU}(2) \subset_{\mathrm{diag}} \mathrm{Pin}(5)$ a non-Abelian finite subgroup of $\mathrm{SU}(2)$,
we have $\mathrm{dim}\big( \left( S^4\right)^{G_{\mathrm{ADE}}}\big) = 1$
and hence we discover that $G_{\mathrm{ADE}}$-equivariant 4-cohomotopy
classifies codimension-1 submanifolds inside MK6-ADE-singularities
of $D = 11$ space-times.
By the above discussion, this is exactly what is needed for
realizing the M5-brane at ADE-singularities in M-theory.
This and other aspects of the formulation``microscopic M-theory''
suggested by the brane bouquet are discussed in more detail in
\cite{Braunack-Mayer:2019ip}

\bibliography{allbibtex}

\bibliographystyle{prop2015}

\end{document}